\documentclass[a4paper,11pt,twoside,openright]{article}

\usepackage[utf8]{inputenc}                         % Ecrire avec les conventions du français.
\usepackage[dvips,pdftex]{graphicx}%\usepackage{graphicx}
\usepackage[left=2.5cm,top=3cm,right=2.5cm,bottom=3cm]{geometry}    % Modification des marges
\usepackage{amsmath,amsthm,amssymb}                             % Utilisation de certains packages de AMS
\usepackage{fancyhdr}
\usepackage{color}

\usepackage{url}

\title{Obliquity of Mercury: influence of the precession of the pericenter and of tides}
\author{Rose-Marie Baland, Marie Yseboodt, Attilio Rivoldini, Tim Van Hoolst\\
\textit{Royal Observatory of Belgium, Ringlaan 3, B-1180 Brussels, Belgium.}\\Email: Rose-Marie.Baland@oma.be}
\date{March 2017\\ Paper accepted for publication in Icarus}

\begin{document}

   \maketitle

\tableofcontents
\newpage

\begin{abstract}

Mercury is expected to deviate from the classical Cassini state since this state is defined for a uniformly precessing rigid planet. We develop an extended Cassini state model that includes the variations (or nutations) in obliquity and deviation induced by the slow precession of the pericenter. The model also describes the constant shift over time in mean obliquity and deviation associated with the short-periodic tidal deformations of Mercury, characterized by the tidal love number $k_2$ and by the ratio $k_2/Q$ of the tidal Love number over the tidal quality factor, respectively. This model is then used to interpret Mercury's orientation, including the deviation from the classical Cassini state, in terms of parameters of Mercury's interior. 

We determine and solve analytically the angular momentum equation, highlighting the respective roles of the pericenter precession and tidal deformations on the spin precession behavior. We also show explicitly that Peale's equation is sometimes wrongly cited in the literature, resulting in wrong estimates of the polar moment of inertia, and review the importance of many effects that change the determination of the polar moment of inertia from obliquity measurements. 

From the observed orientation of Stark et al. (2015b), we estimate that $C/M_{m}R^2=0.3433\pm 0.0134$, which is $\sim 0.9\%$ smaller than the estimate by Stark et al. (2015b) themselves. That difference is due to our refinements of the Cassini state model ($0.1\%$) and to their wrong use of Peale's equation ($0.8\%$). The difference is smaller than the actual precision ($3-4\%$) on the polar moment of inertia but may be of the order of precision that can be reached with BepiColombo mission ($\leq 0.3\%$).

The parameter $k_2$ cannot be estimated from the spin axis orientation, because of its correlation with the polar moment of inertia, which is much more important in determining the obliquity in our improved model. However, it is necessary to include its effect in the model to avoid a systematic error of $0.3\%$ on the determination of the polar moment of inertia. \textcolor{black}{The parameter $k_2/Q$ can be estimated from the spin orientation, since its effect can be easily separated from the effect of the polar moment of inertia on the deviation, as this latter parameter is already well determined by its contribution to the obliquity. Given the actual precision on the spin axis orientation, we place an upper limit of about $0.02$ on the ratio $k_2/Q$ and of about $350$ on $Q$ (assuming $k_2=0.5$) at the $1\sigma$ level. In the future, the relative precision on the determination of $k_2/Q$ from the spin axis orientation could be as good as $30\%$ with BepiColombo, so that the non-elastic parameter of Mercury could be estimated for the first time.}

\end{abstract}

\newpage

\section{Introduction}
% \linenumbers

Following the radar determination of its unusual 3:2 spin-orbit resonance (Pettengill and Dyce, 1965, Colombo, 1965), it was assumed that Mercury is in an equilibrium state called a \textit{Cassini state}, in which the spin axis and the orbit normal precess together about the normal to the Laplace plane at the same constant rate, corresponding to a period of about $300$ kyr, so that the three axes remain coplanar. In that state, the angular separation, called obliquity, from the orbit normal to the spin axis is expected to remain constant over time (\textcolor{black}{Colombo, 1966, Peale, 1969}). While the measured orientation and obliquity of Mercury, observed during the last decade, tend to confirm those predictions within the limits of measurements precision (Earth-based radar observations, see Margot et al. (2007, 2012); MESSENGER data, see Mazarico et al. (2014), Stark et al. (2015b) and Verma and Margot (2016)), the detection of a possible small deviation with respect to the coplanarity allows to question this usual view of the Cassini state. For instance, Stark et al. (2015b) find a deviation of $1.7$ arcsec, for an obliquity of about $2$ arcmin.

The measurements of the spin axis orientation and obliquity can be used to define an orientation model including a non-zero obliquity, as done in Margot (2009), using the determination of the spin location by Margot et al. (2007). This new orientation model has been adopted by the IAU Working Group on Cartographic Coordinates and Rotational Elements of the Planets and Satellites (Archinal et al. 2011).

More importantly, together with the measured gravity field coefficients $C_{20}$ and $C_{22}$, the measured obliquity determines the value of the polar moment of inertia $C$, which is an essential constraint on Mercury's interior structure (e.g. Dumberry and Rivoldini, 2015). To do so, an accurate model of the Cassini state that connects the polar moment of inertia to the measured orientation of the spin axis is needed. If Mercury's rotation axis deviates from the coplanarity, as may be indicated by the measurements, the classical model and the resulting relation between the obliquity and the polar moment of inertia, best known as \textit{Peale's equation} (Peale et al. 1981), may not be accurate enough. A model able to account for the observed deviation needs to be elaborated and consequences to be analyzed in terms of Mercury's interior structure.

Peale (1974) mentioned (see second paragraph of page 727 of the cited paper) that a small periodic deviation from the coplanarity should arise because of the precession of Mercury's pericenter, meaning that the instantaneous precession rate and obliquity should vary with time. He considered the effect to be negligible and did not investigate it further at that time. Recently, Peale et al. (2014, 2016) have numerically demonstrated the effect of pericenter precession. They have concluded that the maximal deviation induced by the pericenter precession with respect to the coplanarity is considerably smaller than the $1\sigma$ measurement uncertainty ($0.87$ arcsec versus $5$ arcsec), and they have not investigated the impact on the determination of the polar moment of inertia. 

\textcolor{black}{Here, we investigate the effects of the precession of the pericenter and of tides on the Cassini state. In a first step to develop further such a detailed model of the obliquity of Mercury, we assume that the rotational response to gravitational forcing from the Sun can be described as a solid body rotation and do not consider differential rotational motion of the inner core and the outer core with respect to the silicate part of Mercury. Such an approach allows to better clarify the effects of tides and pericenter precession by isolating them from other effects like differential rotation, which may even be more important. }

\textcolor{black}{Peale et al. (2014) have studied the evolution of the spin axis of Mercury's mantle to its current equilibrium state, taking into account a liquid core below the solid crust and mantle. They find that the equilibrium orientation of the mantle spin axis is mainly controlled by the external solar torque and by the internal pressure torque at the CMB (core-mantle boundary) and is in agreement with the equilibrium orientation of the spin axis of a solid Mercury, whereas the orientation of the core spin axis is displaced from the mantle's spin by a few arcmin. A deviation of the mantle spin axis is also possible due to the viscous coupling between mantle and core and ranges from $0.016$ to $0.055$ arcsec, depending on the core viscosity and on the shape of the CMB. If an inner core is present, the gravitational torque between the inner core and the silicate outer shell dominates the pressure torque (Peale et al. 2016), and increases the equilibrium obliquity, depending on the solid inner core size, shape and density. In the most realistic cases (small inner core, with a radius smaller than about one third of the total radius, see e.g. Dumberry and Rivoldini 2015), the polar moment of inertia would currently be overestimated by up to $4\%$, which is the current uncertainty of the polar moment of inertia determined from the measured obliquity. However, given a sufficiently large viscous torque, all the inner core, outer core and silicate mantle would precess together as a solid body because of the friction at the boundaries between the core and the mantle and between the inner and outer core, meaning that the polar moment of inertia could be less overestimated than stated in Peale et al. (2016). } 

Noyelles and Lhotka (2013) have already studied the influence of a complex orbital dynamics and of tides on the obliquity of a solid Mercury. They find that the tides and secular variations of the orbital elements affect the modeled obliquity by $0.03$ arcsec over a revolution period and by $\sim 0.01$ arcsec over an interval of 20 years, respectively. The resulting effects on the determination of the polar moment of inertia, $\sim 0.03\%$ and $0.01\%$, are smaller than the actual uncertainty ($\sim 4\%$).
 
Here, we revisit the influence of the precession of the argument of the pericenter (period of about $130$ kyr) on the spin orientation and obliquity. We demonstrate that not only this effect is about half the observed nominal deviation in Margot et al. (2012) or in Stark et al. (2015b), who provide the two most precise determinations of the spin axis orientation, it also induces a deviation just in the observed direction of deviation, which strongly supports the relevance of adding this effect to the Cassini state model. We also find that the effect of the pericenter precession on the determination of the polar moment of inertia is $\sim 0.1\%$, one order of magnitude larger than the effect found by Noyelles and Lhotka (2013).

Secondly, we reassess the effect of tidal periodic deformations. We show that the effect of the tides on the obliquity is $\sim 0.3\%$, one order of magnitude larger than previously estimated by Noyelles and Lhotka (2013), and has to be taken into account to interpret future precise measurements with the BepiColombo mission. We also demonstrate that the spin is lagging behind the expected coplanar orientation because of the non-elasticity, which might explain part of the observed deviation. \textcolor{black}{This lag is at least one order of magnitude larger than the contribution of viscous coupling at the CMB to the deviation of the mantle spin axis found in Peale et al. (2014)}

Finally, we re-estimate Mercury's polar moment of inertia by fitting recent data to the improved model for the Cassini state that takes into account perihelion precession and tidal deformations and assess the perspectives of spin orientation measurements regarding the determination of tidal parameters (Love number and quality factor). \textcolor{black}{Strictly speaking, the results are valid only if the rotational dynamics leading to Mercury's equilibrium spin axis orientation is similar to that of a solid body, but they will serve as a solid foundation for further examination of the case with fluid outer and solid inner cores.}

The plan of the paper is as follows. In Section 2, we develop a model for the Cassini state of a solid and rigid Mercury undergoing a perihelion precession, based on the angular momentum method. In addition to extend the Cassini state model in the form of a compact analytical solution, this independent derivation help sorting the different and mutually incompatible versions of Peale's equation (derived with the Hamiltonian approach) which coexist in the literature, leading to incorrect estimates of the polar moment of inertia. We address the effect of tidal deformations in Section \ref{Section3}. We refer to a model including a classically neglected effect as an \textit{improved Cassini state}, in opposition to the \textit{classical Cassini state}. In Section 4, we describe the existing observations of the spin orientation. In Section 5, we constrain the polar moment of inertia $C$ and the ratio $k_2/Q$ of the tidal Love number $k_2$ over the tidal quality factor $Q$, from recent data and using an improved Cassini state model. In Section 6, we compare our new Cassini state model and our results with the literature. We present a discussion and concluding remarks in Section 7.

% \nolinenumbers

\section{The Cassini state of a rigid Mercury}
\label{Section2}

\subsection{From the Hamiltonian formalism towards the angular momentum formalism}
\label{Section21}

The classical relation between the polar moment of inertia $C$ and the equilibrium obliquity $\varepsilon$ was established from the study of the Hamiltonian of a solid Mercury rotating on itself and orbiting around the Sun, while locked in the Cassini state. Following Eq. (12) of Yseboodt and Margot (2006), this relation can be expressed as
\begin{equation}\label{CCS}
\varepsilon=\frac{-C\dot\Omega \sin i}{C\dot\Omega \cos i+2nM_{m}R^2 G_{201}(e) C_{22}-nM_{m}R^2G_{210}(e) C_{20}}
\end{equation}
where 
\begin{eqnarray}
 G_{210}(e)&=&\left(1-e^2\right)^{-3/2}\\
 G_{201}(e)&=&\frac{7}{2}e-\frac{123}{16}e^3+\frac{489}{128}e^{5}+\mathcal{O}(e^7)
\end{eqnarray}
are eccentricity functions defined by Kaula (1966). $e$ is the orbital eccentricity and $n$ the mean motion. $\Omega$ and $i$ are the longitude of the ascending node and orbital inclination with respect to the Laplace plane. The Laplace plane is the plane that minimizes the variation in orbital inclination. $M_{m}$ and $R$ are the mass and the mean radius of Mercury. \textcolor{black}{The derivation of the numerical values we use for the orbital parameters present in Eq. (\ref{CCS}) and in the rest of this paper is described in Appendix \ref{AppParamOrb}. They are also listed in Tab. \ref{tab0}.} 

\begin{figure}[!htb]
\begin{center}
\includegraphics[width=8cm]{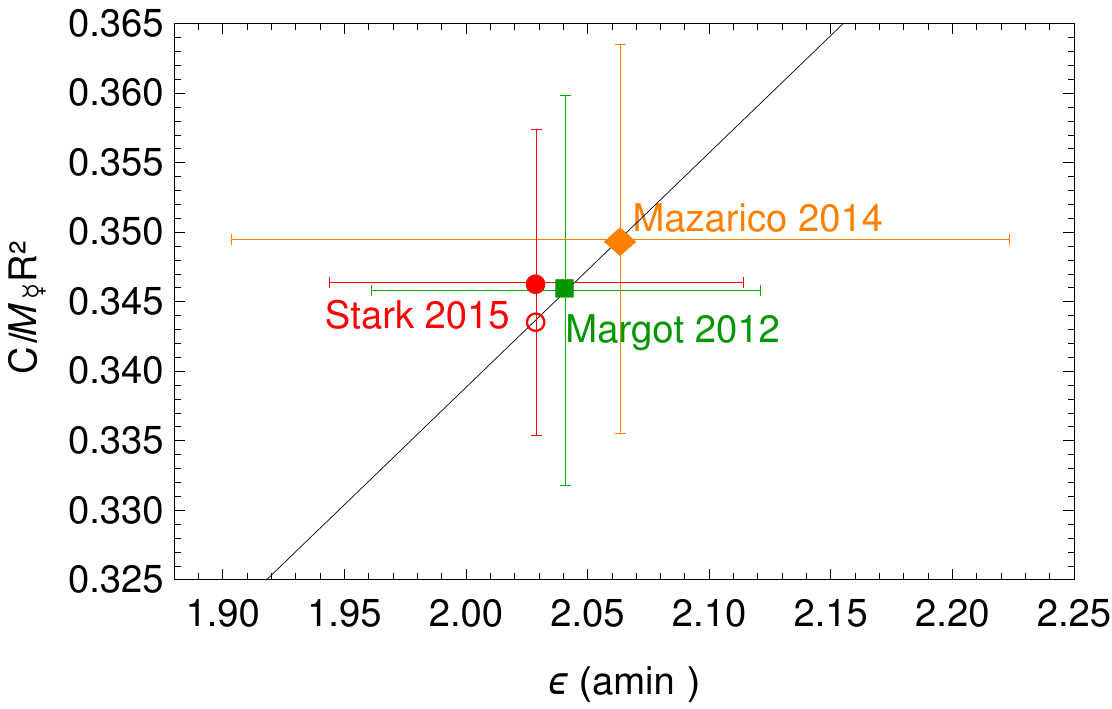}
\caption{\label{Fig3}Normalized polar moment of inertia as a function of the equilibrium obliquity. The solid line illustrates the quasi linear relation between $C/M_{m}R^2$ and $\varepsilon$, resulting from Eq. (\ref{CCS}) for the classical Cassini state model. The colored filled markers and associated errors bars represent the estimates of $C/M_{m}R^2$ reported by Margot et al. (2012), Mazarico et al. (2014) and Stark et al. (2015b), using the classical Cassini state model and their respective estimates of the obliquity $\varepsilon$. Stark's value for the polar moment of inertia must be corrected for their improper use of Peale's equation (see details in Section \ref{section5}), so that it becomes consistent with the solid line (see empty circle marker). }
\end{center}
\end{figure} 

Because of the long period of the orbital node precession ($\sim 326$ kyr), the term $C\dot\Omega \cos i$ in the denominator is smaller than the terms proportional to the gravity field coefficients $C_{20}$ and $C_{22}$, so that $C$ increases almost linearly with the obliquity $\varepsilon$ (see Fig. \ref{Fig3}). Since the largest source of uncertainty in estimating $C$ comes from the uncertainty on the obliquity measurements (Milani et al. 2001, Pfyffer et al. 2011, Margot et al. 2012), and since the uncertainty on the obliquity reported by Mazarico et al. (2014), $0.16$ arcmin, is about twice those reported by Margot et al. (2012) and Stark et al. (2015b), the uncertainty on the polar moment of inertia reported by Mazarico ($0.014$), very similar to those reported by Margot and Stark, might be underestimated by about a factor $2$. 

Equation (12) of Yseboodt and Margot (2006) is presented as the solution of Eq. (4) of Peale (1981), expanded to the first order in obliquity. Note that in Yseboodt and Margot (2006), $G_{201}(e)$ is truncated at third order in $e$. This truncation introduces a systematic error of $-0.035\%$ on the determination of $C/M_{m}R^2$ from a measured obliquity of about $2$ arcmin, two orders of magnitude below the current precision ($\sim4\%$). The first order approximation in the development of $\varepsilon$ leads to an insignificant systematic error of $+4\times 10^{-6}\%$ on $C$. The sources of systematic errors on the determination of $C/M_{m}R^2$ identified throughout this study are summarized in Table \ref{tab4}.

In the remainder of this section, we derive an expression for the obliquity of a rigid Mercury, subject to the gravitational torque exerted by the Sun, by using the angular momentum approach. We assess the effect of the pericenter precession on the analytical solution for the orientation of the spin and assume that Mercury behaves rotationally as one block.

Classically, the plane containing the Laplace and orbit normals (the \textit{Cassini plane}) contains all possible solutions for the orientation of the spin axis in the Cassini state, each solution corresponding to a possible couple of values for $C$ and $\varepsilon$. Therefore, the concepts of \textit{coplanarity} and of \textit{Cassini state} are usually assimilated to each other, which is not the case anymore with the \textit{improved Cassini state model}. We show that the nutations triggered by the pericenter precession can explain in large part the deviation observed e.g. by Stark et al. (2015b).

\subsection{Angular momentum equation}
\label{Section22}

The following demonstration consists in an adaptation of the demonstration given in Appendix A of Baland et al. (2012) for a rigid solid synchronous satellite. The angular momentum equation is written in an inertial reference frame as
\begin{equation}\label{A11}
\frac{d \vec L}{dt}=\vec \Gamma
\end{equation}
with $\vec L$ the angular momentum of Mercury and $\vec\Gamma$ the torque exerted on Mercury by the Sun. We will show that the torque can be defined as the sum of a main term driving the main spin precession, $\vec\Gamma^{prec}$, and a secondary part driving a small nutation about the main precession, $\vec\Gamma^{nut}$. Neglecting the secondary torque would amount to retrieve the classical Cassini state, its coplanarity, and its constant obliquity. 

\begin{figure}[!htb]
\begin{center}
\includegraphics[width=10cm]{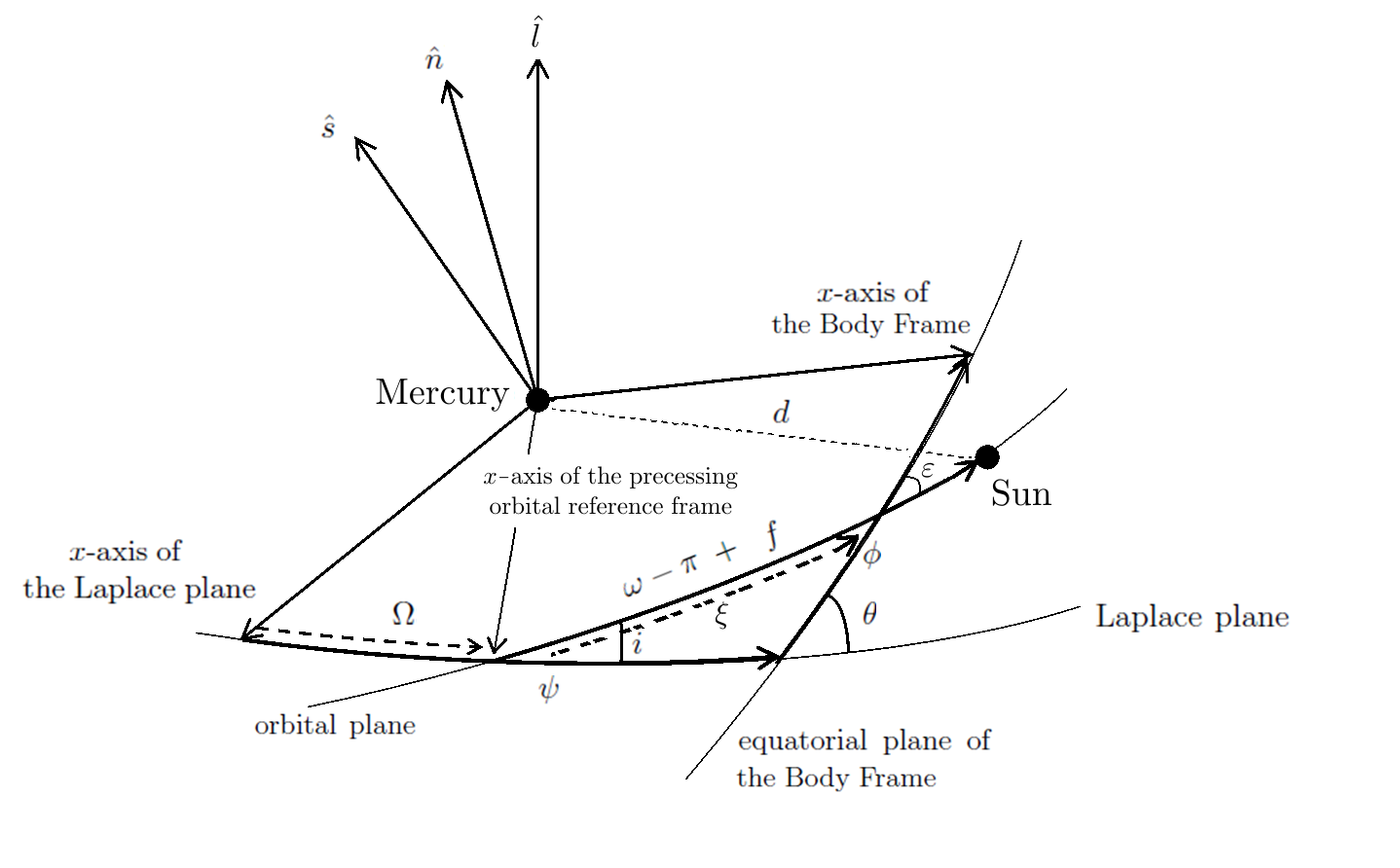}
\caption{\label{appendixfig}Orientation of the Body Frame of Mercury with respect to its orbital plane and to its Laplace plane. \textcolor{black}{The x-axis of the Laplace plane is set here as the ascending node of the Laplace plane on the equator of the ICRF.} $\Omega$, $i$, and $f$ are the longitude of the orbital ascending node, inclination, and true anomaly with respect to the Laplace plane. $\psi$, $\theta$, and $\phi$ are the longitude of the ascending node of the equatorial plane, inertial obliquity, and rotation angle. $\varepsilon$ is the obliquity. $d$ is the distance to the Sun. The unit vectors along the spin, orbit, and Laplace poles are denoted $\hat s$, $\hat n$, and $\hat l$, respectively. $\xi$ is the node of the equatorial plane with respect to the precessing orbital plane. }
\end{center}
\end{figure}

The inertial reference frame is here taken to be a frame attached to the Laplace plane and centered at the center of mass of Mercury \textcolor{black}{(see Fig. \ref{appendixfig}). Note that we set the x-axis of the Laplace plane as the ascending node of the Laplace plane on the equator of the ICRF (International Celestial Reference Frame). This is a natural choice since the orientation of the Laplace plane, as well as spin orientation, are usually given as right ascension and declination with respect to the ICRF (see Section \ref{Sec42} and Appendix \ref{AppParamOrb}.)}

The expression of the torque in the frame attached to the principal axes of inertia of Mercury (Body Frame) is (Murray and Dermott 1999, Eqs. 5.43-5.45):
\begin{equation}\label{TorqueBF}
\vec \Gamma_{BF}=\left(
\begin{array}{c}
 3 n^2 a^3 (C-B) Y Z /d^5 \\
 3 n^2 a^3 (A-C) Z X /d^5 \\
 3 n^2 a^3 (B-A) X Y /d^5
\end{array}\right)
\end{equation}
with $A<B<C$ the principal moments of inertia of Mercury, $n$ its mean motion, $a$ its semi-major axis, $d$ the distance between Mercury and the Sun, and $(X,Y,Z)$ the position of the Sun in the Body Frame of Mercury in Cartesian coordinates. 

If we note $(\psi,\theta,\phi)$ the Euler angles orienting the Body Frame with respect to the inertial frame (see Fig. \ref{appendixfig}), the position of the Sun in the Body Frame is 
\begin{equation} \label{sunposition}
\left(
\begin{array}{c}
 X \\
 Y \\
 Z
\end{array}\right)=R_z(\phi).R_x(\theta).R_z(\psi-\Omega).R_x(-i).R_z(-(\omega-\pi)-f)\left(
\begin{array}{c}
d \\
 0 \\
 0
\end{array}\right)
\end{equation}
with the rotation matrices defined as
\begin{equation}\label{RxRz}
R_x(\theta) = \left(
\begin{array}{ccc}
 1 & 0 & 0 \\
 0 & \cos{\theta} & \sin{\theta} \\
 0 & -\sin{\theta} & \cos{\theta}
\end{array}
\right), \, R_z(\theta) = \left(
\begin{array}{ccc}
 \cos{\theta} & \sin{\theta} & 0 \\
 -\sin{\theta} & \cos{\theta} & 0 \\
 0 & 0 & 1
\end{array}
\right).
\end{equation}
$\Omega$, $i$, and $f$ are the longitude of the ascending node, inclination, and true anomaly of Mercury with respect to the Laplace plane. \textcolor{black}{As explained in Appendix \ref{AppParamOrb}, we use $\Omega(t)=23.73^\circ-0.1106^\circ\, T$, with $T$ the time interval in Julian centuries from J2000, and $i=8.53^\circ$. The period of the ascending node precession, $|2\pi/ \dot \Omega|$, is about $326$ kyr.} The angle $\theta$ can be seen as an \textit{inertial obliquity}, while the angle $\varepsilon$ is the \textit{orbital obliquity}. Unless specified otherwise, \textit{obliquity} in this paper refers to the orbital obliquity $\varepsilon$. The angle $\psi$ is the ascending node longitude of the equatorial plane over the Laplace plane, and $\phi$ is the rotation angle of the planet. 

\textcolor{black}{The angle $\omega$, the argument of the pericenter of Mercury around the Sun measured from the intersection of the orbital plane and the Laplace plane (the pericenter of the Sun seen as in orbit around the planet is then $\omega-\pi$), can be expressed as (see Appendix \ref{AppParamOrb})
\begin{equation} \label{eq8}
\omega(t) = 50.38 ^\circ + 0.2689^\circ \, T.
\end{equation}
The trend of $\omega$ is about $134$ kyr. Note that the sign of the perihelion precession rate $\dot \omega$ is positive, while the sign of the node precession rate $\dot\Omega$ is negative.} 

On the long timescales considered here, we neglect the wobble (the motion of the polar axis with respect to the spin axis), so that the unit vectors along the spin axis ($\hat s$) and the orbit normal ($\hat n$) can be expressed in the Laplace reference frame as
\begin{equation}
\label{hats}\hat s=\left(
\begin{array}{c}
 \sin{\theta} \cos{(\psi-\pi/2)} \\
 \sin{\theta} \sin{(\psi-\pi/2)} \\
\cos{\theta}
\end{array}\right)
\end{equation}
and
\begin{equation}
\label{hatn}\hat n=\left(
\begin{array}{c}
 \sin{i} \cos{(\Omega-\pi/2)} \\
 \sin{i} \sin{(\Omega-\pi/2)} \\
\cos{i}
\end{array}\right).
\end{equation}

Since the eccentricity of Mercury is large ($e=0.2056$), we expand $f$ and $d$ correct up to (at least) the third order in orbital eccentricity $e$. We have (Murray and Dermott 1999, Eq. 2.88 and Eq. 2.104)
\begin{eqnarray}
\label{f} f&=&M-\frac{1}{4} e \left(e^2-8\right) \sin M+\frac{5}{4} e^2 \sin 2M+\frac{13}{12} e^3 \sin 3M+\mathcal{O}(e^4)\\
\label{d} d&=&\frac{1}{8} a \left(8+4 e^2+e \left(3 e^2-8\right) \cos M-4 e^2 \cos 2M-3 e^3 \cos 3M\right)+\mathcal{O}(e^4)
\end{eqnarray}
with $M$ the mean anomaly. 

The next step in deriving an explicit expression for the torque $\vec\Gamma_{BF}$ is the use the spin-orbit resonance to express the rotation angle $\phi$ as a function of the other angles. In order to keep the analytical treatment efficient, we consider approximations that are different for computation of the main torque $\vec\Gamma^{prec}$ than for the secondary torque $\vec\Gamma^{nut}$. Coplanarity is assumed for the computation of $\vec\Gamma^{prec}$, but not for $\vec\Gamma^{nut}$. Conversely, the small angles approximation is considered to compute $\vec\Gamma^{nut}$, but not for $\vec\Gamma^{prec}$.

\subsubsection{Main precessional torque}
\label{mainprectorque}

Contrary to the Galilean satellites, the inclination of Mercury is not very small ($i=8.53^\circ$), so that the small angle approximation in $i$ and $\theta$ made in Baland et al. (2012) to simplify the torque expression is not appropriate here. We consider $\psi=\Omega=\Omega_0+\dot\Omega t$ in Eq. (\ref{sunposition}), with $\dot\Omega$ the orbital node longitude variation rate and $\Omega_0$ the orbital node longitude at J2000, to get a compact expression for the main torque in the Body Frame, $\vec\Gamma^{prec}_{BF}$, computed with Eq. (\ref{TorqueBF}). This assumption, which has not been made in Baland et al. (2012), expresses here that the main torque leads to the classical Cassini state where the spin axis and orbit normal are coplanar with the normal to the Laplace plane, while the equatorial and orbital ascending nodes on the Laplace plane are aligned. As a result, the angle $\phi$ is measured in the equatorial plane from the same point (the aligned orbital and equatorial nodes) as the angle $(\omega-\pi+f)$ is measured in the orbital plane (see Fig. \ref{appendixfig}). Since the equilibrium obliquity $\varepsilon$ is expected to be very small, we can write (e.g. Peale 1969) 
\begin{equation}\label{phi1}
 \phi\simeq \omega-\pi+\frac{3}{2}M,
\end{equation}
correct up to the first order in $\varepsilon$, and neglecting the small longitudinal librations (the variations of the rotation rate). The factor $\frac{3}{2}$ in front of the mean anomaly $M$ arises because of the 3:2 resonance.

The expression of the torque $\vec \Gamma^{prec}_{BF}$ is straightforward to compute, but too long to be reproduced here. The torque can further be expressed in the inertial frame with the appropriate rotations:
\begin{equation}
 \vec \Gamma^{prec}=R_z(-\psi).R_x(-\theta).R_z(-\phi).\vec \Gamma_{BF}^{prec}.
\end{equation}
Finally, the torque is averaged over an orbit period with the slowly varying $\Omega$ held constant. It is also necessary to average over the precessing angle $\omega$, to get rid of any small nutation induced by the pericenter precession that would not be consistent with the aligned nodes assumption. Thanks to the averaging, the torque can now be written in the following compact form:
\begin{eqnarray}\label{TIN}
\nonumber \vec \Gamma^{prec}&=&\frac{3}{2} M_{m}R^2n^2 \left\lbrace- C_{20} G_{210}(e)(\hat s.\hat n)+\right.\\
&&\left.2 C_{22} G_{201}(e)\left(\frac{1+\hat s.\hat n}{2}\right)\right\rbrace (\hat s \wedge \hat n)
\end{eqnarray}
with
\begin{equation}
 C_{20}=-\frac{C-\frac{A+B}{2}}{M_{m}R^2}\quad \textrm{and}\quad C_{22}=\frac{B-A}{4M_{m}R^2}.
\end{equation}
The torque $\vec \Gamma^{prec}$ leads to the expression for a constant equilibrium obliquity over time, as we will see later. 

\subsubsection{Secondary nutational torque}

Because of the small nutations induced by the pericenter precession, $\psi$ is different from $\Omega$, and the rotation angle is expressed as
\begin{equation}
 \phi\simeq -\psi+\Omega+\omega-\pi+\frac{3}{2}M,
\end{equation}
instead of expression (\ref{phi1}) used for the precessional torque. 

In contrast to what we have done for the precessional torque, and to be able to find a compact form for the secondary nutational torque, we now consider in Eq. (\ref{sunposition}) the small angle approximation on $i$ and $\theta$. We also truncate the true anomaly $f$ and the distance $d$ to the third order in $e$. Since the secondary torque is small compared to the main torque, these approximations do not significantly affect the accuracy of the final solution but significantly simplify the torque's expression. As previously, the torque is transformed to the inertial reference frame, before being averaged over an orbital period with the slow angles $\Omega$ and $\omega$ held constant. Note that according to this procedure, we find a torque which is the sum of the main torque (correct only up to the first order in $i$ and $\theta$, though) and of the secondary torque we are looking for.

By introducing the unit vector along the direction of the pericenter:
\begin{eqnarray}
 \nonumber \hat p&=&\left(\begin{array}{c}
 p_x \\
 p_y \\
 p_z
\end{array}\right)=R_z(-\Omega).R_x(-i).R_z(-(\omega-\pi))\left(\begin{array}{c}
 1 \\
 0 \\
 0
\end{array}\right)\\
&\simeq&\left(\begin{array}{c}
 -\cos (\Omega+\omega) \\
 -\sin (\Omega+\omega) \\
 -i \sin \omega
\end{array}\right)
\end{eqnarray}
the secondary torque $\vec \Gamma^{nut}$ is expressed in the inertial reference frame as
\begin{eqnarray} 
 \nonumber \vec \Gamma^{nut}&=&\frac{3}{16} n^2 M_{m} R^2\, 53 C_{22} e^3 \\
&&
\label{LIN2}\left(
\begin{array}{c}
 \left(p_x^2-p_y^2\right)(n_y-s_y)+ \,2 p_x p_y(s_x-n_x)\\
 \left(p_x^2-p_y^2\right)(n_x-s_x)+ 2 p_x p_y (n_y-s_y) \\
 0 \\
\end{array}
\right)
\end{eqnarray}
It depends on $C_{22}$ and on the eccentricity $e$, but only at third order. We will see later that it is responsible for time-variable slight modifications of the equilibrium obliquity (which may be called latitudinal librations or nutations in obliquity). 

\subsection{Solving the angular momentum equation}

Neglecting the small wobble and longitudinal librations, the spin angular momentum in the Body Frame is given by
\begin{eqnarray}
\label{LBF}\vec L_{BF}&=&\left(
\begin{array}{c}
 0 \\
 0 \\
 C (\dot\phi+\dot\psi \cos{\theta}) 
\end{array}\right)
\end{eqnarray}
where $\dot\phi+\dot\psi \cos{\theta}=\frac{3}{2}n-\dot\psi+\dot\Omega+\dot \omega+\dot\psi \cos{\theta}$. In the inertial frame, the spin angular momentum can be written as
\begin{eqnarray}\label{LIN}
\vec L&=& \left(\frac{3}{2}n-\dot\psi+\dot\Omega+\dot \omega+\dot\psi \cos{\theta}\right) \, C\, \hat s.
\end{eqnarray}
with $\hat s$ given by Eq. (\ref{hats}). 

Including Eqs. (\ref{TIN}), (\ref{LIN2}) and (\ref{LIN}), the angular momentum equation becomes 
 \begin{eqnarray}\label{AFinal}
\nonumber \tilde n C \frac{d \hat s}{dt}&=& n\left\lbrace\kappa_{20} (\hat s.\hat n)+\kappa_{22}\left(\frac{1+\hat s.\hat n}{2}\right) \right\rbrace(\hat s \wedge \hat n)\\
 && +n\kappa_{\omega} \left(
\begin{array}{c}
 \left(p_x^2-p_y^2\right)(n_y-s_y)+ \,2 p_x p_y(s_x-n_x)\\
 \left(p_x^2-p_y^2\right)(n_x-s_x)+ 2 p_x p_y (n_y-s_y) \\
 0 \\
\end{array}
\right)
\end{eqnarray}
with 
\begin{eqnarray}
 && \tilde n=n-\frac{2}{3}\dot\psi+\frac{2}{3}\dot\Omega+\frac{2}{3}\dot\omega+\frac{2}{3}\dot\psi\cos \theta\\
 && \kappa_{20}=- M_{m}R^2 n C_{20} G_{210}(e)\\
 && \kappa_{22}= 2 M_{m}R^2 nC_{22} G_{201}(e)\\ 
\label{kappaomega} && \kappa_{\omega}=\frac{ 53}{8}M_{m}R^2\,nC_{22}e^3.
 \end{eqnarray}
 
We have compare a numerical integration of our angular momentum equation (\ref{AFinal}) with a numerical integration of the angular momentum equation of Peale et al. (2014), to assess the possible loss of accuracy resulting from the two-step process used to derive our equation (see Appendix \ref{App3}). We find a difference of maximum $0.01$ arcsec in $s_x$ and $s_y$, and of $0.0018$ arcsec ($0.0015\%$) and of $0.00027$ ($0.03\%$) arcsec in obliquity and deviation, respectively, around J2000. This confirms that the successive assumptions made to obtain our angular momentum equation are adequate. 
 
We now aim to derive an analytical solution, which can be used instead of Peale's equation to interpret rotation measurements with a least squares inversion. An analytical solution also helps to understand how the precession of the pericenter affects the precession of the spin. We use a perturbative approach, with $\kappa_{\omega}$, the strength of the secondary torque, as the perturbative parameter. This approach is valid because $\kappa_{\omega}$ is more than two orders of magnitude smaller than $\kappa_{20}$ and $\kappa_{22}$. We determine the precession amplitude with the order zero approximation and the nutation amplitude with the first order approximation. 

\subsubsection{Main precession amplitude (order zero in $\kappa_{\omega}$)}
\label{appApoint4}

For the first step of the resolution of Eq. (\ref{AFinal}), we set $\kappa_{\omega}=0$. Since we are just left with the main precession torque, we also set that $\psi=\Omega$, $\theta=i+\varepsilon_{\Omega}$, and $\frac{d \theta}{dt}=\frac{d i}{dt}=0$, so that $\tilde n=n+\frac{2}{3}\dot\omega+\frac{2}{3}\dot\Omega\cos (i+\varepsilon_{\Omega})$. Here, $\varepsilon_{\Omega}$ is the constant amplitude over time associated with the orbital node precession angle $\Omega$. The angular momentum equation reduces to
\begin{equation}\label{step1}
\tilde n C \frac{d \hat s_\Omega}{dt}=n \left(\kappa_{20} (\hat s_\Omega.\hat n)+\kappa_{22}\left(\frac{1+\hat s_\Omega.\hat n}{2}\right) \right)(\hat s_\Omega \wedge \hat n)
\end{equation}
for the spin vector $\hat s_{\Omega}$ in the case of precession only. 

By substituting Eq. (\ref{hats}) in the explicit form of
\begin{eqnarray}
\label{s0}\hat s_\Omega&=&\left(
\begin{array}{c}
 \sin(i+\varepsilon_{\Omega})\cos{(\Omega-\pi/2)} \\
 \sin(i+\varepsilon_{\Omega})\sin{(\Omega-\pi/2)} \\
 \cos(i+\varepsilon_{\Omega})
\end{array}\right)
\end{eqnarray}
into Eq. (\ref{step1}), we have that 
\begin{eqnarray}
\nonumber &&-\tilde n C\dot\Omega \sin(i+\varepsilon_{\Omega})=\\
\label{peale} &&n\left\lbrace \kappa_{20}\cos\varepsilon_{\Omega}+\kappa_{22}\left(\frac{1+\cos\varepsilon_{\Omega}}{2}\right)\right\rbrace \sin \varepsilon_{\Omega}.
\end{eqnarray}
Both sides of Eq (\ref{peale}) can be further expanded at first order in $\varepsilon_{\Omega}$ (as we have seen in Section \ref{Section21}, this development leads to an insignificant systematic error on the determination of $C/M_{m}R^2$), and we find that
\begin{eqnarray}
 \label{eq26}\varepsilon_{\Omega}&=&-\frac{ \left(n+\frac{2}{3}\dot\omega+\frac{2}{3}\dot \Omega\cos i\right) C\,\dot\Omega \sin i}{n \kappa+ C\,\dot\Omega \left[\left(n+\frac{2}{3}\dot\omega\right)\cos i +\frac{2}{3}\dot\Omega \cos 2 i\right]}
\end{eqnarray}
with 
\begin{equation}
\kappa=\kappa_{20}+\kappa_{22}.
\end{equation}

Because the precession period of the node and of the pericenter are about six orders of magnitude larger than the orbital period, we neglect $\dot \Omega$ and $\dot \omega$ in front of $n$ (and so $\tilde n=n$), so that 
\begin{eqnarray}\label{Yseboodt}
 \varepsilon_{\Omega}&\simeq& -\frac{ C\,\dot\Omega \sin i}{\kappa+C\,\dot\Omega\,\cos i }
\end{eqnarray}
The consequence of this approximation on the determination of $C/M_{m}R^2$ from a measured obliquity of about $2$ arcmin is a systematic error of $+0.00007\%$. 

Equation (\ref{Yseboodt}) is the same as Eq. (\ref{CCS}) obtained in a Hamiltonian approach, where we have just changed the notation for the obliquity (now $\varepsilon_{\Omega}$) to emphasize the fact that, once the nutations is included, this angle is no longer the separation between the orbit normal and the spin axis, but the amplitude of the main precession. This equality confirms the correctness of Eq. \textcolor{black}{(\ref{CCS})} and shows that several conflicting versions in the literature are wrong. For instance, the estimate of $C/M_{m}R^2$ by Stark et al. (2015b) must be revised from $0.346$ to $0.344$ (see Section \ref{starkerror} for more details).

In the limit of small inclination (which, remember, is not a satisfactory approximation for Mercury), Eq. (\ref{Yseboodt}) becomes
\begin{eqnarray}
\label{wf}\varepsilon_{\Omega}\simeq -\frac{ i\, \dot\Omega }{\omega_f+\dot\Omega}
\end{eqnarray}
which is the classical form of a response to a forcing, where the denominator can become very small, and the response very large, if the forcing period is close to the free period. Here, the denominator is the sum of the positive free precession frequency $\omega_f=\kappa/C$ and of the negative forcing precession frequency $\dot\Omega$. The free precession period, $T_f=2\pi/\omega_f$, is between 1200 and 1350 years for $C/M_{m}R^2$ ranging between 0.32 and 0.36, too far from the forcing period ($326$ kyr) to trigger a resonant excitation of the forced precession.

\subsubsection{Nutation amplitude (first order in $\kappa_{\omega}$)}

For the second step of the resolution, we replace $\hat s$ by $\hat s=\hat s_\Omega+\hat s_{\omega}$ in Eq. (\ref{AFinal}), develop at first order in both small quantities $\hat s_{\omega}$ and $\kappa_{\omega}$, subtract Eq. (\ref{step1}), and set $\tilde n=n$ to find the remaining equation 
 \begin{eqnarray}\label{step2}
C \frac{d\hat s_{\omega}}{dt}&=&\kappa (\hat s_{\omega} \wedge \hat n)\\
&&+ \kappa_{\omega} \left(
\begin{array}{c}
 \left(p_x^2-p_y^2\right)(n_y-s_{\Omega,y})+ \,2 p_x p_y(s_{\Omega,x}-n_x)\\
 \left(p_x^2-p_y^2\right)(n_x-s_{\Omega,x})+ 2 p_x p_y (n_y-s_{\Omega,y}) \\
 0 \\
\end{array}
\right)
\end{eqnarray}
which, neglecting the small variations ($\propto i\, \varepsilon_{\omega}$) in the third component, has for solution 
\begin{eqnarray}\label{Eq29}
\hat s_{\omega}&=&\left(
\begin{array}{c}
 \varepsilon_{\omega}\cos{(2\omega+\Omega-\pi/2)} \\
 \varepsilon_{\omega}\sin{(2\omega+\Omega-\pi/2)} \\
 0
\end{array}\right)
\end{eqnarray}
with 
\begin{equation}
 \label{eomega} \varepsilon_{\omega}=\frac{\varepsilon_{\Omega}\kappa_{\omega}}{\kappa+C(2\dot \omega+\dot\Omega)}.
\end{equation}

\begin{figure}[!htb]
\begin{center}
\includegraphics[width=6cm]{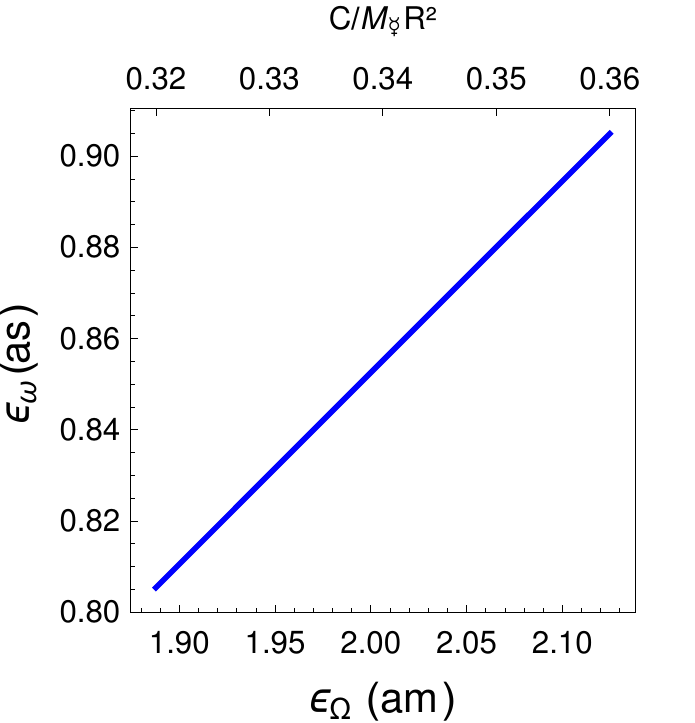}
\caption{\label{Fig5}Nutation amplitude $\varepsilon_{\omega}$ as a function of the precession amplitude $\varepsilon_{\Omega}$, for $C/M_{m}R^2$ ranging from $0.32$ on the left, to $0.36$ on the right.}
\end{center}
\end{figure} 

The amplitude $\varepsilon_{\omega}$ of the spin nutation is proportional to the amplitude $\varepsilon_{\Omega}$ of the main precession and to the torque strength $\kappa_{\omega}$. For $\varepsilon_{\Omega}\simeq 2$ arcmin, $\varepsilon_{\omega}\simeq 0.85$ arcsec (see Fig. \ref{Fig5}). 

\subsubsection{Full solution, obliquity and deviation}
\label{Section233}

For each value of $C$, there exists only one possible value for each amplitude $\varepsilon_\Omega$ and $\varepsilon_\omega$, and so only one solution
\begin{equation}\label{full}
\hat s=(s_x,s_y,s_z)=\hat s_\Omega+\hat s_{\omega}
\end{equation}
for the spin axis orientation at a given time. As a result of the perturbative approach, $\hat s$ is not exactly a unit vector ($|\hat s_\Omega|=1$ and $|\hat s_{\omega}|<<1$). We replace $s_z$ by $\sqrt{1-s_x^2-s_y^2}$, to restore the unit norm of the spin vector, adding small time variations to the third component of the spin vector. We have compared our analytical solution with a numerical integration of the angular momentum Eq. (\ref{AFinal}) to assess the accuracy of the perturbative approach to derive the analytical solution. We find that the analytical solution is $0.008\%$ and $0.03\%$ accurate in obliquity and deviation, respectively, about J2000 (see Appendix \ref{App3}). 

\begin{figure}[!htb]
\begin{center}
\includegraphics[width=7cm]{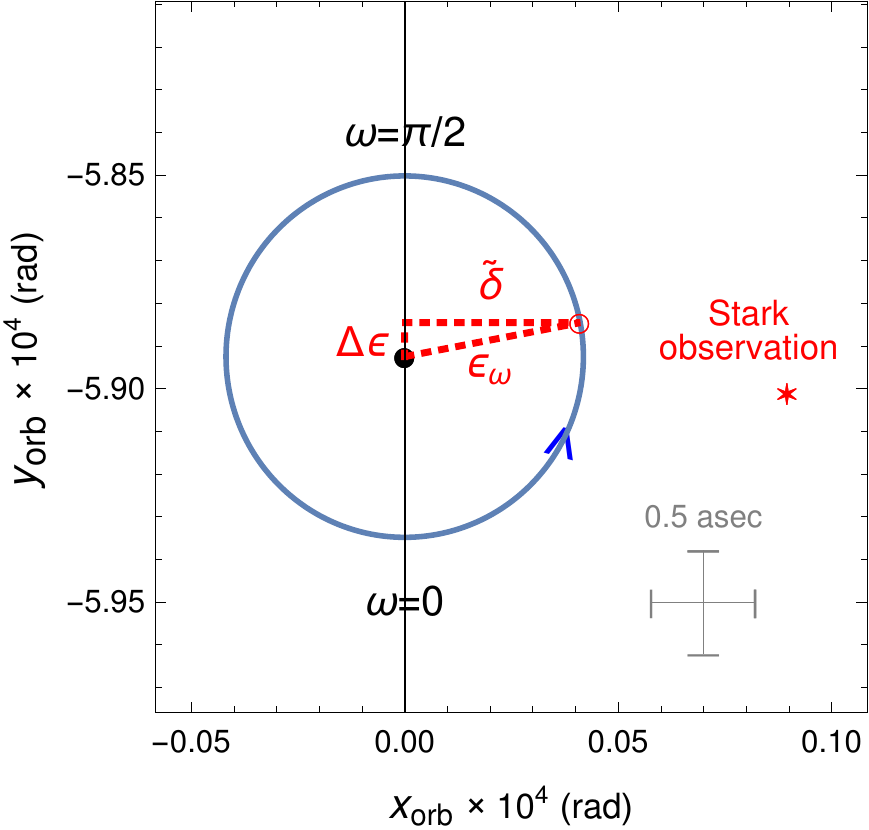}
\caption{\label{nut}Motion of the spin axis of a solid and rigid Mercury, projected onto the precessing orbital plane. The $x_{orb}$-axis points to the node of the orbit on the Laplace plane (see Fig. \ref{appendixfig}), while the $y_{orb}$-axis (black vertical line) corresponds to the projection of the classical Cassini state. The black dot is the constant position of the classical Cassini state for the obliquity amplitude $\varepsilon_{\Omega}$ given in Table \ref{Tab2}. The blue curve is the nutation about this mean position, driven by the perihelion precession, with the amplitude $\varepsilon_{\omega}$ also given in Table \ref{Tab2}. The arrow indicates the direction of the nutational motion. When $\omega=0$ or $\pi/2$, the projected spin axis lies on the $y$-axis, below or above the mean position. The red star represents the spin orientation of Stark et al. (2015b) at the measurement epoch (about 13 years after J2000). The deviation $\tilde \delta$ and the variation in obliquity $\Delta \varepsilon$ with respect to the mean obliquity, at that epoch, are well approximated by the sine and cosine of the projected motion (red circle), with respect to the $y$-axis. At that epoch, the deviation is almost maximal, while the variation in obliquity is almost minimal. The deviation due to the nutation explains about half the observed deviation. }
\end{center}
\end{figure}

The period of the spin nutation $\hat s_{\omega}$ of Eq. (\ref{Eq29}), $2\pi/(2\dot\omega+\dot\Omega)$, is of the order of $84$ kyr (remember that the sign of $\dot \Omega$ is negative, whereas the sign of $\dot\omega$ is positive), while the period of the spin precession $\hat s_\Omega$ of Eq. (\ref{s0}) is $326$ kyr. We can isolate the nutation signal associated with $\dot\omega$ by transforming the full solution $\hat s$ of Eq. (\ref{full}) to a precessing orbital reference frame with the successive rotations $R_{z}(\Omega)$ and $R_{x}(i)$ (see Eq. \ref{RxRz}). In that reference frame, the spin axis nutates about a fixed orientation with the amplitude $\varepsilon_{\omega}$, but with a frequency of $2\dot\omega$ corresponding to a period of about $67$ kyr (see Fig. \ref{nut}). 

Since the time-variable obliquity $\varepsilon$ 
\begin{equation}\label{epsexact}
\cos\varepsilon=\hat n.\hat s
\end{equation}
is measured from the orbit, its variation rate is $2\dot\omega$ too. This can be understood by approximating the obliquity as the norm of the difference between the spin and orbit unit vectors projected onto the Laplace plane (see Baland et al. 2011):
\begin{eqnarray}
\label{epapp}\varepsilon(t)&\simeq&\sqrt{(s_x-n_x)^2+(s_y-n_y)^2}\simeq \varepsilon_{\Omega} +\varepsilon_{\omega} \cos 2\omega(t).
\end{eqnarray}

The mean solution $\hat s_\Omega$ lies in the Cassini plane. Because of the nutation $\hat s_{\omega}$, the spin axis orientation $\hat s$ deviates by an angle $\tilde\delta$ from that plane. The exact definition for $\tilde\delta$ is (see Yseboodt and Margot 2006, Eq. 19. Note that the deviation is defined here with the opposite sign)
\begin{eqnarray}\label{deltaexact}
\sin\tilde\delta&=&-\frac{(\hat n\wedge\hat l) \,.\, \hat s}{\sqrt{1-(\hat n.\hat l)^2}},
\end{eqnarray}
with $\hat l$ is the unit vector along the Laplace plane normal defined as
\begin{eqnarray}
 \hat l&=&\left(\begin{array}{c}
 \cos \delta_{LP} \cos \alpha_{LP} \\
 \cos \delta_{LP} \sin \alpha_{LP} \\
 \sin\delta_{LP}
 \end{array}\right).
\end{eqnarray}
$\alpha_{LP}$ and $\delta_{LP}$ are the equatorial coordinates (right ascension and declination) of the Laplace plane pole with respect to the ICRF at J2000. From the analysis of the DE431 ephemeris (see Appendix \ref{AppParamOrb}), we infer that $\alpha_{LP}=273.81^\circ$ and $\delta_{LP}=69.46^\circ$. 

The deviation can be approximated by the distance from the Cassini plane to the spin unit vector, projected onto the Laplace plane (see Baland et al. 2011):
\begin{equation}
\label{deapp}\tilde\delta(t)\simeq(n_x s_y-n_y s_x)/{\sqrt{n_x^2+n_y^2}}= \varepsilon_{\omega} \sin 2\omega(t)
\end{equation}

\begin{figure}[!htb]
\begin{center}
\includegraphics[width=9cm]{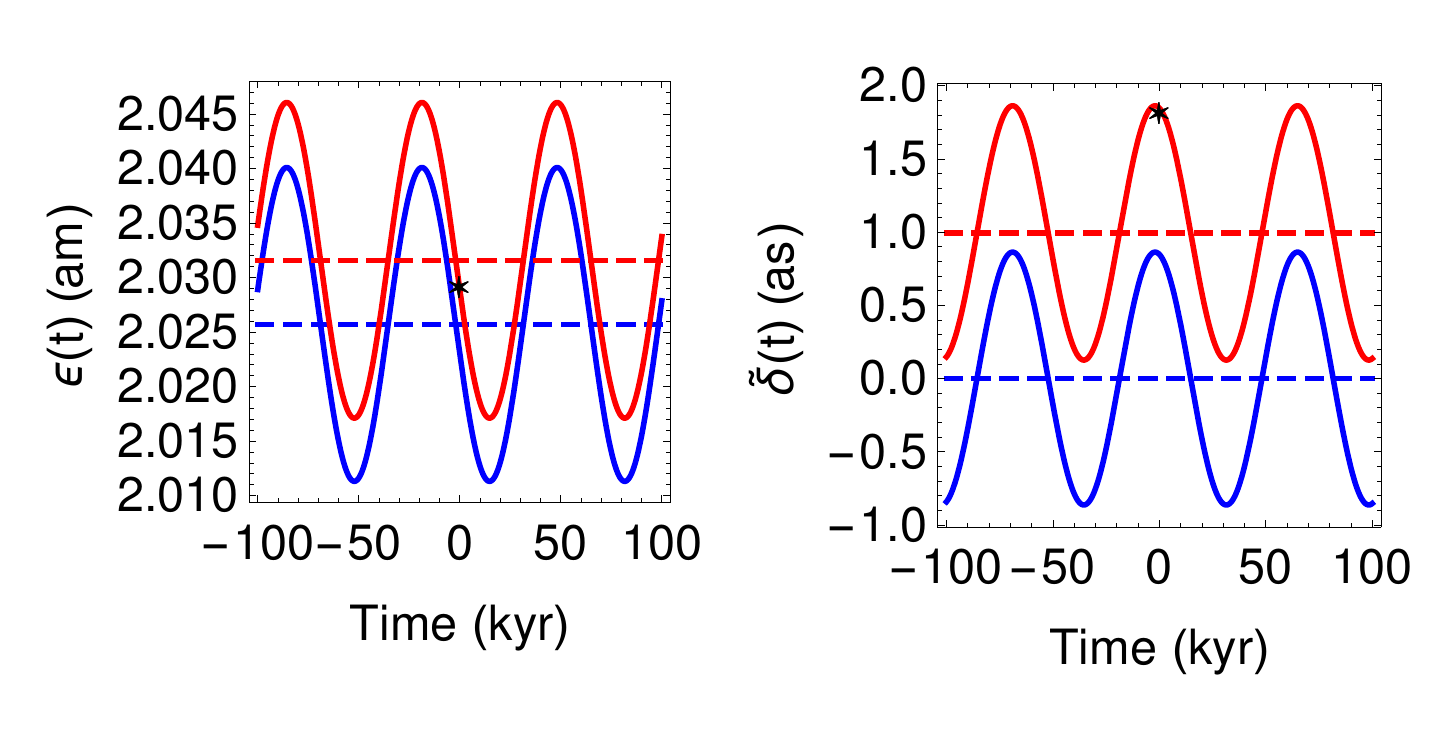}
\caption{\label{Ged} Time-variable obliquity $\varepsilon(t)$ and deviation $\tilde \delta (t)$ as a function of the time, centered about J2000, for the amplitudes given in Table \ref{Tab2}, which correspond to our best fit parameters to the measurements of Stark et al. (2015b). The blue and red curves are for the rigid and the non-rigid cases, respectively. The dashed lines represent the mean values. The black stars represent our estimates of $\varepsilon_{J2000}$ and $\tilde \delta_{2000}$, also given in Table \ref{Tab2}.}
\end{center}
\end{figure}

The obliquity variations, $\Delta \varepsilon = \varepsilon_{\omega} \cos 2\omega$, and the time varying deviation $\tilde\delta$ have the same amplitude $\varepsilon_{\omega}$, but are out of phase by $\pi/2$, since they are measured along two directions orthogonal to each other, in the precessing orbital reference frame (see Fig. \ref{nut} and \ref{Ged}). Over short timescales, as spatial mission durations, the obliquity and deviation can be seen as constant. 

\section{Spin axis orientation of a non-rigid Mercury}
\label{Section3}

\begin{figure}[!htb]
\begin{center}
\includegraphics[width=7cm]{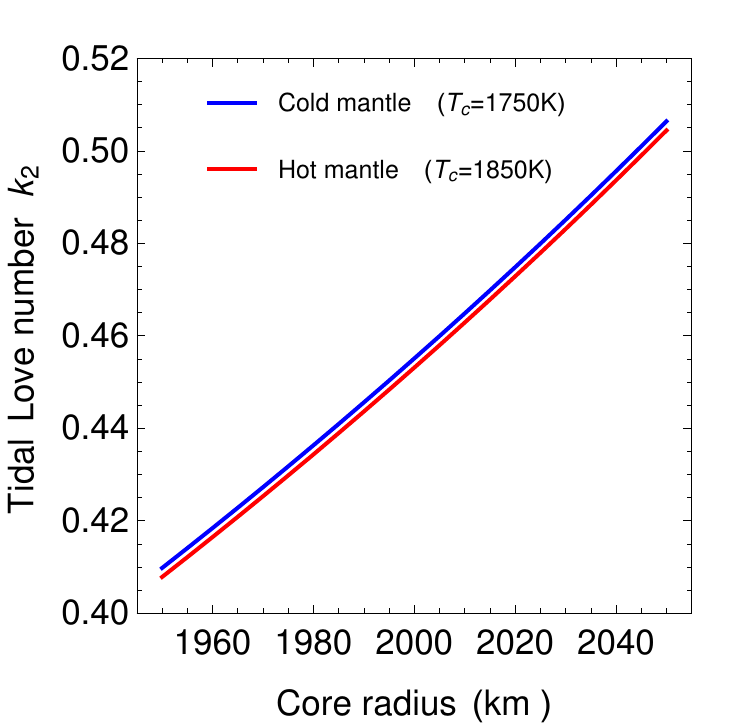}
\caption{\label{k2QTilio} Love number $k_2$ as a function of the core radius for two mantle temperatures. The mantle rheology is based on Jackson et al. (2010). We assume a grain size of $10$ mm for the cold model and of $1$ mm for the hot model.}
\end{center}
\end{figure}

Since Mercury is not a rigid planet, it deforms as a result of the tidal potential, with a tidal period equal to the orbital period. The tidal love number $k_2$, describing the changes in the external gravitational potential resulting from the tidal deformations, is in the range $[0.4,0.52]$ (see Fig. \ref{k2QTilio}, and Padovan et al. 2014). These theoretical predictions are confirmed by results based on analysis of MESSENGER data such as the determinations of Mazarico et al. (2014), $k_2=0.451\pm0.014$, and of Verma and Margot (2016), $k_2=0.464\pm0.023$. The fact that the liquid outer core is explicitly accounted for in the tides with period equal to the orbital period does not preclude the fact that we can and do consider that its spin axis precesses as the one of a solid body on very long timescales. 

We will show that the short timescale tidal deformations have an influence on the average gravitational torque driving the spin precession and on the angular momentum, and therefore, on the equilibrium orientation (obliquity and deviation) of the spin axis. Therefore, we adapt the demonstration given in appendix A of Baland et al. (2016) for an elastic solid synchronous satellite to the 3:2 spin-orbit resonance of Mercury. We derive the contribution of the periodic bulge to the solar torque and determine the correction to be made to take into account the effect of the delay in the response of Mercury to the tidal potential of the Sun. To be able to derive a compact expression of the torque for the 3:2 resonance, the large inclination, and the large eccentricity, several approximations will be introduced. We only consider the main precessional torque (so that $\psi=\Omega$), and assume that $\theta$ and $i$ are small enough for a first order approximation, contrary to what we have done in Section \ref{mainprectorque}. That latter approximation is not entirely justified, but we correct for its effect in a second step.

As in Section \ref{Section22}, we first express the torque in the Body Frame of Mercury in order to include both the effects of the static and of the time varying parts of the components of the inertia tensor. Eq. (\ref{TorqueBF}) is replaced by Eq. (A.13) of Baland et al. (2016), which is adapted from Murray and Dermott, 1999, Eqs. (5.43-5.45):
\begin{equation}\label{BFtides}
\vec \Gamma_{BF}=\frac{3 n^2 a^3}{d^5}\left(
\begin{array}{c}
 (C- B) Y Z - D(Y^2-Z^2)-E\, XY+ F \,XZ \\
 (A- C) Z X +D X Y+ E (X^2-Z^2)- F\, YZ \\
 (B- A) X Y -D X Z+ E \,YZ- F(X^2-Y^2)
\end{array}\right).
\end{equation}
The components of the inertia tensor
\begin{eqnarray}\label{I}
\bar{\bar I}=\left(
\begin{array}{ccc}
 A &- F & -E \\
 - F& B & - D \\
- E & - D& C
\end{array}
\right),
\end{eqnarray}
can be expressed as functions of the coefficients of the time-variable gravity field of Mercury (see Eqs. \ref{C20hstat}-\ref{S21peri} for a demonstration). They are time-delayed because of the non elasticity of the tidal response of Mercury to the attraction of the Sun. This delay is modeled as a phase shift $\zeta$ of the mean anomaly $M$, which can be expressed as $1/Q$, with $Q$ the tidal quality factor. $\zeta$ is the range $[0.16^\circ,1.91^\circ]$ for $Q$ in $[30,360]$, a range that covers the current estimates for the Earth, the Moon and Mars (see Lainey 2016 and references therein). 

We substitute Eqs. (\ref{sunposition},\ref{f},\ref{d},\ref{phi1}) in the torque Eq. (\ref{BFtides}), before transforming it to the inertial reference frame and averaging over the mean motion $M$ and the pericenter argument $\omega$. We find that $\vec \Gamma^{prec}$ can be divided into two parts: 
\begin{eqnarray}\label{TINelas}
 \vec \Gamma^{prec}&=&\vec \Gamma^{prec,k_2}+\vec \Gamma^{prec,\zeta}\\
\nonumber \vec \Gamma^{prec,k_2}&=& \frac{1}{2}k_f M_{m}R^2n^2 q_r \left(1+\frac{3}{2}e^2\right) (\hat s \wedge \hat n)\\
\label{Eq72} &&-\frac{1}{16} k_f \left(\frac{k_f-k_2}{k_f}\right) M_{m}R^2n^2 q_t \left(4+61e^2\right) (\hat s \wedge \hat n)\\
\label{Eq73} \vec \Gamma^{prec,\zeta}&=&\frac{3}{2}n\kappa_{\zeta,\hat n} \hat n +\frac{3}{2}n\kappa_{\zeta,\hat s} \hat s\\
 \kappa_{\zeta,\hat n} &=&- \frac{1}{12}k_2\zeta M_{m}R^2 q_t n (2+63 e^2)\\
 \kappa_{\zeta,\hat s} &=& \frac{1}{4}k_2\zeta M_{m}R^2 q_t n (2+15e^2),
\end{eqnarray}
with $q_r$ the ratio of the centrifugal acceleration to the gravitational acceleration (Eq. \ref{qr}) and $q_t$ the tidal parameter (Eq. \ref{qt}). $k_f$ is the fluid Love number. 

The main part, $\vec \Gamma^{prec,k_2}$, is related to the magnitude of the static and periodic tidal bulges and its terms are proportional to $k_f$ or $k_2$. The second part, $\vec \Gamma^{prec,\zeta}$, is a correction related to the out of phase response of the periodic tidal bulge, and is proportional to $k_2 \zeta=k_2/Q$. $\vec \Gamma^{prec,\zeta}$ is a very small torque compared to $\vec \Gamma^{prec,k_2}$ (four orders of magnitude smaller). 

When we set $k_2=0$ in $\vec \Gamma^{prec,k_2}$, we retrieve the torque on a rigid Mercury (Eq. \ref{TIN}), expressed at first order in $\theta$ and $i$ and third order in $e$, and where $C_{20}$ and $C_{22}$ are reduced to their static hydrostatic counterparts (\ref{C20hstat}-\ref{C22hstat}). We have
\begin{eqnarray}
 \nonumber \vec \Gamma^{prec}&\simeq& \frac{1}{2}k_f M_{m}R^2n^2 q_r \left(1+\frac{3}{2}e^2\right) (\hat s \wedge \hat n)\\
\label{Eq76} &&-\frac{1}{16} k_f M_{m}R^2n^2 q_t \left(4+61e^2\right) (\hat s \wedge \hat n)
\end{eqnarray}

Comparing Eq. (\ref{Eq72}) and Eq. (\ref{Eq76}), we see that the effect of the periodic tidal deformations is to multiply the part of the torque due to the tidal static bulge (proportional to $q_t$) by a factor $\frac{k_f-k_2}{k_f}$, leading to a decrease of the total torque. This follows from the general form of the tidal potential, as demonstrated for the librations (Van Hoolst et al., 2013). Indeed, if we neglect the delay in the response of Mercury, the external gravity field of Eq. (\ref{A2}) can be expressed as
\begin{eqnarray}
\nonumber V^{l=2}(r,\varphi,\lambda,t)&=&k_2 \left(\frac{R}{r}\right)^3 V_t(R,\varphi,\lambda,t)\\
\nonumber &&+k_f \left(\frac{k_f-k_2}{k_f}\right) \left(\frac{R}{r}\right)^3 V_t^{stat}(R,\varphi,\lambda)\\
\label{Eq48}&&+k_f \left(\frac{R}{r}\right)^3 V_c^{stat}(R,\varphi,\lambda).
\end{eqnarray}
where $V_t$ is the tidal potential ($V_t^{stat}$ is its static part) and $V_c^{stat}$ is the (static) centrifugal potential. The first term in Eq. (\ref{Eq48}) does not contribute to the torque since it is symmetric with respect to the planet-Sun line at each instant, while the second term is multiplied by the factor $\frac{k_f-k_2}{k_f}$, compared to the rigid case (with $k_2=0$). In the limit case where Mercury would deform on the short timescales as on the long timescales ($k_2=k_f$), the gravitational torque would vanish.

From this, it follows that the extension of the torque (Eq. \ref{Eq72}) for the rigid case to the non-rigid case is obtained by replacing $\kappa_{20}$ and $\kappa_{22}$ by 
\begin{eqnarray}
\label{kappa20} \kappa_{20}^{k_2}&=&\kappa_{20}+\frac{1}{6}k_2M_{m}R^2q_t n (1+3e^2)\\
\label{kappa22}  \kappa_{22}^{k_2}&=&\kappa_{22}+\frac{49}{24}k_2M_{m}R^2q_t n e^2
\end{eqnarray}
in the unapproximated precession torque $\vec \Gamma^{prec}$ of Eq. (\ref{TIN}), instead of using the approximated torque of Eq. (\ref{Eq72}). The resulting equation also extends Eq. (\ref{TINelas}) derived assuming the small angle (unjustified) approximation. Nevertheless, that approximation has allowed us to find the correction to the torque due to the delayed response (the correction is small, so the small angle approximation is valid for this part). $\kappa_{\omega}$, and so the additional nutational torque $\vec \Gamma^{nut}$ of Eq. (\ref{LIN2}), is negligibly affected by the tidal deformations (effect of the fourth order in eccentricity only, which can be safely neglected as the amplitude of the nutation is already very small). 
 
\subsection{Angular momentum equation}

The periodic tidal bulge also has an effect on the angular momentum. Here, Eq. (\ref{LBF}) for the rigid case must be replaced by 
\begin{eqnarray}
\vec L_{BF}&=&\frac{3}{2}\tilde n \left(
\begin{array}{c}
 -E \\
 -D \\
 C 
\end{array}\right).
\end{eqnarray}
Expressed in the inertial reference frame, and averaged over $M$, it reads as
\begin{eqnarray}\label{LINelas}
\vec L&=& \frac{3}{2}\tilde n \, \tilde C\, \hat s-\frac{3}{2}\tilde n (\tilde C-C)\hat n-\frac{3}{2}\tilde n C_{\zeta} (\hat s\wedge \hat n)\\
\tilde C&=&C+\frac{1}{6}k_2M_{m}R^2q_t\left(1+\frac{3}{2}e^2\right)\\
 C_{\zeta}&=&\frac{3}{2}(\tilde C-C)\zeta
\end{eqnarray}
The notation $\tilde C$ is introduced for the sake of shortening the equations in the following. The difference $\tilde C-C$ is not the periodic part of $C$ due to periodic tides. 

The angular momentum equation becomes
 \begin{eqnarray}\label{AFinalelas}
\nonumber \tilde n \tilde C \frac{d \hat s}{dt}&=& \tilde n (\tilde C-C) \frac{d \hat n}{dt}+n\left\lbrace\kappa_{20}^{k_2} (\hat s.\hat n)+\kappa_{22}^{k_2}\left(\frac{1+\hat s.\hat n}{2}\right) \right\rbrace(\hat s \wedge \hat n)\\
\nonumber && +n\kappa_{\omega} \left(
\begin{array}{c}
 \left(p_x^2-p_y^2\right)(n_y-s_y)+ \,2 p_x p_y(s_x-n_x)\\
 \left(p_x^2-p_y^2\right)(n_x-s_x)+ 2 p_x p_y (n_y-s_y) \\
 0 \\
\end{array}
\right)\\
&&+n\kappa_{\zeta,\hat n} \hat n +n\kappa_{\zeta,\hat s} \hat s+\tilde n C_{\zeta}\frac{d(\hat s\wedge \hat n)}{dt}
\end{eqnarray}

\subsection{Solution}

We solve the angular momentum equation by a perturbative approach, neglecting the small variations in the third component. By also using the excellent approximation $\tilde n=n$, as in Section \ref{Section2}, the solution for the
spin in the non-rigid case can be expressed as
\begin{eqnarray}
 \label{sel}\hat s&=&\hat s_\Omega^{k_2}+\hat s_{\omega}^{k_2}+\hat s_{\zeta}\\
 \label{Eq61}\hat s_\Omega^{k_2}&=&\left(
\begin{array}{c}
 \sin(i+\varepsilon^{k_2}_{\Omega})\cos{(\Omega-\pi/2)} \\
 \sin(i+\varepsilon^{k_2}_{\Omega})\sin{(\Omega-\pi/2)} \\
 \cos(i+\varepsilon^{k_2}_{\Omega})
\end{array}\right)\\
 \hat s_{\omega}^{k_2}&=&\left(
\begin{array}{c}
 \varepsilon^{k_2}_{\omega}\cos{(2\omega+\Omega-\pi/2)} \\
 \varepsilon^{k_2}_{\omega}\sin{(2\omega+\Omega-\pi/2)} \\
 0
\end{array}\right)\\
\label{Eq63} \hat s_{\zeta}&=&\left(
\begin{array}{c}
 \varepsilon_{\zeta}\cos{\Omega} \\
 \varepsilon_{\zeta}\sin{\Omega} \\
 0
\end{array}\right)
\end{eqnarray}
with
\begin{eqnarray}
\label{Eq88} \varepsilon_{\Omega}^{k_2}&=& -\frac{ C\,\dot\Omega \sin i}{\kappa^{k_2}+\tilde C\,\dot\Omega\,\cos i }=\varepsilon_{\Omega}+\Delta\varepsilon_{\Omega}\\
\label{Eq89} \varepsilon_{\omega}^{k_2}&=&\frac{\varepsilon_{\Omega}^{k_2}\kappa_{\omega}}{\kappa^{k_2}+\tilde C(2\dot \omega+\dot\Omega)}\\
\label{Eq90} \varepsilon_{\zeta} &=&-\frac{\varepsilon_{\Omega}^{k_2}(\kappa_{\zeta,\hat s}+C_{\zeta}\dot\Omega)\cos i+(\kappa_{\zeta,\hat n}+\kappa_{\zeta,\hat s})\sin i}{\kappa^{k_2}+\tilde C\dot \Omega}
\end{eqnarray}
and 
\begin{equation}
 \kappa^{k_2}=\kappa_{20}^{k_2}+\kappa_{22}^{k_2}.
\end{equation}

Eqs. (\ref{Eq88}) and (\ref{Eq89}) are formally identical to Eqs. (\ref{Yseboodt}) and (\ref{eomega}) for the rigid case, and only differ in the use of quantities $\kappa^{k_2}$ and $\tilde C$ in the denominators.

Compared to a numerical integration of Eq. (\ref{AFinalelas}), the solution Eq. (\ref{sel}) is $0.014\%$ and $1.5\%$ accurate in obliquity and deviation, respectively, about J2000 (see Appendix \ref{App3}).

By using similar approximations as in Section \ref{Section233}, the obliquity and deviation can be expressed as
\begin{eqnarray}
\varepsilon^{k_2}(t)&\simeq& \varepsilon_{\Omega}^{k_2}+\varepsilon_{\omega}^{k_2} \cos 2\omega(t)\\
\tilde\delta^{k_2}(t)&\simeq& \varepsilon_{\omega}^{k_2} \sin 2\omega(t)+\varepsilon_{\zeta}.
\end{eqnarray}

\begin{figure}[!htb]
\begin{center}
\includegraphics[width=5.4cm]{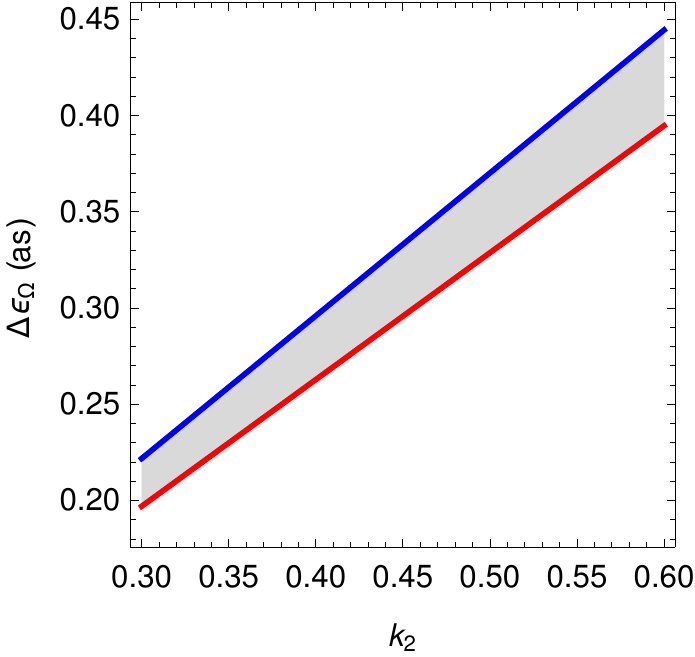} \\
\vspace{0.3 cm}
 \includegraphics[width=5.6cm]{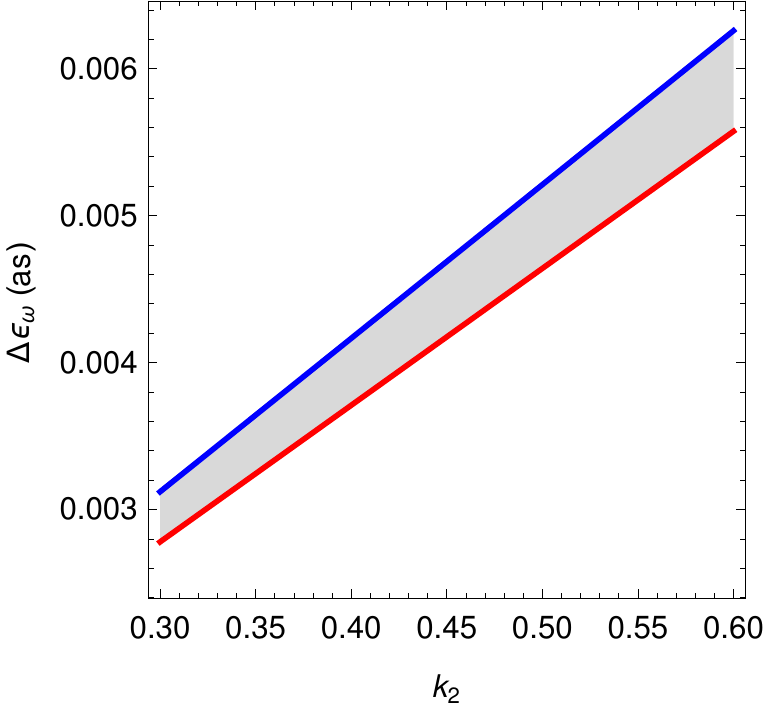} \\
\vspace{0.3 cm}
\includegraphics[width=5.2cm]{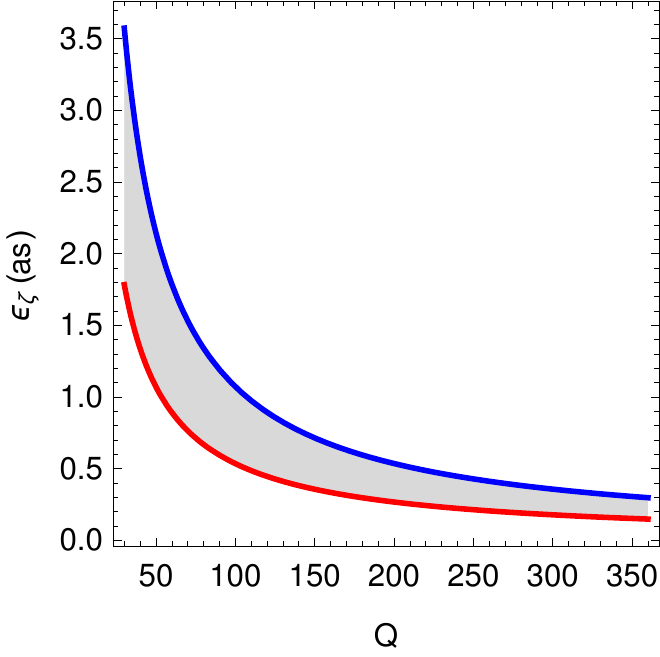}
\caption{\label{Fig6}Effect of tidal deformations on the precession amplitude $\varepsilon^{k_2}_{\Omega}$ (top), on the nutation amplitude $\varepsilon^{k_2}_{\omega}$ (middle), and on $\varepsilon_{\zeta}$, the amplitude of the part of the solution related to delay in the response of Mercury to the tidal potential (bottom panel); as a function of the real part $k_2$ of the tidal Love number or as a function of the tidal quality factor $Q$. In the top and middle panels, the ratio $C/M_{m}R^2$ varies from $0.32$ (bottom curve) to $0.36$ (top curve). The effect of tidal deformations on $\varepsilon^{k_2}_{\omega}$ is very small, while the difference between $\varepsilon^{k_2}_{\Omega}$ and $\varepsilon_{\Omega}$, though larger, is still under the actual precision on the spin position measurement ($5$ arcsec), but may be relevant for interpretation of future measurement. The deviation $\varepsilon_{\zeta}$ is of the order of the arcsec for realistic values of the tidal Love number ($k_2\simeq 0.45$, e.g.). $\varepsilon_{\zeta}$ does not noticeably depend on $C/M_{m}R^2$, but is larger for larger $k_2$ ($k_2=0.3$ and $0.6$ for the bottom and top curves, respectively).} 
\end{center}
\end{figure} 

The difference $\Delta\varepsilon_{\Omega}=\varepsilon_{\Omega}^{k_2}-\varepsilon_{\Omega}$ in the main precession amplitude between the non-rigid and rigid cases is constant over time but depends on Mercury's interior through $k_2$. As the term proportional to $\dot\Omega$ in the denominator of $\varepsilon_{\Omega}^{k_2}$ and $\varepsilon_{\Omega}$ is very small, $\Delta\varepsilon_{\Omega}$ is approximately proportional to the difference $\kappa-\kappa^{k_2}$ (which is proportional to $k_2$), and positive, since tidal deformations decrease the torque. As can be seen in Fig. \ref{Fig6}, $\Delta\varepsilon_{\Omega}=0.3$ arcsec for realistic values of $C/M_{m}R^2$ $(\simeq 0.346,$ see Table \ref{tab1}) and of $k_2$ $(\simeq0.45, $ see Mazarico et al., 2014 and Verma and Margot, 2016). The nutation amplitude is not significantly affected (effect smaller than 0.01 arcsec).

\begin{figure}[!htb]
\begin{center}
\includegraphics[width=7cm]{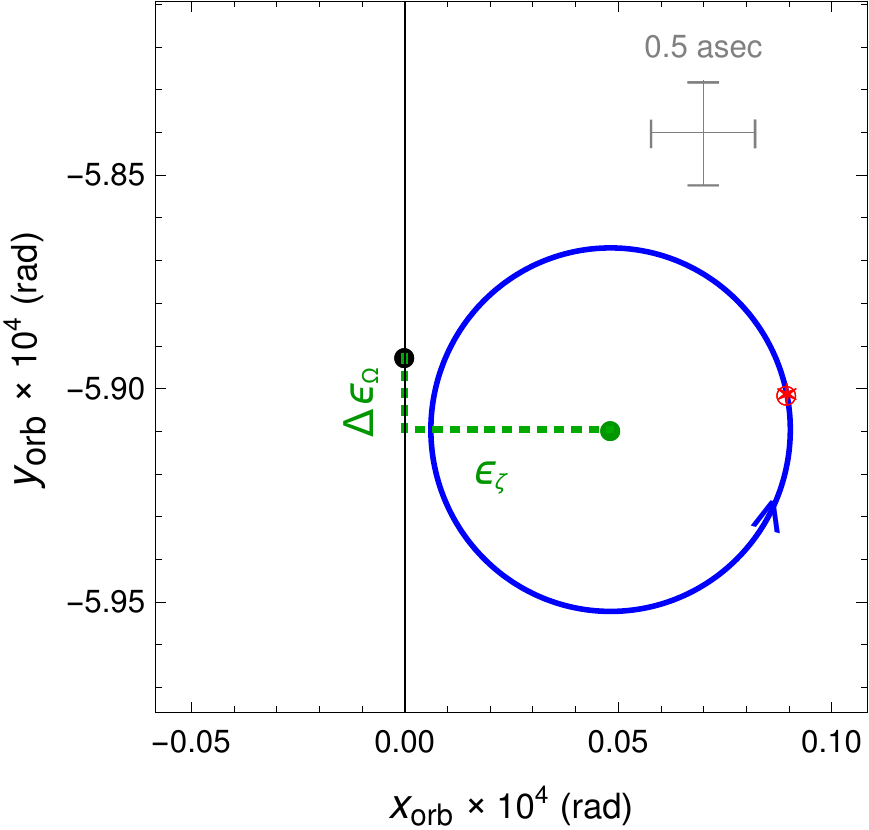}
\caption{\label{nuttides}Motion of the spin axis of a non-rigid Mercury, projected onto the precessing orbital plane. The amplitudes of each component of the solution correspond to our best fit to the observation of Stark et al. (2015b), as given in Table (\ref{Tab2}). Compared to Fig. \ref{nut}, the effect of the tidal deformations adds to the nutation, so that the center of motion is shifted from the black dot to the green dot. Along the $y$- axis, one can see the constant addition $\Delta \varepsilon_{\Omega}$ to the mean obliquity, due to the effect of the tidal deformations. Along the $x$- axis, one can see the constant additional deviation $\varepsilon_\zeta$ due to the non-elasticity. Cumulated effect of the nutation and non-elasticity allows to explain the observed deviation by Stark et al. (2015b). The red circle and star have the same meaning as in Fig. \ref{nut}.}
\end{center}
\end{figure}

The spin vectors $\hat s_\Omega^{k_2}$ and $\hat s_{\zeta}$ precess with the same period, but they are out of phase by $\pi/2$ (see Eqs. \ref{Eq61} and \ref{Eq63}). Seen from the precessing orbital reference frame, the effect of $\hat s_{\zeta}$ is then constant. As shown in Fig. \ref{nuttides}, the non-elasticity introduces a constant deviation over time (let us call it \textit{tidal deviation}), given by the amplitude $\varepsilon_{\zeta}$ of $\hat s_{\zeta}$, with respect to the coplanarity, that adds to the time-varying deviation introduced by the nutation. As the spin lags behind as a result of the delay in response introduced by the non-elasticity, the deviation is positive. The tidal deviation $\varepsilon_{\zeta}$ is approximately proportional to $k_2\zeta=k_2/Q$, as can be seen from the following approximation resulting from an order of magnitude analysis:
\begin{eqnarray}
 \nonumber \varepsilon_{\zeta} &\simeq& -\frac{(\kappa_{\zeta,\hat n}+\kappa_{\zeta,\hat s})\sin i}{\kappa}\\
\label{approxezeta} &=&\left(-\frac{1}{3}+\frac{3}{2}e^2\right)\frac{k_2}{Q}\frac{M_{m}R^2 n q_t \sin i}{\kappa}.
\end{eqnarray}
$\varepsilon_{\zeta}$ is of the order of the arcsec for $Q\simeq 100$ (a reasonable value for a terrestrial body) and $k_2\simeq 0.45$. 

\begin{figure}[!htb]
\centering
\includegraphics[width=6cm]{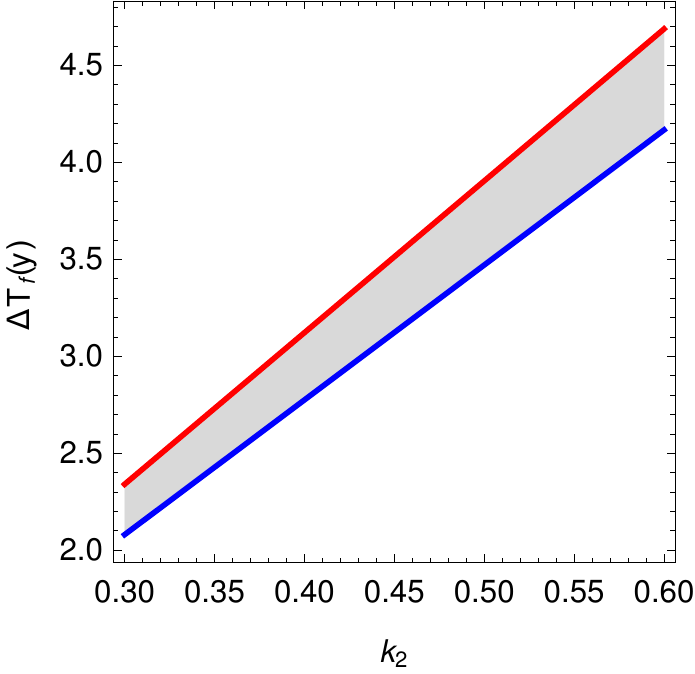} 
\caption{Difference $\Delta T_f=T_f^{k_2}-T_f$ in free precession period between the non-rigid and the rigid cases, as a function of the tidal Love number $k_2$. The ratio $C/M_{m}R^2$ varies from $0.32$ (top curve) to $0.36$ (bottom curve).}
\label{TfEl}
\end{figure}

In the limit of small inclination, Eq. (\ref{Eq88}) can be written as 
\begin{eqnarray}
\varepsilon_{\Omega}^{k2}\simeq -\frac{C}{\tilde C}\frac{ i \, \dot\Omega}{\omega_f^{k_2}+\dot\Omega}
\end{eqnarray}
where the free precession frequency of a non-rigid solid Mercury is $\omega_f^{k_2}=\kappa^{k_2}/\tilde C$. The difference $\Delta T_f=T_f^{k_2}-T_f$, is $5$ years at most (see Fig. \ref{TfEl}), so that the free period is of the order of $1300$ years, like the free period for the rigid Mercury and much shorter than period of the node precession of $326$ kyr. Therefore, tidal deformations cannot help to trigger a resonant excitation of the forced precession.

\section{Spin axis orientation measurements and previous data inversion}

\subsection{Orientation measurements}
\label{Section41}

\begin{figure}[!htb]
\begin{center}
\includegraphics[width=7cm]{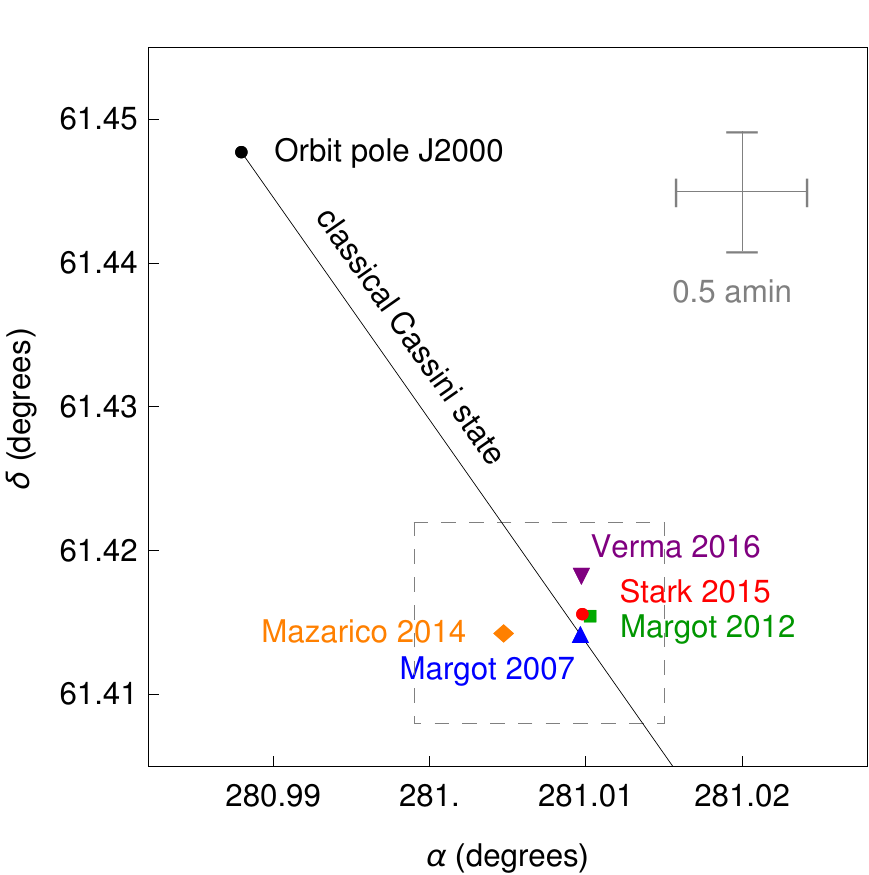}
\caption{\label{fig1} Right ascension ($\alpha$) and declination ($\delta$), in the ICRF, of the orbit pole of Mercury and of its spin pole according to the estimates of Margot et al. (2007, 2012), Mazarico et al. (2014), Stark et al. (2015b) and Verma and Margot (2016). The solid line represents the location of the Cassini state, at J2000, for all possible positive values of the equilibrium obliquity in the frame of the classical Cassini state model. Here, we have chosen the same plot limits as in Fig. 2 of Margot et al. (2007) and Fig. 1 of Margot (2009), to facilitate comparisons. The dashed rectangle represents the plot limits used in Fig. 2 of Stark et al. (2015b), and that will be used in Fig. \ref{Fig4}.}
\end{center}
\end{figure}

The spin axis orientation has been estimated by means of three different techniques so far. First, Margot et al. (2007, 2012) have observed, using Earth-based antennas, the variations of radar echoes of the surface of Mercury, which are tied to the planet's rotation, or more specifically to the rotation of its surface solid layer. Secondly, Stark et al. (2015b) have used images and laser altimeter data from MESSENGER to investigate the rotation of surface. The results of Margot et al. (2012) and Stark et al. (2015b) are in a very good agreement (see Tab. \ref{tab1} and Fig. \ref{fig1}). Thirdly, Mazarico et al. (2014) and Verma and Margot (2016) have analyzed radio tracking data of MESSENGER, which are affected by the orientation of the gravity field of the planet. Those two determinations, although based on similar approaches, are only marginally consistent, and the one of Mazarico is further away from those of Margot et al. (2012) and Stark et al. (2015b). Mazarico et al. (2014) have argued that their estimate may be influenced by a differential rotation of a solid inner core with respect to the surface solid layer. However, Verma and Margot (2016) tend to rule out such differential rotation.

\subsection{Equatorial and Cartesian coordinates}
\label{Sec42}

In the five studies mentioned in Section \ref{Section41}, Mercury's spin orientation is defined with respect to a Mercury-centered ICRF by the values of the right ascension $\alpha$ and declination $\delta$ of the spin pole. In our improved models for the Cassini state (Sections \ref{Section2} and \ref{Section3}), the spin orientation is defined with respect to the Laplace reference frame by the Cartesian coordinates $(s_x,s_y,s_z)$ of a unit vector $\hat s$, see Eq. (\ref{hats}). In the same way, the orientation of the orbit pole can be expressed by the equatorial coordinates ($\alpha_{orb},\delta_{orb}$) with respect to the ICRF, or by the Cartesian coordinates $(n_x,n_y,n_z)$ of the unit vector $\hat n$ with respect to the Laplace reference frame, see Eq. (\ref{hatn}). 

The relations between the two representations are:
\begin{eqnarray}
\label{representations}\left(\begin{array}{c}
 s_x \\
 s_y \\
 s_z
\end{array}\right)&=&R_x(\pi/2-\delta_{LP}).R_z(\pi/2+\alpha_{LP})\left(\begin{array}{c}
 \cos\delta \cos\alpha \\
 \cos\delta \sin\alpha \\
 \sin\delta
\end{array}\right)\\
\label{unapproximated}\left(\begin{array}{c}
 n_x \\
 n_y \\
 n_z
\end{array}\right)&=&R_x(\pi/2-\delta_{LP}).R_z(\pi/2+\alpha_{LP})\left(\begin{array}{c}
 \cos\delta_{orb} \cos\alpha_{orb} \\
 \cos\delta_{orb} \sin\alpha_{orb} \\
 \sin\delta_{orb}
\end{array}\right)
\end{eqnarray}
where $\alpha_{LP}$ and $\delta_{LP}$ are the orientation parameters, with respect to the ICRF, of the unit vector perpendicular to the Laplace plane.

\subsection{Approximations and previous results}

The spin orientation parameters are expressed at the J2000 epoch in all five papers. The choice of J2000 as the reference epoch complies well with the current IAU conventions where it is customary to express $\alpha$ and $\delta$ around J2000 (e.g. Archinal et al. 2011). Besides, it allows an easy comparison between the different studies which use measurements from different epochs. However, it requires a prior assumption on the spin precession behavior to correct for the motion of the spin between J2000 and the measurement epoch. 

To estimate the values of $\alpha_{J2000}$ and $\delta_{J2000}$ at J2000 from the values of $\alpha_{ep}$ and $\delta_{ep}$ measured at epoch $T_{ep}$, the authors of the five studies use the following linear approximation for the spin orientation parameters $\alpha$ and $\delta$: 
\begin{eqnarray}
\label{alJ2000} \alpha&=&\alpha_{J2000}+\dot\alpha\, T\\
\label{delJ2000}\delta&=& \delta_{J2000}+\dot\delta\, T,
\end{eqnarray}
where $T$ is the time expressed in Julian centuries from J2000 and $\dot \alpha$ and $\dot \delta$ are linear variations rates. They assume that $\dot\alpha$ and $\dot\delta$ are equal to the linear variation rates, $\dot\alpha_{orb}$ and $\dot\delta_{orb}$, of the orbit orientation parameters $\alpha_{orb}$ and $\delta_{orb}$, as a result of their choice of the classical Cassini state model, and the associated equality between the precession rates ($\dot\psi=\dot \Omega$) of the equatorial and orbital planes, to interpret the measurement. However, they do not impose the coplanarity ($\psi=\Omega$) associated with this particular state.

The orbital variation rates $\dot \alpha_{orb}$ and $\dot\delta_{orb}$ correspond to a linear approximation around J2000 of a uniform orbital precession with respect to the Laplace pole with a period of about $326$ kyr (longitude of the ascending node at J2000 $\Omega_{J2000}=23.73^\circ$) and an inclination $i$ of about $8.53$ degrees. From our analysis of the ephemeris, and by linearizing the solution of Eq. (\ref{unapproximated}) where $\hat n$ is replaced by Eq. (\ref{hatn}), 
we find that (see Appendix \ref{AppParamOrb})
\begin{eqnarray}
\label{alapp} \alpha_{orb}&=&280.9879^\circ+\dot\alpha_{orb}\, T,\\
\label{delapp} \delta_{orb}&=&61.4478^\circ+\dot\delta_{orb}\, T,\\
 \dot\alpha_{orb}&=&-0.0328^\circ/\textrm{cy},\\
 \dot\delta_{orb}&=&-0.0048^\circ/\textrm{cy}.
\end{eqnarray}
This representation of the orbital precession only stands for the 1000 years' time span covered by the analyzed ephemeris.

With their own determination of the orbital precession rates $\dot \alpha$ and $\dot\delta$, Stark et al. (2015b) estimate the spin orientation parameters $\alpha_{J2000}=281.0098^\circ$ and $\delta_{J2000}=61.4156^\circ$ from Eqs. (\ref{alJ2000}-\ref{delJ2000}) using the spin orientation parameters $\alpha_{ep}=(281.00548\pm0.00088)^\circ$ and $\delta_{ep}=(61.4150\pm0.0016)^\circ$ for the reference measurement epoch MJD56353.5, which is the midterm point of the MESSENGER orbital mission phase ($4809$ days, or $13.17$ years after the J2000 epoch). Note that neither the estimate of $\alpha_{ep}$ and $\delta_{ep}$, nor the reference epoch $T_{ep}$ that we could use to retrieve them from the published $\alpha_{J2000}$ and $\delta_{J2000}$, are provided in the other studies.

Once $\alpha_{J2000}$ and $\delta_{J2000}$ are determined, the obliquity $\varepsilon_{J2000}$ is evaluated in the different studies as the solution of (see Eq. \ref{epsexact})
\begin{eqnarray}
 \label{84}\cos\varepsilon_{J2000}&=&\hat n_{J2000}.\hat s_{J2000}
 \end{eqnarray}
while the deviation $\tilde\delta_{J2000}$ can be computed as (see Eq. \ref{deltaexact})
\begin{eqnarray}
\label{85} \sin\tilde\delta_{J2000}&=&-\frac{(\hat n_{J2000}\wedge\hat l)}{\sqrt{1-(\hat l.\hat n_{J2000})^2}}.\hat s_{J2000},
\end{eqnarray}
where $\hat n_{J2000}$ is obtained by introducing the $(\alpha_{orb},\delta_{orb})$ at J2000 of Eqs. (\ref{alapp}-\ref{delapp}) in Eq. (\ref{unapproximated}) and $\hat s_{J2000}$ is obtained by introducing ($\alpha_{J2000},\delta_{J2000}$) of Eqs. (\ref{alJ2000}-\ref{delJ2000}) in Eq. (\ref{representations}). 

Except for Verma and Margot (2016), all studies converge to an obliquity $\varepsilon_{J2000}$ a few arcsec larger than 2 arcmin (see Tab. \ref{tab1} and Fig. \ref{fig1}). \textcolor{black}{Making use of Peale's equation (for instance, Eq. \ref{peale} with $\dot \Omega$ and $\dot \omega<<n$), where the precession amplitude $\varepsilon_{\Omega}$ is approximated by $\varepsilon_{J2000}$, the estimate of the normalized polar moment of inertia 
\begin{eqnarray}
\nonumber \frac{C}{M_{m}R^2}&\cong&\frac{-n \sin\varepsilon_{J2000}}{\dot\Omega \sin(i+\varepsilon_{J2000})}\big(G_{201}(e) C_{22}(1+\cos\varepsilon_{J2000})\\
&&-G_{210}(e) C_{20} \cos\varepsilon_{J2000}\big)
\end{eqnarray}
is between} $0.34$ and $0.36$ for all studies but Verma and Margot (2016), where the smaller obliquity ($1.9$ arcmin) implies a smaller normalized polar moment of inertia ($0.32$). 

In Margot et al. (2007), the coplanarity is almost perfectly satisfied, while the spin axis lags behind its expected position by $2.3$ arcsec (still consistent with a zero deviation within the limits of the measurements precision) in Margot et al. (2012). In Mazarico et al. (2014), the spin axis tends to be ahead of its expected position by $7.9$ arcsec. In Stark et al. (2015b) and Verma and Margot (2016), the deviation is $1.7$ arcsec and $4.4$ arcsec (lagging behind), respectively. The results of our computations of the deviation are summarized in Table (\ref{tab1}). 

\subsection{Measurements uncertainties}
\label{Section44}

In the best cases (Margot et al., 2012, Stark et al., 2015b), the uncertainty on the obliquity is about $4\%$ ($0.08$ arcmin) and results almost entirely from the uncertainty on the spin measurements. More specifically, it is the uncertainty on $\delta_{J2000}$ that dominates the uncertainty on $\varepsilon_{J2000}$, while the uncertainties related to $\alpha_{J2000}$ and the orientation parameters $\alpha_{orb}$ and $\delta_{orb}$ of the orbit pole do not significantly contribute to the obliquity uncertainty budget. The uncertainties on the orientation of the Laplace pole do not contribute to the obliquity uncertainty budget neither, since the definition of the obliquity (Eq. \ref{84}) is independent of the Laplace pole orientation.

The uncertainty on $\varepsilon$ propagates to the uncertainty on $C/M_{m}R^2$ by using the quasi linear relation between the obliquity and the polar moment of inertia of the classical Cassini state model, Eq. (\ref{CCS}). As noted in Margot et al. (2012), the uncertainties on the gravitational coefficients and on the orbital parameters, that intervene also in Eq. (\ref{CCS}), contribute to a smaller extent to the uncertainty budget on the polar moment of inertia than the uncertainty on the obliquity. With the current uncertainty on the gravitational coefficients (Mazarico et al., 2014, Verma and Margot 2016), their contribution to the uncertainty on $C/M_{m}R^2$ is $0.005\%$.

\textcolor{black}{The uncertainty on the polar moment of inertia related to the orbital parameters is dominated by the uncertainty on the determination of $\dot \Omega$ and $i$ (or equivalently, on the determination of the products $\dot \Omega \sin i$ and $\dot \Omega \cos i$) and is less than $1\%$ (Yseboodt and Margot 2006). Using the uncertainty on $\dot \Omega \sin i$ and $\dot \Omega \cos i$ of Stark et al. (2015a) and their very low correlation, we find an effect of $0.07\%$ on the determination of $C/MR^2$ (Stark et al., 2015a report an effect of $6.1 \times 10^{-5}$, or $0.02\%$). In Appendix A, using our own determination of $\dot \Omega \sin i$ and $\dot \Omega \cos i$, we find that this effect is $0.016\%$, while the uncertainties on $n$ and $e$ have an effect of $0.003\%$ and $0.007\%$, respectively. The difference between the different estimations of the effect of the uncertainties on the orbital parameters is due to the choice of ephemeris, but above all to the time-interval used to fit the orbital parameters, which influences the value of the uncertainty on $\dot \Omega \sin i$, and to the method chosen to evaluate the orbital parameters.}

The uncertainty on the deviation is a few arcsec (see Table \ref{tab1}) and will affect the determination of $k_2/Q$ as we do in Section \ref{Section4}. It is mainly due to the uncertainty on the spin pole orientation parameters. For the spin orientation of Stark et al. (2015b), the uncertainty is very large ($3$ arcsec, or $180\%$). The contribution of the uncertainties on the orbit pole orientation is smaller. The uncertainty on the Laplace plane determination certainly affects the determination of the deviation too. However, Stark et al. (2015) find that the Cassini plane has a $1 \sigma$ thickness of $0.18$ arcsec, an order of magnitude smaller than the few arcsec uncertainty on $\tilde \delta$ due to the uncertainties on the spin orientation parameters. \textcolor{black}{With our approach to determine the orientation of the Laplace plane, we find a $1 \sigma$ thickness of $0.025$ arcsec (two orders of magnitude smaller than the uncertainty on the deviation $\tilde \delta$, see Appendix \ref{AppParamOrb}).} 

\subsection{Systematic errors}
\label{Section45}

Besides the effect of measurement uncertainties on the determination of the polar moment of inertia, some approximations considered in the five studies \textcolor{black}{may biais the data interpretation. In this section, we demonstrate that the systematic errors resulting from the linearization of the precession are smaller than the current uncertainty on the determination of the obliquity and of the deviation.}

Equations (\ref{alJ2000}-\ref{delJ2000}) are linear approximations around J2000 of the spin precession, where the linear variations rates are chosen equal to those of the linearized orbital precession of Eqs. (\ref{alapp}-\ref{delapp}). Because of the non-zero obliquity of the spin axis, its equatorial coordinates have to decrease slightly faster than those of the orbit normal, for the spin to precess at the same rate as the orbit (see Appendix 3 of Stark et al. 2015a). For a uniform spin precession with respect to the Laplace pole with a period of $325,\!513\pm10,\!713$ years and an inertial obliquity $\theta=8.53^\circ+\varepsilon_{\Omega}$, the correct form for the linearized spin parameters, at first order in $\varepsilon_{\Omega}$ (expressed in degrees in the equations hereafter) is
\begin{eqnarray}
 \label{eq103}\alpha&=&\alpha_{J2000}+(\dot\alpha_{orb}-0.0029065\, \varepsilon_{\Omega})\, T,\\
\label{eq104} \delta&=&\delta_{J2000}+(\dot\delta_{orb}-0.0004154\, \varepsilon_{\Omega})\, T.
\end{eqnarray}
and is obtained from Eq. (\ref{representations}) where $\hat s$ is replaced by $\hat s_\Omega$ of Eq. (\ref{s0}) and $\Omega_{J2000}$ is left unknown. For $\varepsilon_{\Omega}=2$ arcmin, Eqs. (\ref{alJ2000}-\ref{delJ2000}) leads to a difference in rates of about $0.3\%$ with respect to those of Eqs. (\ref{eq103}-\ref{eq104}) and of about $\pm 1.7$ and $\pm 0.3$ arcsec in $\alpha$ and $\delta$, respectively, $500$ years before or after J2000, at the limits of the time span covered by the analyzed ephemeris. 

Even if the obliquity was zero, the linear approximation by itself would introduce a difference with respect to the uniform precession. This is the same difference as the one between the linearized orbital parameters of Eqs. (\ref{alapp}-\ref{delapp}) and parameters obtained from Eq. (\ref{unapproximated}) where $\hat n$ is replaced by Eq. (\ref{hatn}). These additional errors on $\alpha$ and $\delta$ are about $0.4$ and $-1$ arcsec, respectively, $500$ years after J2000.

\begin{figure}[!htb]
\begin{center}
\includegraphics[width=6cm]{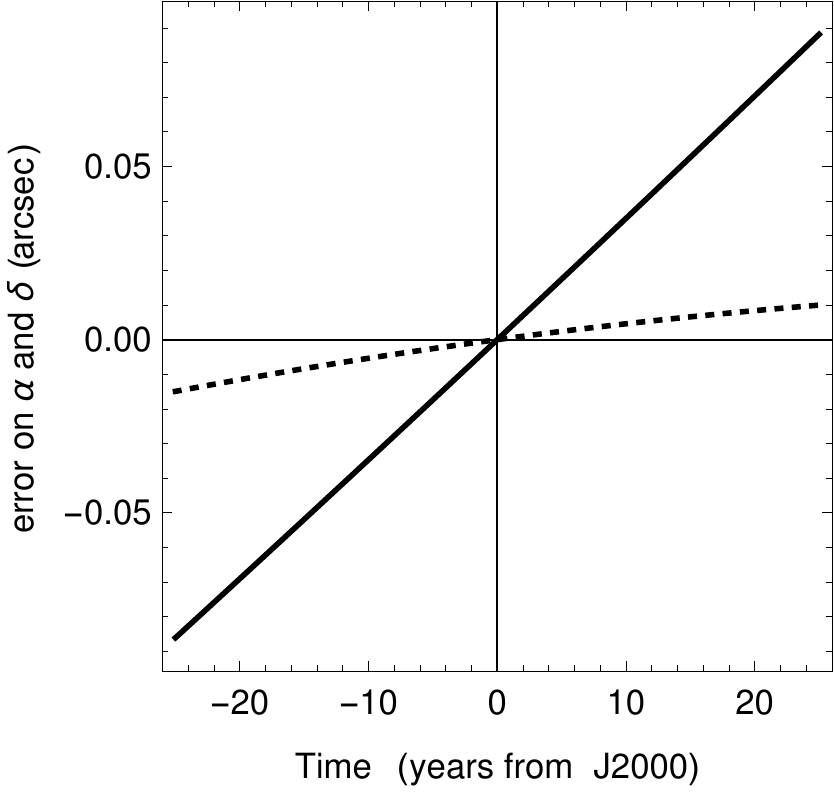}
\caption{\label{fig2} Effect of different approximations on the computation of the spin orientation parameters $\alpha$ (solid lines) and $\delta$ (dashed lines) as a function of the time. The top and middle panels depict the effect of neglecting the obliquity in the linearization of the parameters and the effect of linearization of the orbit precession, respectively. The bottom panel shows the cumulated effects, on a shorter time scale. }
\end{center}
\end{figure} 

Close to J2000, the cumulated differences are dominated by the effect of neglecting the obliquity in the linear variations rates. $4809$ days, or $13$ years, after J2000, these differences are $+0.05$ and $+0.006$ arcsec (see Fig. \ref{fig2}). They propagate oppositely ($-0.05$ and $-0.006$ arcsec) to the determination of $\alpha_{J2000}$ and $\delta_{J2000}$ from the measured $\alpha_{ep}$ and $\delta_{ep}$ and are well below the observational precisions which are $3.2$ and $5.8$ arcsec.

The consequences of these approximations on the determination of the obliquity and of the deviation are about $-0.0008\%$ ($-0.0009$ arcsec) and $-1.3\%$ ($-0.023$ arcsec), respectively, well below the current uncertainties of a few arcsec. Using Peale's equation, the polar moment of inertia is therefore underestimated by only $0.0008\%$. 

Although the linear approximation does not introduce large systematic errors, it introduces an inherent inconsistency since it implicitly relaxes the condition of coplanarity which is used nevertheless in the classical Cassini state model (and Peale's equation) to interpret the estimated obliquity. This compromise is made to allow an easy interpretation of data that are not necessarily consistent with the chosen model. A more consistent approach would consist in finding the coplanar orientation closest to the measured orientation at $T_{ep}$ and to use it for interpretation. But an even better approach is to use a model with which the data can be consistent, a model that is not characterized by coplanarity and uniform spin precession rate. We have shown in Section 3 that the improved Cassini state model taking into account the pericenter precession and non-elastic tidal deformations meets these criteria. We will use this model in the next section to interpret the data, and show that a systematic error of $+0.16\%$ on the determination of the polar moment of inertia can be avoided.

\section{A new inversion of the data}
\label{Section42}
\label{Section4}

Using our improved Cassini state model developed in Section \ref{Section3}, we now aim to inverse the observed $\alpha_{ep}$ and $\delta_{ep}$ of Stark et al. (2015b) in terms of the normalized polar moment of inertia $C/M_{m}R^2$ and of the tidal parameters $k_2$ and $k_2/Q$. These three parameters are sufficient to describe the whole model, and are directly connected to the internal properties of Mercury. In brief, the precession and nutation amplitudes are quasi proportional to $C/M_{m}R^2$, the tidal Love number $k_2$ increases the value of the precession amplitude, while the deviation due to the non-elasticity is quasi proportional to $k_2/Q$ and independent of $C/M_{m}R^2$. Note that for numerical evaluations, the coefficients $C_{20}$ and $C_{22}$ contained in $\kappa_{20}^{k_2}$ and $\kappa_{22}^{k_2}$ (Eqs. \ref{kappa20}-\ref{kappa22}), are replaced by the measured gravity coefficients, which are the sum of the static hydrostatic terms $C_{20}^{h,stat}$ and $C_{22}^{h,stat}$, and of non-hydrostatic contributions.

With the solution, we correct $\alpha_{ep}$ and $\delta_{ep}$ for the spin motion, in order to find the values of $\alpha_{J2000},{\delta_{J2000}}$, but also of $ \varepsilon_{J2000}$ and of $\delta_{J2000}$, that are needed to define Mercury's orientation model according to IAU conventions. 

We chose the data of Stark et al. (2015b) for two reasons. \textcolor{black}{First, they provide the numerical values for $\alpha_{ep}$ and $\delta_{ep}$, which are essentially not affected by any assumptions on the spin precession behavior. Indeed, the systematic errors due to the assumption that the spin precession is equivalent to the linearized orbital precession (see Section \ref{Section45}) are on average close to zero when the reference epoch is set to the midterm point of the observation interval, instead of J2000.} Secondly, their estimation for the spin orientation, from images and altimeter data, is in good agreement with the estimate by Margot et al. (2012), obtained from a different technique (Earth-based radar measurements). \textcolor{black}{In Margot et al. (2012), the observation interval is 2002/05/13 to 2012/07/04, so that we could use the middle date MJD54259.5 ($2715$ days, or $7.43$ years after the J2000 epoch) as a proxy for the reference epoch, and compare the resulting solution for the geophysical parameters with the one obtained using the data of Stark et al. (2015b). The spin orientation parameters corresponding to that 'reference epoch' are $\alpha_{ep}=(281.0079\pm0.0015)^\circ$ and $\delta_{ep}=(61.4151\pm0.0013)^\circ$. The error bars are computed from the covariance matrix for the spin axis orientation at J2000 (Jean-Luc Margot, personal communication).} For now, we consider that the estimates of Mazarico et al. (2014) and Verma and Margot (2016), based on the same technique (analysis of radio tracking data of MESSENGER) are not reliable enough since they are marginally in agreement with each other and with the estimates of Stark et al. (2015b) and of Margot et al. (2012).

In the equatorial plane of the ICRF, the equatorial coordinates $\alpha_{ep}$ and $\delta_{ep}$ correspond to the Cartesian coordinates
\begin{eqnarray}
 x_{ep}&=&\cos\delta_{ep}\cos\alpha_{ep}\\
 y_{ep}&=&\cos\delta_{ep}\sin\alpha_{ep}.
\end{eqnarray}

For our improved model and associated spin vector $\hat s$ (Eq. \ref{sel}), at the reference epoch, the theoretical equatorial coordinates ($\alpha_{th},\delta_{th}$) are given by Eq. (\ref{representations}) (here, we do not use the linearized approximations for the spin and orbit parameters rates). They correspond to a set of Cartesian coordinates in the equatorial plane of the ICRF
\begin{eqnarray}
 x_{th}&=&\cos\delta_{th}\cos\alpha_{th}\\
 y_{th}&=&\cos\delta_{th}\sin\alpha_{th}.
\end{eqnarray}

\textcolor{black}{Using an iterative least squares approach, we adjust $C/M_{m}R^2, k_2$ and $k_2/Q$ to minimize the difference
\begin{eqnarray}
\chi^2&=&\left(\frac{x_{ep}-x_{th}}{\sigma_{x}}\right)^2+\left(\frac{y_{ep}-y_{th}}{\sigma_{y}}\right)^2
\end{eqnarray}
between $(x_{th},y_{th})$ and $(x_{ep},y_{ep})$, with $\sigma_{x}=3.4\times10^{-6}$ and $\sigma_{y}=25.4\times10^{-6}$. These uncertainties were propagated from the uncertainties on $\alpha_{ep}$ and $\delta_{ep}$, assuming a correlation of $0.92$ between them.}

 \subsection{Inversion}
\label{Inversion}

We chose the prior value for $C/M_{m}R^2$ and $k_2$ among the range of values reported in the different observational studies ($0.35$ and $0.50$, respectively) and large prior uncertainties ($0.1$ for both parameters, see Table \ref{Tab2}). We chose a reasonable value ($100$) for $Q$, so that $k_2/Q=0.005$, and we associate a large uncertainty ($0.05$) to the ratio $k_2/Q$. After inversion, we find $C/M_{m}R^2= 0.3433\pm0.0134$, $k_2=0.50\pm0.1$, and $k_2/Q=0.00563\pm0.01651$. The prior uncertainties of the parameters have been chosen larger than what can reasonably be assumed, to ensure that if the uncertainty on the solution is of the order of the present one or smaller, it reflects the measurements precision and not the prior uncertainty. This way, it is possible to know if the measurement of the spin orientation contains or not information about each parameter.

The nominal value for the moment of inertia differs from the (corrected) nominal value of Stark et al. (2015) by only $-0.00045$ ($-0.13\%$). It means that neglecting the effect of pericenter precession and tidal deformations, truncating $G_{210}(e)$ at third order in $e$ (in practice, we use a tenth order development in our calculations), and linearizing the orbital orientation parameters lead to an overestimation of the polar moment of inertia by $0.13\%$. This is mainly because of the tidal deformations ($+0.29\%$) and of the nutations ($-0.13\%$), while the effect of the $G_{210}(e)$ truncation and of the orbit linearization are $-0.035\%$ and $-0.0008\%$, respectively. Table \ref{tab4} summarizes the individual effects of these approximations, and also includes an estimation of the effect of other approximations which can affect the accuracy of the determination of $C$. Our uncertainty on the polar moment of inertia ($\sim4\%$) is similar as the one of Stark et al. (2015b).

The solution for $k_2$ is equal to its prior nominal value and uncertainty (see Table \ref{Tab2}), meaning that it is not possible to constrain $k_2$
from a measurement of the spin orientation. The small effect of tidal deformations in the mean obliquity cannot be separated from the major contribution of the polar moment of inertia (the tidal effect is about $0.35$ arcsec in obliquity whereas the uncertainty on the obliquity is $0.08$ arcmin $=5$ arcsec). Even with a better measurement precision, a determination of $k_2$ from the spin measurement would not be possible because that parameter and $C/M_{m}R^2$ have an effect of the same nature, but of different amplitudes, on the orientation of the spin axis. A large variation in the prior value of $k_2$ of $-0.1$ (so that $k_2=0.4$, the bottom value of a realistic range, see Fig. \ref{k2QTilio}) leads to a change of $+0.06\%$ in the determination of the polar moment of inertia. To find the same moment of inertia as Stark et al. (2015b), we would need to have $k_2\simeq0.3$, which is too low to be realistic. This stresses the need to include tidal deformations in the Cassini state model. The solution for the polar moment of inertia is not robust to a change in the prior value of $k_2$, but we have some knowledge on the value of $k_2$. We estimate that the actual precision on $k_2$, which is about $\pm 0.02$ (Mazarico et al., 2014; Verma and Margot, 2016), induces a contribution of $\pm 0.01\%$ on the measurement uncertainty of $C$, which is below the $4\%$ due to the uncertainty on the determination on $(\alpha_{ep},\delta_{ep})$. 

The solution for the ratio $k_2/Q$, even though not tightly constrained (uncertainty of $300\%$), has a better precision than the prior. This means that, with a better precision on the measurement of the spin orientation than the current one, it would be possible to constrain $k_2/Q$. From $k_2=0.50\pm0.1$ and $k_2/Q=0.00563\pm0.01651$, it follows that $Q=89\pm261$. \textcolor{black}{The solutions for $k_2/Q$ and $Q$ are affected by a change in the prior value of $k_2$. We estimate that the actual uncertainty of $\pm0.02$ on $k_2$ induces a contribution of $\pm 0.02\%$ to the uncertainty on $k_2/Q$ and of $\pm 0.4\%$ to the uncertainty on $Q$, well below the $300\%$ uncertainty to the uncertainty on measurement of the spin orientation. We place an upper limit of about $0.02$ on the ratio $k_2/Q$ and of about $350$ on $Q$ at the $1\sigma$ level ($0.06$ and $900$, respectively, at the $3\sigma$ level). If we consider the spin orientation of Margot et al. (2012), translated to the epoch corresponding to the middle date of their observation interval, we find that $k_2/Q=0.0094\pm0.0148$ and $Q=53\pm85$. The upper limit at the $1\sigma$ level on the ratio $k_2/Q$ is also about $0.02$, but the upper limit on $Q$ is about $140$, lower than the one derived from the data of Stark et al. (2015b).}

The uncertainty on $C/M_{m}R^2$ and $k_2/Q$ reflects only the uncertainty on the determination of $(\alpha_{ep},\delta_{ep})$, or equivalently, the uncertainty on the corresponding obliquity and deviation, and not the uncertainties on the gravity field coefficients or on the orbital parameters that we assumed to be perfectly known and that intervene in the definition of the function for $(x_{th},y_{th})$, through the precession and nutation amplitudes and the tidal deviation. \textcolor{black}{Just as the uncertainty on the obliquity propagates to the determination of the polar moment of inertia, the uncertainty on the deviation ($3$ arcsec, see Section \ref{Section44}) propagates to the determination of $k_2/Q$. The deviation associated to $k_2/Q$ being about $1$ arcsec, the uncertainty on $k_2/Q$ associated to the uncertainty in deviation is logically about $300\%$.}

\textcolor{black}{We have seen that the effect of the uncertainties on the gravity field coefficients and on the orbital parameters contribute very little to the error budget on the polar moment of inertia ($\pm0.005\%$ and $\pm0.03\%$, respectively, small compared to the uncertainty of $\pm4\%$). Similarly, we estimate that the uncertainties on the gravity field coefficients or on the orbital parameters contribute to $\pm0.005\%$ and $\pm3\%$ to the error budget on $k_2/Q$, largely below the uncertainty of $\pm300\%$ (consider Eq. \ref{approxezeta}, and the uncertainties on the gravity coefficient from Mazarico et al. 2014 or Verma and Margot 2016, and the uncertainty $\pm 0.28^\circ$ on the inclination derived in Appendix \ref{AppParamOrb}). If we had used the linearized approximations for the spin and orbit parameters rates, we would have introduced a bias of $-0.023$ arcsec on the deviation (see Section \ref{Section45}), and therefore an error of $-2.3\%$ on the determination of $k_2/Q$, also below the uncertainty of $\pm300\%$. Neglecting the effect of the pericenter nutations would lead to a systematic error of $+86\%$ on the determination of $k_2/Q$, since all the measured deviation would be seen as the tidal deviation. We have neglected the effect of viscous coupling at the CMB on the deviation ($0.016$ to $0.055$ arcsec, according to Peale et al., 2014), meaning that we may have overestimated the tidal deviation and introduced a systematic error of $+2\%$ to $+6\%$ on the determination of $k_2/Q$. The sources of uncertainties and error on the determination of $k_2/Q$ identified throughout this study are summarized in Table \ref{tab5}. }

\subsection{Obliquity, deviation and orientation model}

The solution corresponds to a precession amplitude $\varepsilon_\Omega^{k_2}=(2.032\pm0.080)$ arcmin ($\varepsilon_\Omega=2.026$ arcmin and $\Delta \varepsilon_\Omega=0.006$ arcmin) and a nutation amplitude $\varepsilon_\omega^{k_2}=(0.868\pm0.034)$ arcsec ($\varepsilon_\omega=0.863$ arcsec and $\Delta \varepsilon_\omega=0.005$ arcsec). The nutation accounts for an obliquity variation $\Delta \varepsilon$ of $-0.162$ arcsec and a deviation of $0.853$ arcsec. The additional deviation $\varepsilon_\zeta$ due to the non-elasticity is $(0.995\pm2.914)$ arcsec.

Using Eqs. (\ref{sel}) and (\ref{representations}), we find $\alpha_{J2000}=(281.00981\pm0.00083)^\circ$ and $\delta_{J2000}=(61.41565\pm0.00150)^\circ$. The obliquity at J2000, computed from Eq. (\ref{epsexact}), is $\varepsilon_{J2000}=(2.029\pm0.080)$ arcmin, while the deviation $\tilde \delta_{J2000}=(1.847\pm 2.882)$ arcsec (Eq. \ref{deltaexact}). The uncertainties on the obliquity ($4\%$) and on the deviation ($160\%$) are dominated by the relatively small uncertainty on the polar moment of inertia (through $\varepsilon_\Omega^{k_2}$) and by the relatively large uncertainty on $k_2/Q$ (through $\varepsilon_\zeta$), respectively.

\begin{figure}[!htb]
\begin{center}
\includegraphics[width=8cm]{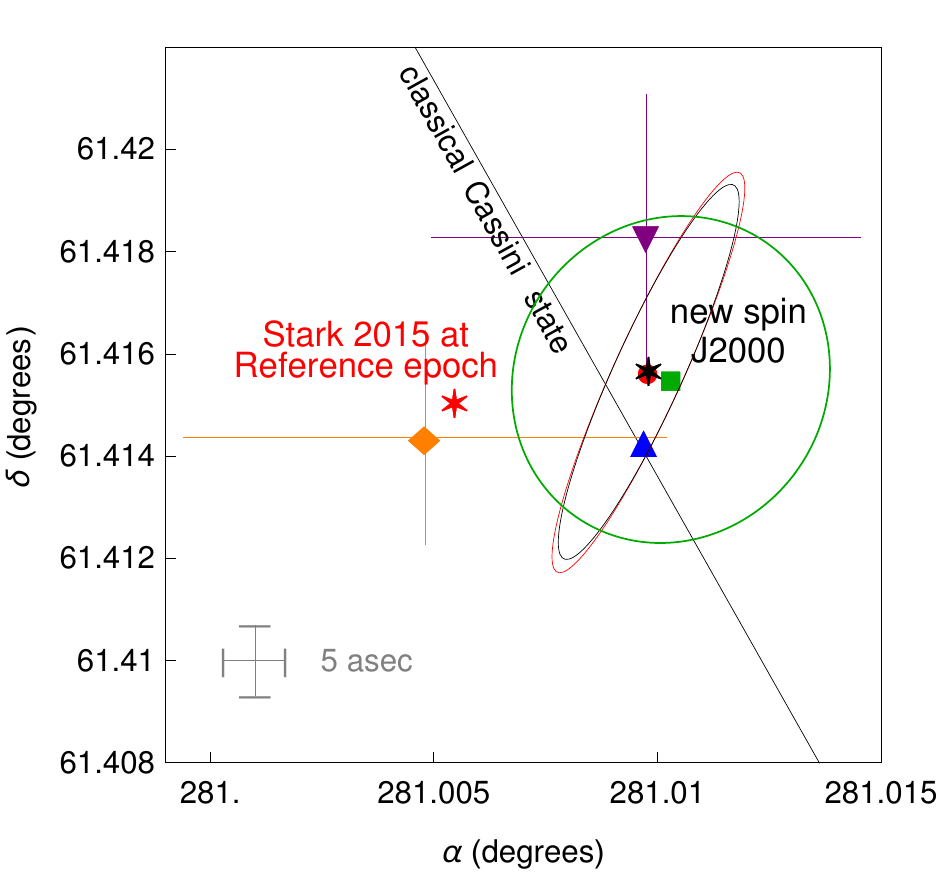}
\caption{\label{Fig4} Right ascension ($\alpha$) and declination ($\delta$) of the rotation axis of Mercury in the ICRF at J2000, according to the five studies mentioned in Section \ref{Section41}. The $2\sigma$ uncertainties of Margot et al. (2012) and Stark et al. (2015b), represented by ellipses\textcolor{black}{, are computed from the covariance matrix for the spin axis orientation at J2000 (Jean-Luc Margot, personal communication) and reproduced from Stark et al. (2015b), respectively.} The uncertainty of Mazarico et al. (2014) and of Verma and Margot (2016) are at the $1\sigma$ level. The location of the spin axis at the measurement reference epoch of Stark et al. (2015b) is also indicated (red star). The black star and the black ellipse stands for our determination of the J2000 position \textcolor{black}{and the corresponding $2\sigma$ uncertainty region}, using our improved Cassini state model where the effect of the pericenter precession and tidal deformations are taken into account. See Fig. \ref{fig1} for legend of the other markers. The solid line represents the location of the classical Cassini state at J2000. We have chosen the same plot limits as in Figure 2 of Stark et al. (2015b) to ease the comparison with their results.}
\end{center}
\end{figure} 

As can be seen directly from Fig. (\ref{Fig4}), this solution is very close to the determination of the spin orientation at J2000 of Stark et al. (2015b). \textcolor{black}{As a result of the propagation of errors, we find uncertainty on $\alpha_{J2000}$ and on $\delta_{J2000}$, as well as the correlation between them, very similar to those of Stark et al. (2015b).} The use of the classical Cassini state model by Stark et al. (2015b), without imposing coplanarity, compensates for the other model approximations (pericenter, tides, \dots) when it comes to determine $(\alpha_{J2000},\delta_{J2000})$, because of a geometrical (but not physical) equivalence. In Stark et al. (2015b), the linearization of the orbital parameters translates the observed deviation from the measurement epoch to J2000, without giving it any physical meaning. Here, the deviation arises because of the physics of the model used in our inversion. However, the compensation does not apply to the determination of $C$. The polar moment of inertia is related to the mean obliquity in our approach, and not to the obliquity at J2000, explaining the $0.13\%$ difference between the two estimates of $C$. 

In principle, improving the Cassini state model implies a slightly different definition for Mercury's orientation model. However, while the effect of the precession of the pericenter and of tidal deformations may become relevant in the future to the determination of the polar moment of inertia and tidal parameters, it does not bring significant improvements to the definition of the orientation model parameters adopted by the IAU (Archinal et al. 2011) on the basis of Margot et al. (2007, 2009) which are
\begin{eqnarray}
 \alpha&=&281.0097^\circ-0.0328^\circ T\\
 \delta&=&61.4143^\circ-0.0049^\circ T.
\end{eqnarray}
Because the effect ($0.3\%$) of obliquity on the rates $\dot \alpha_{orb}$ and $\dot \delta_{orb}$ is already neglected, it will not be necessary to correct those rates for the effects of nutations and tidal deformations on the obliquity. The central values at J2000 could be improved in time if the measurement precision is improved, however, they are not really affected by the addition of nutations and tidal deformations to the Cassini state model.

\section{Comparisons with the literature}
\label{section5}

In Sections \ref{Section2} and \ref{Section3}, we have developed a new analytical model for the Cassini state of a (non) rigid Mercury with a precessing perihelion and we have discussed some similarities and differences with the existing literature. Here, we deepen the comparison on several points.

\subsection{On the relation between the mean equilibrium obliquity and the polar moment of inertia}

Equation (\ref{peale}) and the approximated Eqs. (\ref{eq26}) and (\ref{Yseboodt}) relate the mean equilibrium obliquity to the polar moment of inertia. It is similar to what is found in earlier studies carried out with the Hamiltonian approach. We now discuss similar equations. 

\subsubsection{Peale's equation}
\label{starkerror}
Let us consider Eq. (4) of Peale (1981). Note that Peale has neglected $\dot \Omega$ and $\dot \omega$ in front of $n$, so that $\tilde n=n$, write $\dot\Omega$ as $\mu$, and denotes the orbital obliquity ($\varepsilon_{\Omega}$ here) by $\theta$ (not to be confounded with our angle $\theta$ which is the inertial obliquity). A comparison between Peale's equation and our equation (\ref{peale}) shows straightforwardly that Peale has in fact defined the obliquity as a clockwise angle, against the usual convention. Therefore, the sign of $\theta$ has to be changed in his equation in order to avoid the determination of a negative ratio $C/M_{m}R^2$ from the measured obliquity.

This sign mistake has recently propagated to Eq. (37) of Stark et al. (2015a) who denotes the obliquity by $i_c$. In addition, Stark defines $\mu$ as the absolute value $|\dot\Omega|=-\dot \Omega$ instead of $\dot\Omega$, likely in an attempt to compensate for the sign error on the determination of the polar moment of inertia introduced by the sign mistake in the obliquity definition. The cumulated mistakes lead to a misdetermination of $C/M_{m}R^2$ in Stark et al. (2015b) by $0.8\%$. They find $0.346$, with the correct sign, instead of $0.344$ (see also Fig. \ref{Fig3}).

The fact that the sign of the node precession rate is negative is crucial. It is likely that the different versions of Peale's equation existing in the literature, using the negative rate or its absolute value from one time to the next, contributes to an understandable confusion. Rivoldini and Van Hoolst (2013) use the absolute value of the precession rate (denoted $\dot \psi$) in their version of Peale's equation. They correctly derive a normalized polar moment of inertia of $0.346$ from the measured obliquity of Margot et al. (2012), using a correct version of Peale's equation in their numerical code, while their written version of Peale's equation is affected by a typo (a minus sign is missing in front of the term $\cos i$ of their Eq. 4). Note that this incorrect version of Peale's equation is in fact equivalent to Eq. (37) of Stark et al. (2015a), expanded at first order in obliquity. 

Smith et al. (2012), using $2.06$ arcmin for the obliquity value of Mercury (Margot et al., 2011), deduce that $C/M_{m}R^2=0.353$. However, this value is not correct and should be $0.349$, as can be ascertained by using Eq. (\ref{Yseboodt}) together with the gravity coefficients and orbital parameters used in Smith et al. (2012). \textcolor{black}{The difference is likely due to a problem similar as the one encountered by Stark et al. (2015a,b), since in a previous paper, Smith et al. (2010) present a version of Peale's equation in agreement with Eq. 37 of Stark (2015a), provided that the sign of the precession rate is dropped to ensure a positive polar moment of inertia. }

\subsubsection{Noyelles and Lhotka's equations}
\label{Noyelleserror}

We reproduce below, but according to our notations and after some trigonometric manipulations, Eq. (21) of Noyelles and Lhotka (2013): 
\begin{eqnarray}
\nonumber &&-\left(n+\frac{2}{3}\dot\omega+\frac{2}{3}\dot\Omega\cos (i+\varepsilon_{\Omega})\right)+\frac{2n^2 M_{m}R^2}{C\dot\Omega \sin{(i+\varepsilon_{\Omega})}}\\
\nonumber &&\left\lbrace C_{20}\left(\frac{1}{2}+\frac{3}{4}e^2+\frac{15}{16}e^4\right)\cos{\varepsilon_{\Omega}}-C_{22}\left(\frac{7}{2}e-\frac{123}{16}e^3\right)\right.\\
\nonumber&&\left.\frac{1+\cos{\varepsilon_\Omega}}{2}\right\rbrace \sin{\varepsilon_{\Omega}}-\frac{5n^2 M C_{40}}{32 C \dot\Omega \sin{(i+\varepsilon_{\Omega})}}\frac{R^4}{a^2}\\
&& \left(\frac{1}{2}+\frac{5}{2}e^2+\frac{105}{16}e^4\right)(2\sin(2\varepsilon_{\Omega})+7\sin(4\varepsilon_{\Omega}))=0
\end{eqnarray}
In the limit case where $C_{40}=0$, our equation (\ref{peale}) is in agreement with Noyelles and Lhotka's equation, which has then the advantage to take into account the effect of the coefficient $C_{40}$ of the gravitational potential.

Noyelles and Lhotka (2013) have expanded their Eq. (21) at first order in the small $\varepsilon_{\Omega}$ (denoted $\varepsilon$ in their paper), to determine an explicit expression for $\varepsilon_{\Omega}$ (their Eq. 22). \textcolor{black}{Note that $\varepsilon_{\Omega}$ is defined as a clockwise angle in Noyelles and Lhotka (2013), as can be seen from their unnumbered equation after their Eq. (19), whereas it is defined as a counterclockwise here.} Our Eq. (\ref{eq26}) for $\varepsilon_{\Omega}$ does not match their Eq. (22) with $C_{40}=0$, despite the agreement between our Eq. (\ref{peale}) and their Eq. (21). This is because the development of their Eq. (21) in series of $\varepsilon$ converges slowly to the exact expression (we have checked that numerically: a relative difference of about $0.5\%$ arises at first ordering, for $\varepsilon\simeq1$ arcmin). By first multiplying their Eq. (21) by the factor $\sin(i+\varepsilon)$, the equation has a form closer to our Eq.(\ref{peale}), and the series in $\varepsilon$ converge faster (difference of $0.000005\%$ only). At first order, we then find an expression compatible with our equation and the one of Yseboodt and Margot (2006):
\begin{eqnarray}
 \nonumber \varepsilon_{\Omega}&=&-\left(1+\frac{2}{3}\frac{\dot\Omega}{n} \cos i+\frac{2}{3}\frac{\dot\omega}{n}\right)\left(\frac{C\dot\Omega \sin i}{nM_{m}R^2}\right)\textrm{{\LARGE /}} \\
 \nonumber &&\left[2C_{22}\left(\frac{7}{2}e-\frac{123}{16}e^3\right)-C_{20}\left(1+\frac{3}{2}e^2+\frac{15}{8}e^4\right)\right.\\
 \nonumber &&\left.+\frac{R^2}{a^2}C_{40}\left(\frac{5}{2}+\frac{25}{2}e^2+\frac{525}{16}e^4\right)\right.\\
 \label{Eq91} &&\left.+\left(1+\frac{2}{3}\frac{\dot\Omega}{n} \cos i+\frac{2}{3}\frac{\dot\omega}{n}\right)\frac{C}{nM_{m}R^2}\dot\Omega \cos{i}\right]
\end{eqnarray} 
Note that, instead of a very small term proportional to $(\dot\Omega/n)^2 \sin i$, this corrected solution now contains a larger term proportional to $(\dot\Omega/n) \cos i$ at the denominator. The difference between the two terms is the cause of a $450$ mas offset in obliquity with respect to Yseboodt and Margot's formula, for a given polar moment of inertia. Conversely, compared to the corrected Eq. (\ref{Eq91}), the use of Eq. (22) of Noyelles and Lhotka (2013) to interpret a measured obliquity leads to an overestimate of the polar moment of inertia by $0.4\%$. 

From the corrected Eq. (\ref{Eq91}), we estimate that neglecting $C_{40}$ leads to a negligible overestimate of the polar moment of inertia by $1.5 \times 10^{-7}\%$, in agreement with the conclusions of Noyelles and Lhotka (2013). In their effort to derive an expression for the mean equilibrium obliquity, Noyelles and Lhotka (2013) do not find any contribution from the gravitational coefficient $C_{30}$. However, they do find an effect of $C_{30}$ when they allow the semi-major axis to be a time dependent quantity and integrate numerically the non-averaged Hamilton's equations. Since $C_{30}=-1.26\times 10^{-5}$ (Mazarico et al., 2014), the mean obliquity would be $270$ mas larger than expected from the analytical modeling, for a given polar moment of inertia. This effect would lead in turn to overestimate the polar moment of inertia by $0.22\%$ when interpreting a measured obliquity with Eq. (\ref{Eq91}).

\subsection{On the nutation}

In the Hamiltonian of Peale et al. (1974), there are two terms that would lead to nutations induced by the perihelion precession. Peale (1974) has neglected these terms since the driving torque is proportional to the eccentricity at third order (see our factor $\kappa_{\omega}$, Eq. \ref{kappaomega}). However, the main torque also contains a term of third order in eccentricity which is never neglected in the literature. The first nutation term in the Hamiltonian is proportional to $G_{213}(e)=\frac{53e^3}{16}+\dots$ and corresponds to the secondary torque we find within the angular momentum approach. The other term is proportional to $G_{225}(e)$, which is of the fifth order in $e$. We have not found the corresponding very small torque, since we have considered an expansion up to the third order in eccentricity to derive the secondary torque. Considering expansion of higher order would go against our search for a compact analytical solution, without improving the accuracy of the solution in a useful way. 

More recently, Peale et al. (2016) have showed numerically that the projection of the spin vector onto the orbital plane is time-dependent and describes two circles of 0.87 arcsec of radius for each precession of the argument of pericenter, in agreement with our estimates for $\varepsilon_{\omega}$ (see Fig. (\ref{nut}), Section \ref{Section2}). They have argued that this is considerably less than the current 5 arcsec uncertainty in the spin axis orientation and they have not paid further attention to it. However, this nutation explains about half of the measured deviation. 

Noyelles and Lhotka (2013) have concluded that the secular variations of the orbital elements (including the secular variation of the pericenter argument) alter the obliquity by $10$ mas ($0.008\%$) after an interval of $20$ years. However, it is more relevant to estimate the difference between the equilibrium obliquity (which can be seen as constant on very short timescales) and the mean obliquity, due to the secular variations of the orbital elements, in order to assess their effect. From Section \ref{Section2}, we estimate that the nutations induced by the pericenter precession have an effect $\Delta \varepsilon$ of $\simeq 160$ mas ($0.13\%$) at J2000, one order of magnitude larger than the effect stated by Noyelles and Lhotka (2013).

Yet, some numerical results presented in Noyelles and Lhotka (2013) can be understood in terms of this nutation. From the frequency analysis of their numerical integration of the equations of Hamilton (see their table 2 where the notation $\omega$ is used for the pericenter longitude instead of the usual notation $\varpi$) it can be seen that both the difference between the inertial obliquity and inclination ($\theta-i\simeq\varepsilon$) and between the equatorial and orbital node longitudes ($\psi-\Omega\simeq \tilde \delta/\theta$) have a term with frequency $2\varpi-2\Omega=2\omega$, which is our nutation frequency. The amplitude of those terms also corresponds to our nutation amplitude. Dufey et al. (2009) also report some latitudinal librations induced by the pericenter precession (see the first line of their Tab. 3), but with an amplitude half of ours. 

Besides that nutation, Noyelles and Lhotka (2013) have found other variations of higher and smaller amplitudes. Those terms are related to frequencies issued from an orbital frequency analysis where the node and pericenter precession are assumed to consist of two superimposed motions with respect to the ecliptic plane at J2000, instead of a uniform motion with respect to the Laplace plane. Such a decomposition certainly improves the fit to astrometric measurements, but the additional spin nutations may have no physical meaning in practice, since the time-span of astrometric observation is very short compared to the precession timescales.

\subsection{On the effect of tidal deformations}

\label{noyellestides}
\label{pealeapproach}

Peale (2005) and Peale et al. (2014) introduce a tidal torque to study the evolution of the spin towards its present equilibrium orientation. They only need the part related to the time delay (and proportional to $k_2\zeta$) to do so. Here, as we aim to study the effect of tides on the equilibrium orientation itself, the term proportional to $k_2$ in Eq. (\ref{Eq72}) cannot be ignored. 

Noyelles and Lhotka (2013) aim to study the effect of elastic tidal deformations on the equilibrium obliquity $\varepsilon_{\Omega}$. While we find an effect of $300$ mas (see Section \ref{Section3}), they find a ten times smaller effect of 30 mas (see their table 4). This difference can be explained by a thorough comparison of both approaches. 

They present the effect of elastic deformations on the obliquity as being only variations of small period, with peak-to-peak variations of $30$ mas (because they have neglected $S_{22}$, $C_{21}$ and $S_{21}$, we expect that they have underestimated the peak-to-peak variations by a factor 2). However, as we have seen above, the main effect of elastic tides is to add a constant term to the mean obliquity, shifting it by about $300$ mas. We have chosen to average over the small periods and, as a result, over the variations of small amplitude they introduce, since we have aimed to keep the model analytically manageable. Noyelles and Lhotka (2013) certainly have reproduced that shifting of the mean value while numerically evaluating the obliquity variations (see their Fig. 4), but without realizing it.

An estimation of the periodic variations of the obliquity due to the tidal deformations is relevant, as it may affect the determination of the polar moment of inertia (a variation of $30$ mas in obliquity shifts the polar moment of inertia by $0.025\%$). However, the estimation by Noyelles and Lhotka (2013) is inaccurate for two reasons. First, they limit the effect of tidal deformations on the equilibrium obliquity to the first order in eccentricity, which is inappropriate because of the large eccentricity of Mercury. Secondly, in their expression for Mercury's gravity field (see their unnumbered equation at the beginning of their section 2.4), they neglect the effect of $S_{22}$, $C_{21}$ and $S_{21}$. This is correct in the rigid case where those coefficients are zero. However, in the non-rigid case, this amounts to consider only the radial bulge but not the obliquity and librational bulges. 

For synchronous satellites, only the part of the tidal deformations related to the obliquity and the variations of $C_{21}$ influences the equilibrium obliquity $\varepsilon_{\Omega}^{k_2}$, and not the radial and librational parts (Baland et al. 2016). Here, within the 3:2 resonance, all tidal deformations contribute to the change in equilibrium obliquity. The individual effects of $C_{20}^{h,peri}, C_{22}^{h,peri}, S_{22}^{h,peri}, C_{21}^{h,peri}$ and $S_{21}^{h,peri}$ are $-0.12, -0.46, -0.02, 1.46$ and $0.14\, \Delta \varepsilon_{\Omega}$, respectively. 

An updated numerical estimation of the periodic variations in obliquity due to the tidal deformations is needed, but is beyond the scope of this paper.

\subsection{On the choice of the reference plane}
\label{arg}

Our choice of the reference plane as the Laplace plane is different from the one of Peale (2005) and Peale et al. (2014, 2016) who chose the orbital plane as a reference plane and aim, within the angular momentum formalism, to study the forced precession. A positive aspect of their choice is that the gravitational torque (not averaged over the pericenter angle $\omega$) is written straightforwardly in a compact form (see their Eq. 7), while we need to implement a two-step procedure to reach the compact formulation of the right hand side of Eq. (\ref{AFinal}). On the other side, they need to numerically integrate the resulting angular momentum equation (their Eq. 3), while we managed to obtain an analytical solution which highlights the role of the pericenter precession on the spin behavior.

Peale (2005) aims to study the free precession and assumed that no forced precession occurred and as a result that the orbital plane was inertial. This assumption is questionable, since it leads to a mistake in the computation of the free precession rate of Mercury. In fact, by averaging Eq. (\ref{step1}) over the forcing period, we can find that the frequency of the free precession is given by $\omega_f=\kappa/C$ (see Eq. \ref{wf} ). Peale (2005) finds that (using our notations here) $\omega_f=\sqrt{\kappa^2-\kappa_{\omega}^2}/C$. This is because, at the same time, he sets $\Omega=0$ and $i=0$ and chose to consider that the pericenter direction was the X-axis of the inertial plane, so that $\omega=0$ is a constant angle, while keeping $e\neq0$. However, setting $\Omega=0$ does not amount to the same as averaging over $\Omega$ (for instance, the cosine of the node longitude would be 1 instead of 0 on average). To be convinced of this, one can use those values and wrongly replace $\hat n$ by $(0,0,1)$ and $\hat p$ by $(1,0,0)$ so that Eq. (\ref{AFinal}) becomes $C\hat s=\kappa (s_y,-s_x,0)+\kappa_{\omega} (-s_y,-s_x,0)$, leading to $\omega_f=\sqrt{\kappa^2-\kappa_{\omega}^2}/C$, as found by Peale. This mistake introduces a shift of $-0.03$ years in the free precession period.

\section{Discussion and conclusions}

We have developed an improved model for the Cassini state of Mercury, seen as a solid body, motivated by the possible detection of a small deviation from the classical Cassini state. Our model describes 1) the nutations of the spin axis about its mean orientation induced by the pericenter precession and 2) the changes in mean obliquity and deviation associated to tidal deformations. We have analytically solved the angular momentum equation, averaged over short periods, highlighting the respective roles of the pericenter precession and tidal deformations on the spin precession behavior (obliquity and deviation). 

The model is fully described by three parameters: 1) the normalized polar moment of inertia $C/M_{m}R^2$, which defines the mean obliquity and the nutation amplitude (the nutation induces some time variations in obliquity and in deviation with respect to the Cassini plane), 2) the tidal Love number $k_2$, which determines the constant shift over time in mean obliquity due to the tidal deformations, and 3) $k_2/Q$, the ratio of the Love number and of the tidal quality factor, which induces the additional constant deviation over time due to non-elasticity. 

We have assessed the consequences of our model improvements on the determination of the polar moment of inertia from the recently measured orientation of Stark et al. (2015b). The refinements we propose allow to avoid a systematic error of $+0.1\%$ on $C/M_{m}R^2$, essentially because we do not neglect the pericenter precession and tidal deformations. This systematic error is below the actual precision on the polar moment of inertia ($3-4\%$, corresponding to a $5$ arcsec precision for an obliquity of about $2$ arcmin) but may be of the order of precision that can be reached with the BepiColombo mission ($\leq 0.3\%$, or $\leq 0.5$ arcsec, Cical\`{o} et al. 2016). We also avoid other approximations, such as the truncation of the mean obliquity at third order in eccentricity or the linearization of the orbital and spin orientations parameters, which have smaller effects on the determination of the polar moment of inertia (see Tab. \ref{tab4}).

In our search for a compact analytical solution, we have nevertheless been forced to make some approximations. We have assessed the internal accuracy of our analytical solution, obtained by a perturbative approach from an averaged angular momentum equation derived with a two-step procedure which distinguishes between the main precession and the small nutations and tidal effects. From a comparison to the numerical integration of an alternative form of the averaged angular momentum equation, we estimated that interpreting measurements with our analytical solution implies a systematic error of $+0.015\%$ in the determination of $C/M_{m}R^2$. This is one order of magnitude below the combined effect of nutations and tides and the future expected precision. 

The cumulated effect of other approximations may reach the future precision. The most important is our averaging of the angular momentum equation over short periods. The periodic effect of tidal deformations could alter the determination of the polar moment of inertia by $0.025\%$ (because of a $30$ mas effect on the $2$ arcmin obliquity, see Noyelles et al., 2013). We have also neglected the polar motion and the short-period nutations which could have an impact of $0.07\%$ (Noyelles et al. 2010, $80$ mas) and of $0.02\%$ (Dufey et al., 2009, $20$ mas), respectively. 

Our independent derivation of the solution for the mean equilibrium obliquity has helped clarifying the different and mutually incompatible versions of Peale's equation presented in the literature, allowing us to avoid mistakes of up to $+0.8\%$. The difference of $+0.9\%$ between the estimation $C/M_{m}R^2=0.346\pm0.011$ by Stark et al. (2015b) and our estimation $0.3433\pm0.0134$ is due to that mistake of $+0.8\%$ and to the cumulated error of $+0.1\%$ made by neglecting the pericenter precession and tidal deformations. 

We cannot constrain $k_2$ from the measured spin orientation, since its effect on the mean obliquity is $0.35$ arcsec, one order of magnitude below the actual measurement precision ($5$ arcsec). In the future, even if the measurement precision falls below $0.35$ arcsec, a determination of the tidal Love number from spin measurement would not be possible because of the large correlation between this parameter and $C/M_{m}R^2$, which both participate to the obliquity. However, it is necessary to include its effect in the model to avoid a systematic error of $0.3\%$ on the determination of the polar moment of inertia. Theoretical estimations or actual estimates of $k_2$ from MESSENGER data can be used to that end. 

\textcolor{black}{The effect of $k_2/Q$ (about $1$ arcsec) can be separated from the one of $C/M_{m}R^2$ (about $0.8$ arcsec) in deviation, since $C/M_{m}R^2$ is well determined thanks to its major contribution to the mean obliquity. Because of the limited accuracy on the spin orientation, we only put a loose constraint on the ratio $k_2/Q$ ($0.00563\pm0.01651$), corresponding to an upper limit of about $0.02$ on the ratio $k_2/Q$ and of about $350$ on $Q$ (assuming $k_2=0.5$) at the $1\sigma$ level, in agreement with the upper limit placed by the current estimates for the Earth, Moon, and Mars)}. The large uncertainty $(300\%)$ exceeds the inaccuracy of our analytical solution related to the use of a perturbative approach ($1.5\%$). The possible effects of the averaging over short periods ($30$ mas for the periodic tides or $3\%$ of the $1$ arcsec tidal deviation, $20$ mas or $2\%$ for the short-period nutations, $80$ mas or $8\%$ for the polar motion) are also below the $300\%$ uncertainty. If the precision on the orientation of the spin axis is improved by a factor ten in the future (accuracy on the deviation of $0.3$ arcsec instead of $3$ arcsec), $k_2/Q$ could be constrained to the $30\%$ level, providing essential information about the internal structure of Mercury, as it would be the first determination for this planet. 

\textcolor{black}{We have tested the use of the orbital parameters as derived by Stark et al. (2015a), instead of the orbital parameters derived in Appendix A. The solutions for $C/MR^2$ and $k_2/Q$ decrease by $0.015\%$ and $9.8\%$, respectively, below the expected values of $\pm0.3\%$ and $\pm 30\%$ of the future uncertainties on the measured obliquity and deviation (Cical\`{o} et al. 2016).}

\textcolor{black}{A caveat of the precession model developed in this paper is that we assume Mercury to be entirely solid. Measurements of tides and longitudinal librations at a period of 88 days indicate that the core of Mercury is at least partially liquid, decoupling the solid mantle and crust from the interior (Margot et al. 2007, 2012). Peale et al. (2016) have shown that the internal torques resulting from the presence of a fluid outer core and a solid inner core might increase the equilibrium obliquity and lead to an overestimate of the polar moment of inertia, but they do not properly study the effect of the pericenter precession and of non-elastic tidal deformations. In order to prepare the interpretation of future BepiColombo rotation measurements, it will be necessary to develop a Cassini state model that takes into account all the effects of pericenter precession, tidal deformations, conservative torques related to the presence of an outer fluid core and of an inner solid core, dissipative core-mantle boundary viscous coupling, on the equilibrium orientation of the mantle spin axis, and on the estimation of the polar moment of inertia and of the imaginary part of the tidal Love number.}

\section*{Acknowledgments}

We thank Alexander Stark and Beno\^{\i}t Noyelles for their valuable comments on the first version of the manuscript which helped to improve the paper. Our gratitude also goes to Alexander Stark and Jean-Luc Margot for fruitful discussions. R.-M. Baland is funded by the Interuniversity Attraction Poles Programme initiated by the Belgian Science Policy Office through the Planet Topers alliance. The research leading to these results has received funding from the Belgian PRODEX program managed by the European Space Agency in collaboration with the Belgian Federal Science Policy Office.

\appendix

\section{\textcolor{black}{Orbital parameters and determination of the Laplace plane}}
\label{AppParamOrb}

\textcolor{black}{The values for the orbital parameters of Mercury used in this paper are derived from the DE431 ephemeris provided by the HORIZONS Web-Interface (http://ssd.jpl.nasa.gov/horizons.cgi, see also Folkner et al. 2014) under the form of time series for the orbital elements in the ICRF (International Celestial Reference Frame). We consider a time span of $\pm 500$ years about the J2000 epoch, with a time step of 10 years.}

\textcolor{black}{For the mean motion $n$, the eccentricity $e$, and the semi-major axis $a$, we consider the mean value and standard deviation of their respective time series:
\begin{eqnarray}
 n&=&(4.092345556\pm0.000011374)^\circ\textrm{/day}\\
 e&=&0.2056318\pm 0.0000622\\
 a&=&(5.790907\pm0.000011) \times 10^7\, \textrm{km}
\end{eqnarray}
Even though this simple approach tends to result in uncertainties larger than those obtained by Stark et al. (2015a), who decompose the orbital elements into a quadratic polynomial for the long-term behavior and a sum of periodic terms for the short-term behavior, ensuring a better fit to the data, these uncertainties have little effect ($0.007\%$ for $e$ and $0.003\%$ for $n$) on the determination of the polar moment of inertia from the measured obliquity, using Eq. (\ref{CCS}).}

\textcolor{black}{The orbit pole is moving in space. Ideally, we are looking for the orientation of a fixed axis, called the Laplace pole and perpendicular to the Laplace plane, which is the axis of a cone swept at a constant rate by the orbit normal, the semi-aperture of this cone being the constant inclination of the orbit with respect to the Laplace plane. However, such an axis does not exist, as the orbital precession is not a regular motion in reality (the precession rate and the inclination are not constant). Instead, we search for the fixed axis which minimizes the variations in orbital inclination with respect to Laplace plane, and we use the regular motion defined by the fitted cone as the orbit precession, neglecting the small variations about it. The five quantities needed to define that cone and the regular orbit precession are the Laplace plane equatorial coordinates $\alpha_{LP}$ and $\delta_{LP}$ with respect to the ICRF, the constant inclination $i$ of the orbit with respect to the Laplace plane, the precession rate $\dot\Omega$ and the longitude $\Omega_0$ of the ascending node of the orbit with respect to the Laplace plane at the J2000 epoch (see Fig. \ref{appendixfig}). We find that
\begin{eqnarray}
 \alpha_{LP}&=&(273.811048\pm0.324494)^\circ\\
 \delta_{LP}&=&(69.457475\pm0.259017)^\circ\\
 i&=&(8.533019\pm0.282935)^\circ\\
 \dot\Omega&=&(-0.1105948\pm0.0036399)^\circ\textrm{/cy}\\
 \Omega_0&=&(23.730329\pm0.303925)^\circ.
\end{eqnarray}
Our determination of the Laplace plane and of the orbital precession is consistent with the determinations of Stark et al. (2015a). The precession period is $325,\!513\pm10,\!713$ years. The correlation between $ \alpha_{LP}$ and $\delta_{LP}$ is $-99.998\%$. } 

\textcolor{black}{The orientation of the orbit pole, defined by the equatorial coordinates $\alpha_{orb}$ and $\delta_{orb}$, can be deduced at any time from these five quantities, before to be linearized around J2000. We obtain:
\begin{eqnarray}
\alpha_{orb}&=&(280.987906\pm0.000009)^\circ+\dot\alpha\, T,\\
\delta_{orb}&=&(61.447794\pm0.000006)^\circ+\dot\delta\, T,\\
\dot\alpha_{orb}&=&(-0.0328007\pm0.0000029)^\circ/\textrm{cy},\\
\dot\delta_{orb}&=&(-0.0048484\pm0.0000014)^\circ/\textrm{cy},
\end{eqnarray}
with $T$ the time interval in Julian centuries from J2000. Here, we have a better precision (about one order of magnitude) than Stark et al. (2015a), as a result of the different approach we use for the determination of the Laplace pole orientation and orbit precession. In particular, the uncertainties on the precession rates $\dot\alpha$ and $\dot\delta$ induce a $1\sigma$ thickness of the Cassini plane of about $0.025$ arcsec at the location of the spin axis, thinner than the estimation of $0.18$ arcsec by Stark et al. (2015a).} 

\textcolor{black}{
We can also deduce the products $\dot\Omega \sin i$ and $\dot \Omega \cos i$ that intervene in the definition of the mean obliquity (Eq. \ref{Eq88}):
\begin{eqnarray}
\dot\Omega \sin i&=&(-2.864081\pm 0.000240)\times 10^{6}/\textrm{y},\\
\dot \Omega \cos i&=&(-19.088758\pm 0.642402)\times 10^{6}/\textrm{y}.
\end{eqnarray}
Here, we also have a better precision than Stark et al. (2015a) and we find that the correlation between $\dot\Omega \sin i$ and $\dot \Omega \cos i$ is $5\times 10^{-3}$, which is small. It is generally accepted that the product $\dot \Omega \sin i$ is a quantity that can be estimated with a good precision and slightly depends on the chosen Laplace plane, contrary to $\dot \Omega \cos i$ (Yseboodt and Margot 2006, Stark et al. 2015a). It has to be noted that the determinations of $\dot\Omega \sin i$ and $\dot \Omega \cos i$ are weakly correlated, contrary to the determinations of $\dot \Omega$ and $i$ (Corr$(\dot \Omega,i)=99.9997\%$). The uncertainties on these products are the cause of an uncertainty of $0.016\%$ on the determination of the polar moment of inertia from the measured obliquity, using Eq. (\ref{CCS}). Cumulated to the effect of the uncertainties on $e$ and $n$, we consider that the effect of the uncertainties on the orbital parameters is about $0.03\%$. }

\textcolor{black}{Finally, as we study the long term nutations of Mercury, we need an estimation of $\omega$, the argument of the pericenter of Mercury around the Sun measured from the intersection of the orbital plane and the Laplace plane. This angle can be expressed as
\begin{equation}
\omega = \eta+\omega_{ICRF},
\end{equation}
with $\eta$, the angle from the ascending node of the orbit on the Laplace plane to the ascending node of the orbit on the equatorial plane of the ICRF, and $\omega_{ICRF}$, the argument of the pericenter of Mercury around the Sun with respect to the ICRF equator, as given by the chosen ephemerides. $\omega_{ICRF}$ is equal to $67.56^\circ$ at J2000, and has mainly a linear trend with a period of about $191$ kyr. The exact value of the period is not well known because the ephemeris covers a much shorter time span than one full cycle. Even for a perfectly regular orbit (with the orbital plane precessing regularly around a fixed Laplace pole), $\eta$ is not constant over time. Around J2000, its numerical value is about $17.18^\circ$. We use a linear time dependence relationship for the argument of the pericenter of Mercury with respect to the Laplace plane: 
\begin{eqnarray} 
\omega(t) &=& (50.379554\pm0.001140) ^\circ + \dot\omega \, T,\\
\dot\omega&=&(0.268943\pm0.000391)^\circ/\textrm{cy}.
\end{eqnarray}
The period of $\omega$ is about $134$ kyr. Using their own determination of the Laplace plane, Stark et al. (2015a) evaluate $\omega = 50.3895^\circ + 0.26855 ^\circ \, T$, which is close to our computation, but not entirely consistent, due to the difference in time dependence relationship (linear here, quadratic for the long-term behavior and periodic for the short-term behavior in Stark et al. 2015a).}

\section{Time-variable gravity field and moments of inertia}
\label{App2}

The hydrostatic part of the external degree-two gravitational potential of Mercury $ V^{l=2}$, exerted at a radial distance $r$, colatitude $\varphi$, and longitude $\lambda$, at time $t$, is the result of deformations induced by the tidal and centrifugal degree-two potentials $V_t$ and $V_c$:
\begin{eqnarray}
 \nonumber&& V^{l=2}(r,\varphi,\lambda,t)\\
 \nonumber&&=-\frac{GM_{m}}{r}\left(\frac{R}{r}\right)^2\sum_{m=0}^{2}(C_{2m}^h \cos m\lambda + S_{2m}^h \sin m\lambda)P^m_2(\cos \varphi)\\
 \nonumber&&=k_f \left(\frac{R}{r}\right)^3 (V_t^{stat}(R,\varphi,\lambda)+V_c^{stat}(R,\varphi,\lambda))\\
 \label{A2}&&\quad +k_2 \left(\frac{R}{r}\right)^3 V_t^{peri}(R,\varphi,\lambda,t-\Delta t),
\end{eqnarray}
where $G$ is the universal gravitational constant, $C_{2m}^h$ and $S_{2m}^h$ are second-degree hydrostatic gravity field coefficients, $k_f$ is the fluid Love number and $k_2$ is the real part of the tidal Love number, $P^m_2$ is the Legendre function of degree two and order $m$. 

The centrifugal potential is just static here, as we neglect the longitudinal librations and the polar motion (see Coyette et al. (2016) for complete expressions for a synchronously wobbling and librating satellite):
\begin{eqnarray}
 V_c^{stat}(r,\varphi,\lambda)&=&q_r \frac{GM_{m}}{3 R}\left(\frac{r}{R}\right)^2 P^0_2(\cos \varphi)
\end{eqnarray}
with 
\begin{equation} 
 \label{qr}q_r=\frac{9}{4}\frac{n^2R^3}{GM_{m}}.
\end{equation}

The tidal potential
\begin{eqnarray}
\nonumber V_{t}(r,\varphi,\lambda,t)&=&-\frac{GM_{sun}}{d}\sum_{l=2}^{\infty}\sum_{m=0}^{l}\left(\frac{r}{d}\right)^l(2-\delta_{m0})\frac{(l-m)!}{(l+m)!}\\
\nonumber\label{129}&&P_{l}^{m}(\cos{\varphi})P_{l}^{m}(\cos{\varphi_{sun}})\cos m(\lambda-\lambda_{sun})\\
\label{eq38}&&=V_t^{stat}(r,\varphi,\lambda)+V_t^{peri}(r,\varphi,\lambda,t)
\end{eqnarray}
is here divided into its static and periodic parts. $M_{sun}$ is the mass of Sun. $d$ is its distance to the center of Mercury, see Eq. (\ref{d}). $\lambda_{sun}$ and $\varphi_{sun}$ are the longitude and colatitude of the Sun, respectively, in the Body Frame of Mercury. In terms of Cartesian coordinates of the Sun, see Eq. (\ref{sunposition}), they can be written as $\lambda_{sun}=\arctan (Y/X)\simeq f-\frac{3}{2}M$ and $\cos\varphi_{sun}=Z$.

The periodic part of the tidal potential is evaluated at $t-\Delta t$ in Eq. (\ref{A2}), instead of $t$, to model the delay introduced by non-elasticity in the response of Mercury to the attraction of the Sun (see e.g. Williams et al. 2001 who follows the same approach to get the time-delayed moments of inertia of the Moon). In practice, the time delay is introduced by subtracting a phase shift $\zeta$ from the mean anomaly $M$ appearing in $\lambda_{sun}$, $\varphi_{sun}$ and $d$. $\zeta$ is approximated by the ratio of the imaginary part of the tidal Love number over its real part, or equivalently, by $1/Q$ with $Q$ the tidal quality factor. \textcolor{black}{Note that we assume here that the real and imaginary parts of the tidal Love number do not depend on the frequency of the excitation, which ranges from $n/2$ to $4n$ (see below, Eqs. \ref{c20}-\ref{ss21}).}

Including Eqs. (\ref{sunposition},\ref{f},\ref{d},\ref{phi1}) in Eq. (\ref{eq38}) and assuming $\psi=\Omega$, we find
\begin{eqnarray}
 \nonumber V_t^{stat}(r,\varphi,\lambda)&=&-q_t \frac{GM_{m}}{6R}\left(\frac{r}{R}\right)^2\left[\left(1+\frac{3}{2}e^2\right)P^0_2(\cos \varphi)\right.\\
 &&\left.-\frac{1}{2}\left(\frac{7}{2}e-\frac{123}{16}e^3\right)P_2^2(\cos \varphi)\cos 2\lambda\right],\\
 \nonumber V_t^{peri}(r,\varphi,\lambda,\zeta)&=& q_t \frac{GM_{m}}{3R}\left(\frac{r}{R}\right)^2(c_{20} P^0_2(\cos \varphi)+c_{22} P^2_2(\cos \varphi)\\
 \nonumber &&\cos 2\lambda+s_{22} P^2_2(\cos \varphi)\sin 2\lambda+c_{21} P^1_2(\cos \varphi)\\
 \label{B4} &&\cos \lambda+s_{21} P^1_2(\cos \varphi)\sin \lambda)
 \end{eqnarray}
at third order in orbital eccentricity $e$ and at first order in $\theta$ and $i$, with
\begin{eqnarray}
 \nonumber c_{20}&=&- \left[3e\left(\frac{1}{2}+\frac{9}{16}e^2\right)\cos (M-\zeta)\right.\\
 \label{c20}&&\left.+e^2\left(\frac{9}{4}\cos 2(M-\zeta) +\frac{53}{16}e\cos 3(M-\zeta)\right)\right]\\
 \nonumber c_{22}&=&\left[\left(\frac{1}{4}+\frac{3}{2}e^2\right)\cos (M-\zeta)+e \left(-\frac{1}{8}+\frac{53}{12}e^2\right)\right.\\
 &&\left.\cos 2(M-\zeta)+\frac{1}{192}e^3\cos 4(M-\zeta)\right]\\
 \nonumber s_{22}&=& \left[\left(-\frac{1}{4}+\frac{11}{4}e^2\right)\sin (M-\zeta)+e\left(\frac{1}{8}+\frac{841}{192}e^2\right)\right.\\
 &&\left. \sin 2(M-\zeta)-\frac{1}{96}e^3\cos 3(M-\zeta)\sin (M-\zeta)\right] \\
%  \end{eqnarray}
%  \begin{eqnarray}
 \nonumber c_{21}&=&- (\theta-i) \cos \left(\frac{M-\zeta}{2}\right)\left[\sin (M-\zeta+\omega )\right.\\
 \nonumber &&\left.+ e \left(2 \sin (M-\zeta+\omega )+\frac{3}{2} \sin (2 (M-\zeta)+\omega )\right.\right.\\
 \nonumber &&\left.\left.-\frac{1}{2}\sin \omega\right)+ e^2 \left(-\frac{19}{4} \sin (M-\zeta+\omega )\right.\right.\\
 \nonumber &&\left.\left.+\frac{25}{4} \sin (2 (M-\zeta)+\omega )+\frac{9}{4} \sin (3 (M-\zeta)+\omega )\right.\right.\\
 \nonumber &&\left.\left.+\frac{9}{4} \sin \omega\right)+e^3 \left(-\frac{79}{24}\sin ((M-\zeta)-\omega )\right.\right.\\
 \nonumber &&\left.\left.-\frac{1}{48}\sin (2 (M-\zeta)-\omega )+\frac{59}{12} \sin (M-\zeta+\omega )\right.\right.\\
 \nonumber &&\left.\left. -\frac{605}{48} \sin (2 (M-\zeta)+\omega )+\frac{243}{24} \sin (3 (M-\zeta)+\omega )\right.\right.\\
 &&\left.\left.+\frac{53}{16} \sin (4 (M-\zeta)+\omega )-\frac{155}{48} \sin \omega\right)\right]
 \end{eqnarray}
 \begin{eqnarray}
 \nonumber s_{21}&=& (\theta-i) \sin \left(\frac{(M-\zeta)}{2}\right) \left[\sin (M-\zeta+\omega )\right.\\
 \nonumber &&\left.- e \left(2 \sin (M-\zeta+\omega )-\frac{3}{2} \sin (2 (M-\zeta)+\omega )\right. \right.\\
 \nonumber &&\left.\left.+\frac{1}{2}\sin \omega\right)- e^2 \left(\frac{19}{4} \sin (M-\zeta+\omega )\right.\right.\\
 \nonumber &&\left.\left.+\frac{25}{4} \sin (2 (M-\zeta)+\omega )-\frac{9}{4} \sin (3 (M-\zeta)+\omega )\right.\right.\\
 \nonumber &&\left.\left.+\frac{9}{4} \sin \omega\right)+e^3 \left(\frac{79}{24}\sin ((M-\zeta)-\omega )\right.\right.\\
 \nonumber &&\left.\left.-\frac{1}{48}\sin (2 (M-\zeta)-\omega )-\frac{59}{12} \sin (M-\zeta+\omega )\right.\right.\\
 \nonumber &&\left.\left. -\frac{605}{48} \sin (2 (M-\zeta)+\omega )-\frac{243}{24} \sin (3 (M-\zeta)+\omega )\right.\right.\\
 \label{ss21} &&\left.\left.+\frac{53}{16} \sin (4 (M-\zeta)+\omega )-\frac{155}{48} \sin \omega\right)\right]
\end{eqnarray}
and 
\begin{equation} 
 \label{qt} q_t=-3\frac{GM_{sun}}{GM_{m}}\left(\frac{R}{a}\right)^3.
\end{equation}
Because of Kepler's third law ($GM_{sun}=n^2a^3$), we have $q_t=-\frac{4}{3}q_r$. 

\begin{figure}[!htb]
\begin{center}
\includegraphics[width=17cm]{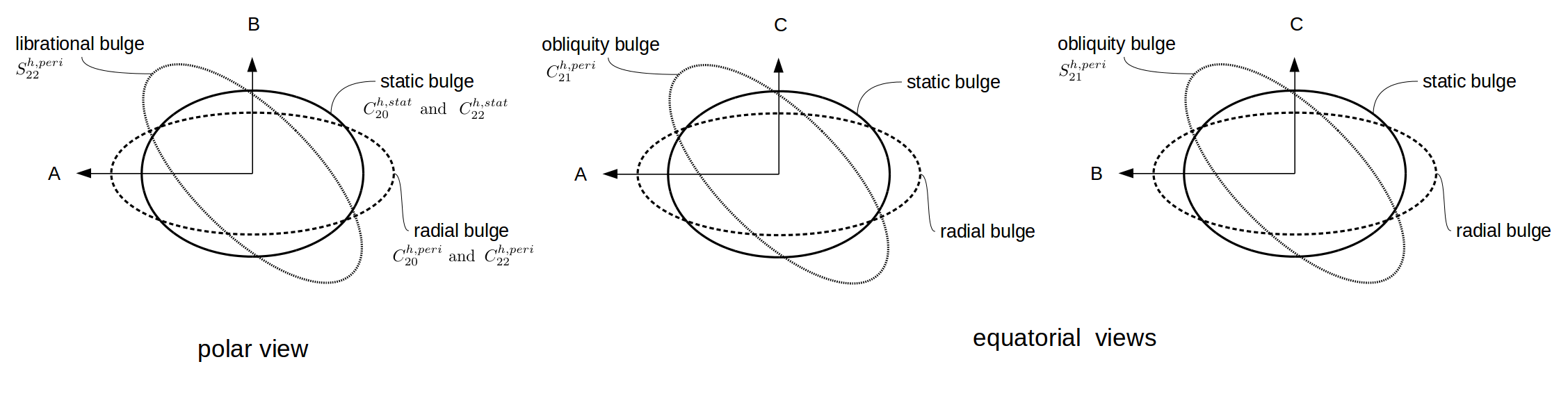}
\caption{\label{FigA2}Orientation of the periodical bulges with respect to the static bulge, seen from the equatorial plane (defined by the A and B axes) and from the planes defined by the moments of inertia $(A,C)$ and $(B,C)$. The radial, librational and obliquity bulges vary in amplitude over time. }
\end{center}
\end{figure}

Using Eqs. (\ref{A2}-\ref{B4}), expressions for the static and periodic parts of the hydrostatic gravity coefficients can be easily found. We find static parts
\begin{eqnarray}
 \label{C20hstat}C_{20}^{h,stat}&=&k_f\left(\frac{-2q_r+q_t\left(1+\frac{3}{2}e^2\right)}{6}\right)\\
 \label{C22hstat}C_{22}^{h,stat}&=&-k_f\frac{q_t}{12}\left(\frac{7}{2}e-\frac{123}{16}e^3\right)\\
 S_{22}^{h,stat}&=& 0\\
 \label{C21} C_{21}^{h,stat}&=&0\\
 \label{S21} S_{21}^{h,stat}&=&0
\end{eqnarray}
which are in agreement with Eq. (A17) of Matsuyama and Nimmo (2009). Below, for conciseness, we write the periodic part at first order in $e$ and with no approximation on $\zeta$, but do use the expansions correct up to order three in eccentricity and order one in $\zeta$ in further developments.
\begin{eqnarray}
 C_{20}^{h,peri}&=&\frac{1}{2}k_2 e q_t\cos (M-\zeta)\\
 C_{22}^{h,peri}&=&\frac{1}{24}k_2 q_t (-2\cos (M-\zeta) +e \cos 2 (M-\zeta))\\
 S_{22}^{h,peri}&=&\frac{1}{24}k_2 q_t (2\sin (M-\zeta) -e \sin 2 (M-\zeta))\\
 \nonumber C_{21}^{h,peri}&=& \frac{1}{6}k_2 q_t (\theta-i) \cos \left(\frac{M-\zeta}{2}\right)\left[-\sin (M-\zeta+\omega )\right.\\
 &&\left. +(2+4e)\sin(M-\zeta+\omega)+3e\sin(2M-2\zeta+\omega)\right]\\
 \nonumber S_{21}^{h,peri}&=& \frac{1}{6}k_2 q_t (\theta-i) \sin \left(\frac{M-\zeta}{2}\right)\left[-\sin (M-\zeta+\omega )\right.\\
\label{S21peri} &&\left. +(2-4e)\sin(M-\zeta+\omega)+3e\sin(2M-2\zeta+\omega)\right]
\end{eqnarray}
The periodic coefficients represent the gravitational contribution of the periodic tidal bulge, which can be divided into four components having fixed directions with respect to the static bulge, but time varying amplitudes. $C_{20}^{h,peri}$ and $C_{22}^{h,peri}$ represent the radial bulge, aligned with the static bulge. $S_{22}^{h,peri}$ represent the librational bulge, which differs by $45^\circ$ in the plane defined by the moment of inertia $A$ and $B$ from the orientation of the static bulge. $C_{21}^{h,peri}$ and $S_{21}^{h,peri}$ correspond to two obliquity bulges, which also differ by $45^\circ$ in the planes defined by the moments of inertia $(A,C)$ and $(B,C)$, respectively, from the orientation of the static bulge (see Fig. \ref{FigA2}).

\newpage

With the static and periodic gravitational coefficients of Eqs. (\ref{C20hstat}-\ref{S21peri}), we can express the time-varying components of the inertia tensor of Eq. (\ref{I}) as
\begin{eqnarray}
 \label{A14} C-A&=&M_e R^2 \left(-(C_{20}^{h,stat}+C_{20}^{h,peri})+2(C_{22}^{h,stat}+C_{22}^{h,peri})\right),\\
 C-B&=&M_e R^2 \left(-(C_{20}^{h,stat}+C_{20}^{h,peri})-2(C_{22}^{h,stat}+C_{22}^{h,peri})\right),\\
 \label{A16} B-A&=&4 M_e R^2 (C_{22}^{h,stat}+C_{22}^{h,peri}), \\ 
 D&=&-M_e R^2 S_{21}^{h,peri},\\
 \label{A18} E&=&-M_e R^2 C_{21}^{h,peri},\\
\label{A19} F&=& 2 M_e R^2 S_{22}^{h,peri}.
\end{eqnarray}
 
\section{Assessment of the accuracy of our analytical developments}
\label{App3}
\label{AppB1}
\label{AppB2}

To assess the accuracy of the perturbative approach, we compare the $x$ and $y$ components of the analytical solution $\hat s$ for the rigid or non-rigid cases (Eq. \ref{full} or \ref{sel}) with a numerical integration of the $x$ and $y$ components of the corresponding angular momentum equation (Eq. \ref{AFinal} or \ref{AFinalelas}, respectively), where $s_z$ is replaced by $\sqrt{1-s_x^2-s_y^2}$, as $\hat s$ is a unit vector. The initial condition of the numerical integrations are chosen to minimize the free precession amplitude around J2000.

\begin{figure}[!htb]
\begin{center}
 \includegraphics[width=8cm]{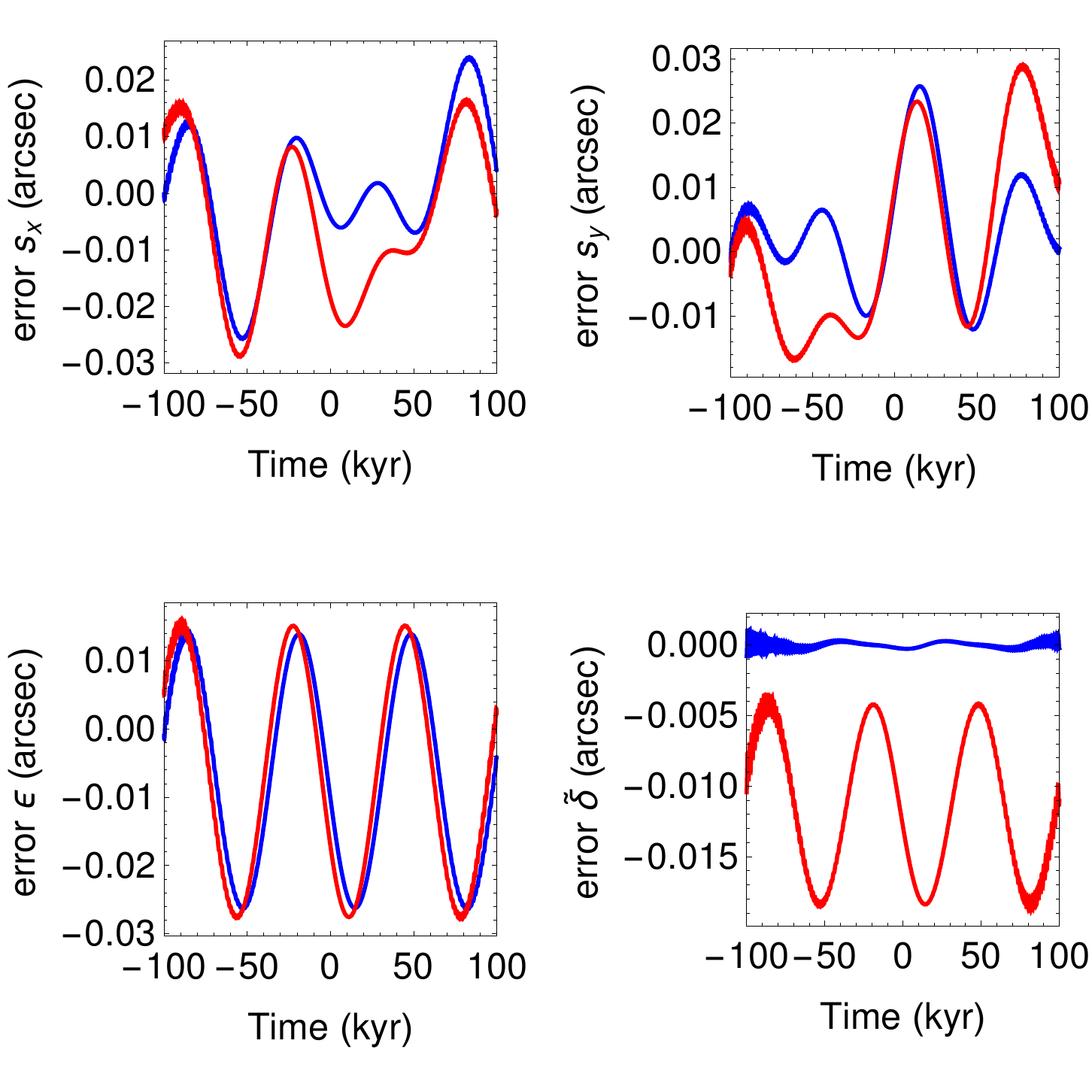}
\caption{\label{FigB1}Difference in the $x$ and $y$ components of the spin vector, obliquity, and deviation, between the analytical solution and a numerical integration of the angular momentum equation for the rigid (blue) and non-rigid (red) cases.}
\end{center}
\end{figure}

We find a maximum difference of $0.03$ arcsec in $s_x$ and $s_y$ over a very long integration time (from -$100$ kyr before J2000 to + $100$ kyr), and of $0.03$ arcsec in the time-variable obliquity. The maximum difference in deviation is $0.001$ arcsec and $0.020$ arcsec, for the rigid and non rigid case, respectively (see Fig. \ref{FigB1}). Around J2000, in the rigid case, the discrepancies in the obliquity and deviation are $-0.01$ arcsec ($-0.008\%$ of the obliquity of about $2$ arcmin) and $-0.00025$ arcsec ($-0.03\%$ of the tidal deviation of about $1$ arcsec), respectively. For the non-rigid case, the discrepancies in the obliquity and deviation are $-0.017$ arcsec ($-0.014\%$) and $-0.013$ arcsec ($-1.5\%$ of the tidal dissipation), respectively. 

Compared to the error of $0.16\%$ in obliquity arising when neglecting the nutations and tides with the classical Cassini state model, and the possible future precision in obliquity measurement with the BepiColombo mission ($\leq 0.3\%$), the $0.014\%$ accuracy in obliquity of our analytical solution is satisfactory. Our analytical solution also demonstrates a sufficient accuracy in deviation ($1.5\%$ in tidal deviation), compared to the classical Cassini state model which does not account for the deviation, and compared to the possible future precision in deviation measurements ($\leq 30\%$). 

As our averaged angular momentum equation (\ref{AFinal}) has been obtained after a two-step process, we also assess the accuracy of the angular momentum equation itself, through a comparison of its numerical integration to an integration of a variation of the averaged angular momentum equations of Peale et al. (2014), adapted to our case. For a non-librating solid rigid body with an external gravitational torque not averaged over the pericenter angle, Eq. (3) of Peale et al. (2014) can be written as
\begin{eqnarray}
\nonumber \frac{3}{2}n \frac{d p(t)}{dt}&=&\left(1-p^2(t)\right)N_x+p(t) q(t) N_y\\
\label{snif} &&-p(t) \sqrt{1-p^2(t)-q^2(t)}N_z\\
 \nonumber\frac{3}{2}n \frac{d q(t)}{dt}&=&-p(t) q(t) N_x-\left(1-q^2(t)\right)N_y\\
 \label{snif2} &&-q(t) \sqrt{1-p^2(t)-q^2(t)}N_z
\end{eqnarray}
where
\begin{eqnarray}
 \left(
\begin{array}{c}
 N_x \\
 N_y \\
N_z
\end{array}\right)&=&\frac{1}{C}\left(
\begin{array}{c}
\langle T_x\rangle\\
\langle T_y\rangle\\
0
\end{array}\right)-\frac{3}{2}n\, \hat \mu \wedge \hat k
\end{eqnarray}
and
\begin{eqnarray}
\nonumber \langle T_x\rangle&=&-\frac{3}{2}M_{m} R^2 n^2\left\lbrace -C_{20} q(t) \sqrt{1-p^2(t)-q^2(t)} G_{210}(e) \right.\\
\nonumber &&+C_{22} q(t) (1+\sqrt{1-p^2(t)-q^2(t)}) G_{201}(e)-C_{22} G_{213}(e)\\
&&\left.(1+\sqrt{1-p^2(t)-q^2(t)}) (q(t)\cos 2\omega +p(t) \sin 2\omega ) \right\rbrace\\
\nonumber \langle T_y\rangle&=&-\frac{3}{2}M_{m} R^2 n^2\left\lbrace -C_{20} p(t) \sqrt{1-p^2(t)-q^2(t)} G_{210}(e) \right.\\
\nonumber &&+C_{22} p(t) (1+\sqrt{1-p^2(t)-q^2(t)}) G_{201}(e)-C_{22} G_{213}(e)\\
&&\left.(1+\sqrt{1-p^2(t)-q^2(t)}) (-p(t)\cos 2\omega +q(t) \sin 2\omega ) \right\rbrace\\
\hat \mu&=&\left(
\begin{array}{c}
0\\
\dot \Omega \sin i\\
\dot \Omega \cos i
\end{array}\right)\\
\hat k &=& \left(
\begin{array}{c}
 p(t) \\
 -q(t) \\
\sqrt{1-p^2(t)-q^2(t)}
\end{array}\right).\\
\end{eqnarray}
Note that we correct here the sign of the third term of the x-component of the torque Eq. (7) of Peale et al. (2014).

The variables to solve for are
\begin{eqnarray}
 p(t)&=&\sin \varepsilon \cos \xi\\
 q(t)&=&\sin \varepsilon \sin \xi.
\end{eqnarray}
which are related to the obliquity $\varepsilon$ (denoted $i$ by Peale, not to be confounded with our angle $i$ for the inclination of the orbit with respect to the Laplace plane) and the deviation $\tilde \delta$ by the following relations
\begin{eqnarray}
 \varepsilon(t)&=&\arccos{\sqrt{1-p^2(t)-q^2(t)}}\\
 \tilde\delta(t)&\simeq& \varepsilon \sin\xi =\frac{\varepsilon(t)p(t)}{\sqrt{p^2(t)+q^2(t)}}.
\end{eqnarray}
where $\xi$ (see Fig. \ref{appendixfig}) is the angle between the node of the orbital plane on the Laplace plane and the node of the equatorial plane on the orbital plane (denoted $\Omega$ by Peale, not to be confounded with our angle $\Omega$ for the longitude of the ascending node of the orbit with respect to the Laplace plane). Since Peale et al. (2014) chose the orbital precessing frame as their reference frame (see Section \ref{arg}), the couple $\varepsilon$ and $\xi$, or equivalently $p$ and $q$, is sufficient to describe the spin precession.
$\hat k$ is the unit vector along the spin expressed in the orbital reference frame, while the unit vector along the spin axis $\hat s$ expressed in the Laplace reference frame is given by
\begin{equation}
 \hat s = (s_x,s_y,s_z)= R_z(-\Omega).R_x(-i). \hat k\,.
\end{equation}

\begin{figure}[!htb]
\begin{center}
 \includegraphics[width=8cm]{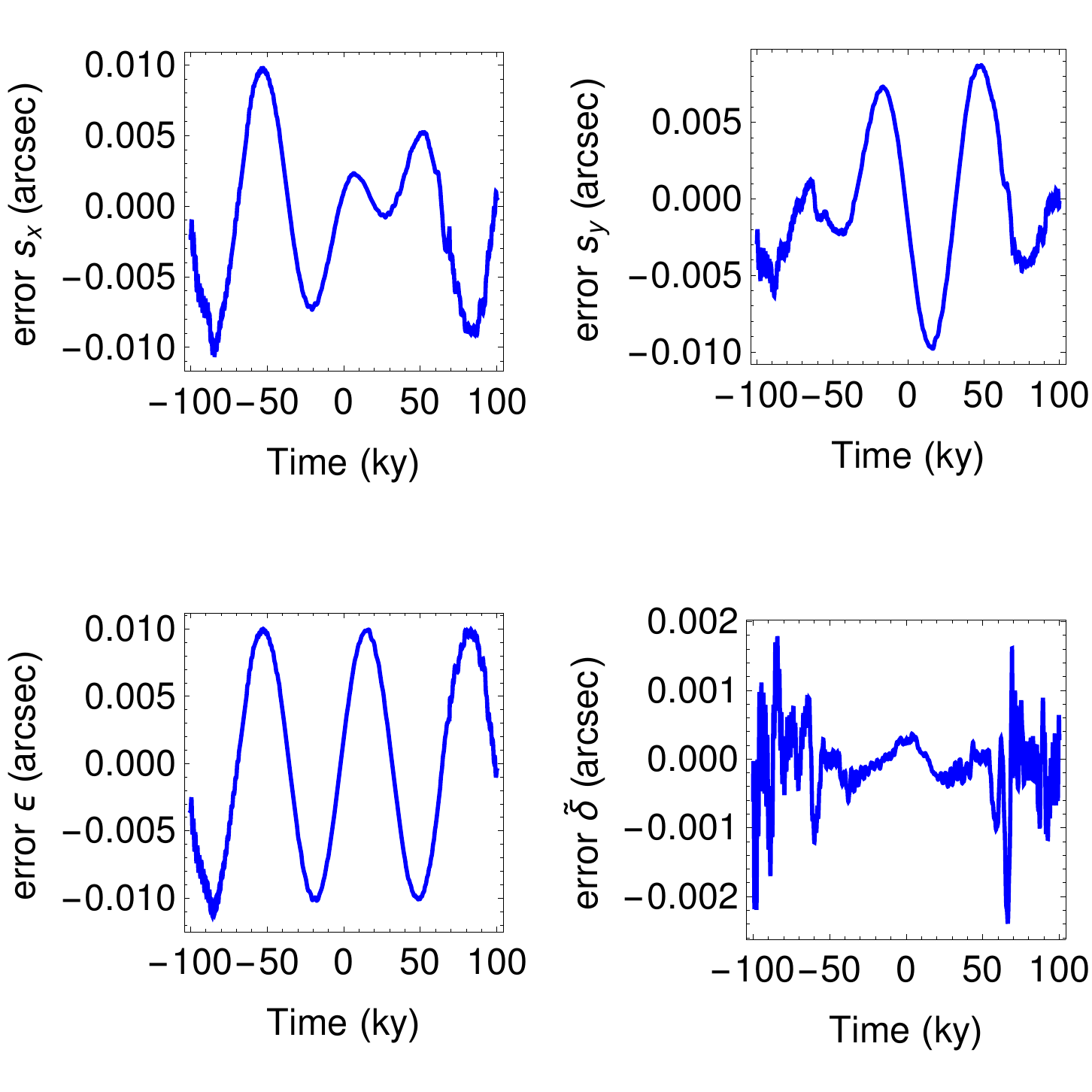}
\caption{\label{FigB3}Difference between the numerical integration of the angular momentum equation (Eq. \ref{AFinal}) and a numerical integration of the equations of Peale et al. (2014), to assess the accuracy of our two-step procedure to derive Eq. (\ref{AFinal}).}
\end{center}
\end{figure}

With respect to the numerical integration of Eq. (\ref{snif}-\ref{snif2}), the numerical integration of our angular momentum equation (\ref{AFinal}) differs by up to $0.01$ arcsec over long timescales (see Fig. \ref{FigB3}). About J2000, the discrepancies in obliquity and deviation are $+0.0018$ arcsec ($+0.0015\%$) and $0.00027$ arcsec ($+0.03\%$ of the tidal deviation), respectively. The two-step process to derive the angular momentum equation is therefore very accurate. We consider that the accuracy of our non-rigid angular momentum equation should be similar to the accuracy of our rigid angular momentum equation. 

We conclude that the internal accuracy of our non-rigid solution (which is the one we use to interpret data) is mainly affected by the use of the perturbative approach to solve analytically the angular momentum equation. We consider that interpreting measurements with our analytical solution implies a systematic error of about $+0.015\%$ in the determination of $C/M_{m}R^2$, which is more than two orders of magnitude below the actual precision on $C$ ($\sim 4\%$), and of about $+1.5\%$ on the ratio $k_2/Q$ (note that a small error on this ratio implies however a larger error on the tidal quality factor $Q$).

\newpage

\begin{table}[htp]
% \tiny
\begin{center}
\begin{tabular}{lccr}
\hline 
Parameter 					& Symbol & Value & \\
\hline 
Mean radius	 				& $R$ 			& $2440$ km			& \\
Mean density 					& $\bar \rho$		& $5430$ kg m$^{-3}$		& \\
Mass 						& $M_{m}$ 	& $3.30414\times 10^{23}$ kg 	&\\
Gravity coefficient degree two order zero 	& $C_{20}$ 		& $-5.03216 \times 10^{-5}$	& \\
Gravity coefficient degree two order two 	& $C_{22}$ 		& $0.80389\times 10^{-5}$	& \\
Gravity coefficient degree three order zero 	& $C_{30}$ 		& $-1.26094 \times 10^{-5}$	& \\
Gravity coefficient degree four order zero 	& $C_{40}$ 		& $-1.75473\times 10^{-5}$	& \\
Mean motion (at J2000)				& $n$ 			& $4.09235^{\circ}$ /day	& \\
Orbital eccentricity (at J2000)			& $e$ 			& $0.20563$			& \\
Semi-major axis					& $a$			& $5.7909\times 10^7$ km & \\
Right ascension of the Laplace pole with respect to ICRF 	& $\alpha_{LP}$ 	& $273.811^\circ$	& \\
Declination of the Laplace pole with respect to ICRF 		& $\delta_{LP}$ 	& $69.457^\circ$	& \\
Orbital inclination wrt the Laplace plane 			& $i$ 			& $8.5330^{\circ}$ 	& \\ 

Ascending node longitude rate (period: $325,\!513\pm10,\!713$ years) 	& $\dot\Omega$ 	& $-0.1105948^{\circ} /\textrm{cy}$& \\ 
Ascending node longitude at J2000 wrt the Laplace plane 		& $\Omega_0$ 	& $23.7303^\circ$	& \\

Pericenter argument rate (period: $134,\!477$ years)			& $\dot\omega$ 		& $0.268943^\circ /\textrm{cy}$	& \\
Pericenter argument at J2000 				& $\omega_0$ 		& $50.3796^\circ$ 	& \\
\hline
 \end{tabular}\\
\end{center}
\caption{\label{tab0}Numerical values of the different parameters used in the paper. The values of the mean radius and density, and therefore of the mass, are taken from Smith et al. (2012). We use unnormalized gravity coefficients, which are obtained by multiplying the coefficients of Mazarico et al. (2014) by $\sqrt{\frac{(2-\delta_{0m})(l-m)!}{(l+m)!}(2l+1)}$. The values for the orbital parameters are derived from an analysis of the DE431 ephemeris over 1000 years centered on J2000, obtained through the HORIZONS Web-Interface. They are consistent with values presented in Stark et al. (2015a). We have chosen the $x-$axis of the Laplace plane as the direction of the ascending node of the Laplace plane over the equatorial plane of the ICRF. }
\normalsize
\end{table}

\newpage

\begin{table}[htp]
 \tiny
\begin{center}
\begin{tabular}{lccccc}
\hline 
			& Margot et al. (2007) 	& Margot et al. (2012) 	& Mazarico et al. (2014) 		& Stark et al. (2015b) 			& Verma and Margot (2016)\\
\hline 
reference epoch ($ep$)	& nm & nm & nm & \textbf{J2000+4809 days}			& nm\\
$\alpha_{ep}$ 		& nm & nm & nm & $\mathbf{(281.00548\pm0.00088)^\circ}$	& nm\\
$\delta_{ep}$ 		& nm & nm & nm & $\mathbf{(61.4150\pm0.0016)^\circ}$	& nm\\
$\alpha_{J2000}$ 	& $\mathbf{281.0097^\circ}$ 	&$\mathbf{(281.0103}\pm0.0015)^\circ$	&$\mathbf{(281.00480\pm0.0054)^\circ}$	&$\mathbf{(281.00980\pm0.00088)^\circ}$	&$\mathbf{(281.00975\pm0.0048)^\circ}$\\
$\delta_{J2000}$	& $\mathbf{61.4143^\circ}$	&$\mathbf{(61.4155}\pm0.0013)^\circ$	&$\mathbf{(61.41436\pm0.0021)^\circ}$	&$\mathbf{(61.4156^\pm0.0016)^\circ}$		&$\mathbf{(61.41828\pm0.0028)^\circ}$\\
$\varepsilon_{J2000}$ (arcmin)		& $\mathbf{(2.11\pm0.1)}$ &$\mathbf{(2.04\pm0.08)}$ &$\mathbf{(2.06\pm0.16)}$ 	&$\mathbf{(2.029\pm0.085)}$ 			&$(1.88\pm0.16)$ \\
$\tilde \delta_{J2000}$	(arcsec)	& $0.08$ 	&$2.34\pm2.83 $ 	&$(-7.92\pm 9.13) $ 		&$(1.73\pm3.07)$ 		&$(4.39\pm8.41)$ \\
$C/M_{m}R^2$ 	& $0.357$			&$\mathbf{0.346\pm0.014}$	&$\mathbf{0.349\pm0.014}$			& $0.3437\pm\mathbf{0.011}$ (*)			& $0.318\pm0.028$	\\
\hline 
 \end{tabular}\\
\end{center}
\caption{\label{tab1} Values of the orientation parameters, obliquity, deviation and polar moment of inertia of Mercury, from five different studies. The values written in bold are taken directly from the studies. The abbreviation \textit{nm} is for the information not mentioned in the studies, and which cannot be retrieved from published information. Other values are derived here, using the published $(\alpha,\delta)$ at J2000 and specific values for the orbital parameters and gravity coefficients, or obtained via personal communication with J.-L. Margot in the case of Margot et al. (2012). For the sake of consistency, in the first four columns, the chosen values for the orbital parameters are the one used by the respective authors (that is to say: those of Margot et al. (2007) and Margot (2009) in the first three columns and those of Stark et al. (2015a) in the fourth column). For the last column, we chose to use the values of Margot at al. (2007) and (Margot 2009) since the authors of the last study did not defined their orbital parameters values. 
The chosen values for the gravity coefficients are those of Smith et al. (2012) for the first two columns, those of Mazarico et al. (2014) for the third and fourth columns, and those of Verma and Margot (2016) for the last column. (*) The nominal value of the estimate of the polar moment of inertia by Stark et al. (2015b) is corrected here (see Section \ref{Section2}). The value chosen for the orbital parameters may slightly affect the determination of the obliquity, deviation, and polar moment of inertia. In particular, it should be noted that if we use the orbital parameters values listed in Tab. \ref{tab0}, we find $\delta_{J2000}=1.83$ arcsec instead of the value ($1.73$ arcsec) reported in the fourth column, whereas $\varepsilon_{J2000}$ and $C/M_{m}R^2$ are affected only to the 5th decimal. This explains why $\delta_{J2000}$ in Tab. \ref{Tab2} is larger than $1.73$ arcsec.} 
\normalsize
\end{table}

\newpage
 
\begin{table}[htp]
\begin{center}
\begin{tabular}{ll}
\hline 
Parameter &Value \\
\hline 
\multicolumn{2}{l}{\textbf{Prior}} \\
\multicolumn{2}{l}{\textit{Interior parameters}} \\
$C/M_{m}R^2$ 		& $0.35\pm 0.1$\\
$k_2$				& $0.50\pm0.1$ \\
$k_2/Q$				& $0.005\pm0.05$ ($k_2=0.50\pm0.1$ and $Q=100\pm1000$)\\
\\
\multicolumn{2}{l}{ \textbf{Solution}} \\
\multicolumn{2}{l}{\textit{Interior parameters}} \\
$C/M_{m}R^2$ 		& $0.3433\pm 0.0134$\\
$k_2$				& $0.50\pm0.1$ \\
$k_2/Q$				& $0.00563\pm0.01651$ ($Q=89\pm261$)\\
\multicolumn{2}{l}{\textit{Corresponding amplitudes of improved non-rigid model (improved rigid model)}} \\
$\varepsilon^{k_2}_{\Omega}$ ($\varepsilon_{\Omega}$)	& $2.032 (2.026)\pm 0.080$ arcmin \\
$\varepsilon^{k_2}_{\omega}$ ($\varepsilon_{\omega}$) 	& $0.868 (0.863)\pm 0.034$ arcsec\\
$\varepsilon_{\zeta}$ 		& $0.995\pm 2.914$ arcsec\\
 \multicolumn{2}{l}{\textit{Corresponding spin orientation at J2000}} \\ 
$\alpha_{J2000}$ 	& $(281.00981\pm 0.00083)^\circ$\\
$\delta_{J2000}$ 	& $(61.41565\pm 0.00150)^\circ$\\
$\varepsilon$ at J2000	& $2.029\pm 0.080$ arcmin\\
$\tilde \delta$	at J2000	& $1.847\pm2.882$ arcsec \\
\hline 
\end{tabular}\\
\end{center}
\caption{\label{Tab2}Summary of our least squares inversion. The first and second parts of the table contain the prior values and the solution, respectively, for the parameters we solved for. Using the solution, we compute the corresponding obliquity and deviation amplitudes for the non-rigid and rigid models (third part). Finally, in the fourth part, we give the orientation parameters, obliquity, and deviation at J2000, corresponding to the solution.}
\normalsize
\end{table}

\newpage

\begin{table}[htp]
 \tiny
\begin{center}
\begin{tabular}{lll}
\hline 
Measurements uncertainties (MU)/ Systematic errors (SE)/ Mistakes (M) & Influence on $C/M_{m}R^2$ & Reference/Section\\
\hline 
 \qquad \qquad Uncertainty due to the obliquity &$ 3-4\%$& Margot et al. (2012), Stark et al. (2015b), \\
 && Section \ref{Inversion}\\
 \qquad \quad $+$ Uncertainty due to the orbital parameters & $ 0.03\%$ &Appendix \ref{AppParamOrb}\\
 \qquad \quad $+$ Uncertainty due to the gravitational coefficients & $0.005\%$ & Section \ref{Section44}\\ 
 \qquad \quad $+$ Uncertainty due to the tidal Love number & $0.01\%$ & Section \ref{Inversion}\\ 
(MU) $=$ Present uncertainty on the determination of $C/M_{m}R^2$ & $ 3-4\%$& \\
(MU) Future obliquity uncertainty (BepiColombo mission) & $\leq 1\%$ & Milani et al. (2001)\\
 & $\leq 0.3\%$ & Cical\`{o} et al. (2016)\\
\\
\textcolor{black}{(SE*) Neglecting the effect of a small inner core} & \textcolor{black}{$+4\%$} & \textcolor{black}{Peale et al. (2016) (new estimation needed)}\\ 
(SE) Neglecting the effect of tidal deformations on the mean obliquity & $+0.29\%$ & Section \ref{Inversion}\\
(SE) Neglecting the effect of nutations on the obliquity & $-0.13\%$ & Section \ref{Inversion}\\
(SE*) Neglecting polar motion & $0.07\%$ & Noyelles et al. (2010)\\ 
(SE) Truncation of $G_{201}(e)$ at fourth order in Eq. (\ref{Yseboodt}) & $-0.035\%$ &Section \ref{Section21}\\
(SE*) Neglecting the periodic effect of tidal deformations on the obliquity & $0.025\%$ & Noyelles and Lhotka (2013) (new estimation needed)\\
(SE*) Neglecting short-period nutations &$0.02\%$ & Dufey et al. (2009)\\
(SE*) Inaccuracy of our analytical solution & $+0.015\%$ &Appendix \ref{App3}\\
(SE) Linearization of the orbit and spin pole orientation parameters & $-0.0008\%$ & Section \ref{Section45}\\
(SE*) Neglecting $\dot \omega$ and $\dot \Omega$ in front of $n$ in Eq. (\ref{eq26}) & $+0.00007\%$ & Noyelles and Lhotka (2013) and Section \ref{appApoint4}\\
(SE*) First order development in $\varepsilon$ in Peale's equation & $+4 \times 10^{-6}\, \%$ & Section \ref{Section21}\\
(SE*) Neglecting $C_{40}$ & $+1.5\times 10^{-7}\, \%$ & Noyelles and Lhotka (2013) and Section \ref{Noyelleserror}\\
(SE*) Neglecting $C_{30}$ & $\simeq0\%$ (or $+0.2\%$?) & Noyelles and Lhotka (2013) (inconsistent results)\\
\\
(M) Using Eq. (37) of Stark et al. (2015a) & $+0.8\%$ & Section \ref{starkerror}\\
(M) Using Eq. (22) of Noyelles and Lhotka (2013) & $+0.4\%$ & Section \ref{Noyelleserror}\\
\hline
\end{tabular}\\
\end{center}
\caption{\label{tab4}Possible systematic errors (resulting from approximations in the modeling process) and mistakes (resulting from errors in the modeling process) on the Cassini state model, classified according to their effect on the determination of the polar moment of inertia from the measured orientation of the rotation axis of Mercury, and compared to actual and future measurement precision. The systematic errors denoted with an asterisk correspond to the approximations we made in our inversion, while we avoid the others. All those systematic errors and mistakes have an effect equal or below the present uncertainty ($4\%$) on the determination of $C/M_{m} R^2$. Their importance for the future depends on the precision that will be reached with the BepiColombo mission. }
\normalsize
\end{table}

% \newpage

\begin{table}[htp]
 \tiny
\begin{center}
\begin{tabular}{lll}
\hline
Measurements uncertainties (MU)/ Systematic errors (SE) & Influence on $k_2/Q$ & Reference/Section\\
\hline 
 \qquad \qquad Uncertainty due to the deviation &$ 300\%$& Section \ref{Inversion}\\
 \qquad \quad $+$ Uncertainty due to the orbital parameters & $ 3\%$ &Section \ref{Inversion}\\
 \qquad \quad $+$ Uncertainty due to the gravitational coefficients & $0.005\%$ &Section \ref{Inversion} \\ 
 \qquad \quad $+$ Uncertainty due to the tidal Love number & $0.02\%$ & Section \ref{Inversion}\\ 
(MU) $=$ Present uncertainty on the determination of $k_2/Q$ & $ 300\%$& \\
(MU) Future deviation uncertainty (BepiColombo mission) & $\leq 30\%$ & Cical\`{o} et al. (2016)\\
\\
(SE*) Neglecting the effect of viscous tidal torque at core boundaries & $+2-6\%$ & Peale et al. (2014) (new estimation needed)\\
(SE) Neglecting the effect of nutations on the deviation & $+86\%$ &Section \ref{Inversion}\ \\
(SE*) Neglecting polar motion & $8\%$ & Noyelles et al. (2010)\\ 
(SE*) Neglecting the periodic effect of tidal deformations on the deviation & $3\%$ & Noyelles and Lhotka (2013) (new estimation needed)\\
(SE) Linearization of the orbit and spin pole orientation parameters & $-2.3\%$ & Section \ref{Inversion}\\
(SE*) Neglecting short-period nutations &$2\%$ & Dufey et al. (2009)\\
(SE*) Inaccuracy of our analytical solution & $+1.5\%$ &Appendix \ref{App3}\\
\hline
\end{tabular}\\
\end{center}
\caption{\label{tab5}\textcolor{black}{Possible systematic errors (resulting from approximations in the modeling process) on the improved Cassini state model, classified according to their effect on the determination of $k_2/Q$ from the measured orientation of the rotation axis of Mercury, and compared to actual and future measurement precision. The systematic errors denoted with an asterisk correspond to the approximations we made in our inversion, while we avoid the others. All those systematic errors have an effect below the present uncertainty ($300\%$) on the determination of $k_2/Q$. Their importance for the future depends on the precision that will be reached with the BepiColombo mission. }}
\normalsize
\end{table}

\end{document}